
%
%
%
%
%
%


\magnification=1200
\hsize=31pc
\vsize=55 truepc
\hfuzz=2pt
\vfuzz=4pt
\pretolerance=5000
\tolerance=5000
\parskip=0pt plus 1pt
\parindent=16pt
%

%
%
\font\fourteenrm=cmr10 scaled \magstep2
\font\fourteeni=cmmi10 scaled \magstep2
\font\fourteenbf=cmbx10 scaled \magstep2
\font\fourteenit=cmti10 scaled \magstep2
\font\fourteensy=cmsy10 scaled \magstep2

%
\font\large=cmbx10 scaled \magstep1

%
\font\sans=cmssbx10

%
\def\bss#1{\hbox{\sans #1}}

%
\font\bdi=cmmib10
\def\bi#1{\hbox{\bdi #1\/}}

%
\font\eightrm=cmr8
\font\eighti=cmmi8
\font\eightbf=cmbx8
\font\eightit=cmti8

\font\eightsy=cmsy8
\font\sixrm=cmr6
\font\sixi=cmmi6
\font\sixsy=cmsy6

%
\def\tenpoint{\def\rm{\fam0\tenrm}%
  \textfont0=\tenrm \scriptfont0=\sevenrm
                      \scriptscriptfont0=\fiverm
  \textfont1=\teni  \scriptfont1=\seveni
                      \scriptscriptfont1=\fivei
  \textfont2=\tensy \scriptfont2=\sevensy
                      \scriptscriptfont2=\fivesy
  \textfont3=\tenex   \scriptfont3=\tenex
                      \scriptscriptfont3=\tenex
  \textfont\itfam=\tenit  \def\it{\fam\itfam\tenit}%
  \textfont\slfam=\tensl  \def\sl{\fam\slfam\tensl}%
  \textfont\bffam=\tenbf  \scriptfont\bffam=\sevenbf
                            \scriptscriptfont\bffam=\fivebf
                            \def\bf{\fam\bffam\tenbf}%
  \normalbaselineskip=20 truept
  \setbox\strutbox=\hbox{\vrule height14pt depth6pt
width0pt}%
  \let\sc=\eightrm \normalbaselines\rm}
\def\eightpoint{\def\rm{\fam0\eightrm}%
  \textfont0=\eightrm \scriptfont0=\sixrm
                      \scriptscriptfont0=\fiverm
  \textfont1=\eighti  \scriptfont1=\sixi
                      \scriptscriptfont1=\fivei
  \textfont2=\eightsy \scriptfont2=\sixsy
                      \scriptscriptfont2=\fivesy
  \textfont3=\tenex   \scriptfont3=\tenex
                      \scriptscriptfont3=\tenex
  \textfont\itfam=\eightit  \def\it{\fam\itfam\eightit}%
  \textfont\bffam=\eightbf  \def\bf{\fam\bffam\eightbf}%
  \normalbaselineskip=16 truept
  \setbox\strutbox=\hbox{\vrule height11pt depth5pt width0pt}}
\def\fourteenpoint{\def\rm{\fam0\fourteenrm}%
  \textfont0=\fourteenrm \scriptfont0=\tenrm
                      \scriptscriptfont0=\eightrm
  \textfont1=\fourteeni  \scriptfont1=\teni
                      \scriptscriptfont1=\eighti
  \textfont2=\fourteensy \scriptfont2=\tensy
                      \scriptscriptfont2=\eightsy
  \textfont3=\tenex   \scriptfont3=\tenex
                      \scriptscriptfont3=\tenex
  \textfont\itfam=\fourteenit  \def\it{\fam\itfam\fourteenit}%
  \textfont\bffam=\fourteenbf  \scriptfont\bffam=\tenbf
                             \scriptscriptfont\bffam=\eightbf
                             \def\bf{\fam\bffam\fourteenbf}%
  \normalbaselineskip=24 truept
  \setbox\strutbox=\hbox{\vrule height17pt depth7pt width0pt}%
  \let\sc=\tenrm \normalbaselines\rm}
\def\today{\number\day\ \ifcase\month\or
  January\or February\or March\or April\or May\or June\or
  July\or August\or September\or October\or November\or
December\fi
  \space \number\year}

%
\newcount\secno      
\newcount\subno      
\newcount\subsubno   
\newcount\appno      
\newcount\tableno    
\newcount\figureno   
%

%
\normalbaselineskip=20 truept
\baselineskip=20 truept

%
%
\def\title#1
   {\vglue1truein
   {\baselineskip=24 truept
    \pretolerance=10000
    \raggedright
    \noindent \fourteenpoint\bf #1\par}
    \vskip1truein minus36pt}
%

%
\def\author#1
  {{\pretolerance=10000
    \raggedright
    \noindent {\large #1}\par}}

%
\def\address#1
   {\bigskip
    \noindent \rm #1\par}

%
\def\shorttitle#1
   {\vfill
    \noindent \rm Short title: {\sl #1}\par
    \medskip}

%
\def\pacs#1
   {\noindent \rm PACS number(s): #1\par
    \medskip}

%
\def\jnl#1
   {\noindent \rm Submitted to: {\sl #1}\par
    \medskip}

%
\def\date
   {\noindent Date: \today\par
    \medskip}

%
\def\beginabstract
   {\vfill\eject
    \noindent {\bf Abstract. }\rm}

%
\def\keyword#1
   {\bigskip
    \noindent {\bf Keyword abstract: }\rm#1}

%
\def\endabstract
   {\par
    \vfill\eject}

%
%
\def\contents
   {{\noindent
    \bf Contents
    \par}
    \rightline{Page}}

%
\def\entry#1#2#3
   {\noindent
    \hangindent=20pt
    \hangafter=1
    \hbox to20pt{#1 \hss}#2\hfill #3\par}

%
\def\subentry#1#2#3
   {\noindent
    \hangindent=40pt
    \hangafter=1
    \hskip20pt\hbox to20pt{#1 \hss}#2\hfill #3\par}
\def\checkforsub{\futurelet\nexttok\decide}
\def\ssf{\relax}
\def\decide{\if\nexttok\ssf\let\endspace=\nospace
                \else\let\endspace=\extraspace\fi\endspace}
\def\nospace{\nobreak\par\nobreak}
%
%
\def\section#1{%
    \goodbreak
    \vskip50pt plus12pt minus12pt
    \nobreak
    \gdef\extraspace{\nobreak\bigskip\noindent\ignorespaces}%
    \noindent
    \subno=0 \subsubno=0
    \global\advance\secno by 1
    \noindent {\bf \the\secno. #1}\par\checkforsub}

%
\def\subsection#1{%
     \goodbreak
     \vskip24pt plus12pt minus6pt
     \nobreak
     \gdef\extraspace{\nobreak\medskip\noindent\ignorespaces}%
     \noindent
     \subsubno=0
     \global\advance\subno by 1
     \noindent {\sl \the\secno.\the\subno. #1\par}\checkforsub}

%
\def\subsubsection#1{%
     \goodbreak
     \vskip20pt plus6pt minus6pt
     \nobreak\noindent
     \global\advance\subsubno by 1
     \noindent {\sl \the\secno.\the\subno.\the\subsubno. #1}\null.
     \ignorespaces}

%
\def\appendix#1
   {\vskip0pt plus.1\vsize\penalty-250
    \vskip0pt plus-.1\vsize\vskip24pt plus12pt minus6pt
    \subno=0
    \global\advance\appno by 1
    \noindent {\bf Appendix \the\appno. #1\par}
    \bigskip
    \noindent}

%
\def\subappendix#1
   {\vskip-\lastskip
    \vskip36pt plus12pt minus12pt
    \bigbreak
    \global\advance\subno by 1
    \noindent {\sl \the\appno.\the\subno. #1\par}
    \nobreak
    \medskip
    \noindent}

%
\def\ack
   {\vskip-\lastskip
    \vskip36pt plus12pt minus12pt
    \bigbreak
    \noindent{\bf Acknowledgments\par}
    \nobreak
    \bigskip
    \noindent}


%

%
\def\tabcaption#1
   {\global\advance\tableno by 1
    \noindent {\bf Table \the\tableno.} \rm#1\par
    \bigskip}

%

%

%

%

%
\def\figures
   {\vfill\eject
    \noindent {\bf Figure captions\par}
    \bigskip}

%
\def\figcaption#1
   {\global\advance\figureno by 1
    \noindent {\bf Figure \the\figureno.} \rm#1\par
    \bigskip}

%

%
\def\numreferences
     {\vfill\eject
     {\noindent \bf References\par}
     \everypar{\parindent=30pt \hang \noindent}
     \bigskip}

%
\def\refjl#1#2#3#4
   {\hangindent=16pt
    \hangafter=1
    \rm #1
   {\frenchspacing\sl #2
    \bf #3}
    #4\par}

%
\def\refbk#1#2#3
   {\hangindent=16pt
    \hangafter=1
    \rm #1
   {\frenchspacing\sl #2}
    #3\par}

%
\def\numrefjl#1#2#3#4#5
   {\parindent=40pt
    \hang
    \noindent
    \rm {\hbox to 30truept{\hss #1\quad}}#2
   {\frenchspacing\sl #3\/
    \bf #4}
    #5\par\parindent=16pt}

%
\def\numrefbk#1#2#3#4
   {\parindent=40pt
    \hang
    \noindent
    \rm {\hbox to 30truept{\hss #1\quad}}#2
   {\frenchspacing\sl #3\/}
    #4\par\parindent=16pt}

%
\def\dash{---{}---}

\def\ref#1{\noindent \hbox to 21pt{\hss
#1\quad}\frenchspacing\ignorespaces}

%
\def\frac#1#2{{#1 \over #2}}

%

%
\def\d{{\rm d}}

%
\def\e{{\rm e}}


\chardef\ii="10

%
\def\case#1#2{{\textstyle{#1\over #2}}}

%
\def\etal{{\sl et al\/}\ }

%

\catcode`\@=11
\def\vfootnote#1{\insert\footins\bgroup
    \interlinepenalty=\interfootnotelinepenalty
    \splittopskip=\ht\strutbox 
    \splitmaxdepth=\dp\strutbox \floatingpenalty=20000
    \leftskip=0pt \rightskip=0pt \spaceskip=0pt \xspaceskip=0pt
    \noindent\eightpoint\rm #1\ \ignorespaces\footstrut\futurelet\next\fo@t}

%
%
\def\eq(#1){\hfill\llap{(#1)}}
\catcode`\@=12
%
%
\def\bhat#1{\hbox{\bf\^{\bdi #1}}}


%
%
\def\CQG{Classical Quantum Grav.}

\def\JPA{J. Phys. A: Math. Gen.}




\def\RPP{Rep. Prog. Phys.}

%
%

\def\APNY{Ann. Phys., NY}

\def\JMP{J. Math. Phys.}

\def\NC{Nuovo Cim.}

\def\NP{Nucl. Phys.}
\def\PL{Phys. Lett.}
\def\PR{Phys. Rev.}
\def\PRL{Phys. Rev. Lett.}

\def\RMP{Rev. Mod. Phys.}

%
%
\def\bnabla{\hbox{\bsy\char'162}}
%
%

%
\def\ms{\noalign{\vskip4pt plus3pt minus2pt}}
%

%

%
\def\gap{\;\lower3pt\hbox{$\buildrel > \over \sim$}\;}
%
%
\def\lap{\;\lower3pt\hbox{$\buildrel < \over \sim$}\;}
\def\tqs{\hbox to 25pt{\hfil}}


%
%
%
\def\LaTeX{L\kern-.26em \raise.6ex\hbox{\fiverm A}%
   \kern-.15em\TeX}%
\def\AmSTeX{%
{$\cal{A}$}\kern-.1667em\lower.5ex\hbox{%
 $\cal{M}$}\kern-.125em{$\cal{S}$}-\TeX}




\def\ie{{i.e.}}
\def\dddot#1{\mathord{\mathop#1^{\cdot\hskip-1pt\cdot\hskip-1pt\cdot}}}
\def\dprime{\mathaccent"707D}
\def\tprime#1{#1\kern-2.9pt\mathaccent 19 {\strut}\kern-1.6pt
   \mathaccent 19{\strut}\kern-1.6pt\mathaccent 19 {\strut}\kern5.1pt }

\def\math{\mathsurround 0pt}
\def\oversim#1#2{\lower.5pt\vbox{\baselineskip0pt \lineskip-.5pt
        \ialign{$\math#1\hfil##\hfil$\crcr#2\crcr{\scriptstyle\sim}\crcr}}}

\def\({\left(} \def\){\right)}
\def\[{\left[} \def\]{\right]}
\def\pa{\partial}
\def\frac#1#2{
  {\mathchoice{{\textstyle{#1\over #2}}}{#1\over #2}{#1\over #2}{#1 \over #2}}}
\def\half{{\mathchoice{{\textstyle{1\over 2}}}{1\over 2}{1\over 2}{1 \over 2}}}

\def\unit#1{\ifinner \; 
            \else \;\; \fi
            {\rm #1}}

\def\overleftrightarrow#1{\vbox{\ialign{##\crcr
    $\leftrightarrow$\crcr\noalign{\kern-1pt\nointerlineskip}
    $\hfil\displaystyle{#1}\hfil$\crcr}}}
\def\dbw{\overleftrightarrow\partial}
\def\Slash#1{\mathord{\not\mathrel{#1}}}

\def\hidehrule#1#2#3{\kern-#1 \hrule
     width #3 height#1 depth#2 \kern-#2}
\def\hidevrule#1#2#3{\kern-#1{\dimen0=#1
     \advance\dimen0 by#2 \vrule height#3 width\dimen0}\kern-#2}
\def\makeblankbox#1#2#3#4#5{\raise #5\hbox{
     \vbox{\hidehrule{#1}{#2}{#3}
     \kern-#1
     \hbox to #3{\hidevrule{#1}{#2}{#4}
     \vbox to #1{}
     \vtop to #1{}
     \hfil\hidevrule{#2}{#1}{#4}}
     \kern-#1\hidehrule{#2}{#1}{#3}}}}

\def\Box{\mathop{
     \makeblankbox{.4pt}{.0pt}{.6em}{.61em}{.2pt}}}

\def\al{\alpha}
\def\be{\beta}
\def\ga{\gamma}
\def\de{\delta}
\def\ep{\epsilon}
\def\ze{\zeta}
\def\et{\eta}
\def\th{\theta}

\def\ka{\kappa}
\def\la{\lambda}
\def\rh{\rho}
\def\si{\sigma}
\def\ta{\tau}

\def\ph{\phi}
\def\ch{\chi}
\def\ps{\psi}
\def\om{\omega}
\def\Ga{\Gamma}
\def\De{\Delta}

\def\La{\Lambda}
\def\Si{\Sigma}

\def\Om{\Omega}
\def\vp{\varphi}

\def\APJ{Ap. J.}
\def\CMP{Comm. Math. Phys.}
\def\GRG{Gen. Rel. Grav.}
\def\JETP{Sov. Phys. JETP}
\def\JETPL{Sov. Phys. JETP Lett.}
\def\MNRAS{Mon. Not. R. Ast. Soc.}

\def\MPL{Mod. Phys. Lett.}
\def\YF{Yad. Fiz.}
\def\ZETF{Zh. Eksp. Teor. Fiz.}
\def\ZETFPR{Zh. Eksp. Teor. Fiz. Pis'ma Red.}

\def\section#1{%
    \goodbreak
    \vskip50pt plus12pt minus12pt
    \nobreak
    \gdef\extraspace{\nobreak\bigskip\noindent\ignorespaces}%
    \noindent
    \subno=0 \subsubno=0
    \global\advance\secno by 1
    \eqnnumber=0                     
    \noindent {\bf \the\secno. #1}\par\checkforsub}


\newcount\eqnnumber
\def\eqqno{\global\advance\eqnnumber by 1
           \eqno(\the\secno.\the\eqnnumber)}
\def\label#1{\global\advance\eqnnumber by 1
             \global\edef#1{{\the\secno.\the\eqnnumber}}
             \eqno{(\the\secno.\the\eqnnumber)}
             }

\def\gobble#1{} 

\def\rf#1{\def\refstring{\expandafter\gobble\string#1}\ignorespaces
\expandafter\ifx\csname\refstring\endcsname\relax\ignorespaces
{\it\expandafter\gobble\string#1}\else{#1}\fi}


\def\ct#1{\csname CiteLabel#1\endcsname}

\def\buildcitelabel#1{                 
       \edef\doit{\global\def\csname CiteLabel#1\endcsname{\the\citeno}}\doit}

\def\stripspace#1 \delim{#1}           

\def\newloop#1#2\dontrepeat{\def\body{#1}\def\condition{#2}\again}
\def\again{\condition\let\next=\relax\else\body\let\next=\again\fi\next}
\let\dontrepeat=\fi

\openin0=allrefs.ref                   
\newcount\citeno
\newloop                               
      {\global\advance\citeno by 1
      \read0 to\temprefname            
      \if\temprefname\endgraf\relax    
      \else                            
         \edef\temprefname{\expandafter\stripspace\temprefname\delim}
         \expandafter\buildcitelabel{\temprefname}
      \fi}
\ifeof0
\dontrepeat
\closein0

\newcount\citeitemno

\def\beginnumreferences{
\immediate\openout0 allrefs.ref
\numreferences
}

\def\citeitem#1{\global\advance\citeitemno by 1
     \immediate\write0{#1}
     \par
   {\parindent=40pt
    \hang
    \noindent
    \rm {\hbox to 30truept{\hss [\the\citeitemno]\quad}}}}

\def\ref#1#2#3#4                 
    {\parindent=40pt
     \hang#1
    {\frenchspacing\it #2\/
     \bf #3}
     #4}

\def\refbk#1#2#3                  
   {\parindent=40pt
    \hang#1
   {\frenchspacing\sl #2}
    #3}

\def\orig#1#2#3#4{                
    \parindent=40pt
    \hang
    [{\frenchspacing \it #1\/ \bf #2} #3 (#4)]}

\def\figcap#1#2
   {\global\advance\figureno by 1
    \noindent {\bf Figure #1.} \rm#2\par
    \bigskip}

\def\<{\langle}
\def\>{\rangle}
\def\cd{{\cdot}}
\def\id{{\bss{1}}}
\def\breaksto{\mathop{\kern0pt \to}}
\def\vev#1{\langle#1\rangle}
\def\bra#1{\langle#1|}
\def\ket#1{|#1\rangle}
\def\diag{{\rm diag}}
\def\ov{\over}
\def\ap{\approx}
\def\rms#1{_{\rm #1}}
\def\pt{\propto}
\def\tr{\mathop{\rm tr}\nolimits}
\def\bnabla{\bf\nabla}  

\def\ba{\bi{a}}
\def\bb{\bi{b}}
\def\bk{{\bi k}}
\def\bn{{\bi n}}
\def\bp{{\bi p}}
\def\bq{{\bi q}}
\def\bt{{\bi t}}
\def\bv{{\bi v}}
\def\bx{{\bi x}}
\def\bX{\bi{X}}

\def\bXd{\dot{\bX}}
\def\bXp{\mathaccent 19 {\bX}}
\def\bap{\mathaccent 19 {\ba}}
\def\bbp{\mathaccent 19 {\bb}}
\def\bXperp{\bi{X}^\perp}
\def\bUp{\bi{U}^\perp}
\def\bVp{\bi{V}^\perp}

\def\L{{\cal L}}
\def\E{{\cal E}}
\def\M{{\cal M}}

\def\Bbb{\rm}
\def\R{{\Bbb R}}

\def\Z{{\Bbb Z}}


\def\gb{\bar\ga}

\def\xb{\bar\xi}
\def\bphi{\bar\phi}
\def\ad{\dot a}

\def\htil{\tilde{h}}
\def\Xd{\dot{X}}
\def\Xp{\mathaccent 19 X}


\def\fa{f_{\rm a}}

\def\Gl{G_{\rm l}}
\def\Gg{G_{\rm g}}
\def\Hl{H_{\rm l}}
\def\Hg{H_{\rm g}}
\def\Ml{\M_{\rm l}}
\def\Mg{\M_{\rm g}}
\def\MPl{M_{\rm Pl}}
\def\ma{m_{\rm a}}
\def\ms{m_{\rm s}}
\def\mv{m_{\rm v}}
\def\muf{\mu_{\rm eff}}
\def\na{n_{\rm a}}
\def\Og{\Om\rms{gr}}
\def\rs{\rho_{\rm s}}
\def\rt{\rho_{\rm tot}}
\def\rc{\rho_{\rm c}}
\def\rt{\rho_{\rm tot}}
\def\SiInf{S^1_\infty}
\def\sg{\si_{\rm 8,gal}}
\def\Tc{T_{\rm c}}
\def\td{t_{\rm d}}
\def\tf{T_{\rm eff}}
\def\xg{x\rms{g}}
\def\xs{x\rms{s}}
\def\zeq{{z\rms{eq}}}
\def\zrec{{z\rms{rec}}}
\def\zg{z\rms{g}}

\rightline{SUSX--TP--94--74}
\rightline{IMPERIAL/TP/94-95/5}
\rightline{NI94025}
\rightline{{\tt hep-ph/9411342}}

\title{Cosmic strings}

\author{M B Hindmarsh\dag\S\ and T W B Kibble\ddag\S}

\address{\dag School of Mathematical and Physical Sciences,
University of Sussex, Brighton BN1 9QH, UK}

\address{\ddag Blackett Laboratory, Imperial College, London SW7 2BZ, UK}

\address{\S Isaac Newton Institute for Mathematical Sciences, 20 Clarkson
Road, Cambridge CB3 0EH, UK}

\vfill

\jnl{\RPP}

\date

\beginabstract
The topic of cosmic strings provides a bridge between the physics of the very
small and the very large.  They are predicted by some unified theories of
particle interactions.  If they exist, they may help to explain some of the
largest-scale structures seen in the Universe today.  They are `topological
defects' that may have been formed at phase transitions in the very early
history of the Universe, analogous to those found in some condensed-matter
systems --- vortex lines in liquid helium, flux tubes in type-II
superconductors, or disclination lines in liquid crystals.  In this review, we
describe what they are, why they have been hypothesized and what their
cosmological implications would be.  The relevant background from the standard
models of particle physics and cosmology is described in section 1.  In
section
2, we review the idea of symmetry breaking in field theories, and show how the
defects formed are constrained by the topology of the manifold of degenerate
vacuum states.  We also discuss the different types of cosmic strings that can
appear in different field theories.  Section 3 is devoted to the dynamics of
cosmic strings, and section 4 to their interaction with other fields.  The
formation and evolution of cosmic strings in the early Universe is the subject
of section 5, while section 6 deals with their observational implications.
Finally, the present status of the theory is reviewed in section 7.

\endabstract

\contents

\entry{1}{Introduction}{}
\subentry{1.1}{Outline}{}
\subentry{1.2}{Symmetry, spontaneous symmetry breaking and symmetry
restoration}{}
\subentry{1.3}{Defect formation}{}
\subentry{1.4}{The Big Bang}{}
\subentry{1.5}{Unification}{}
\subentry{1.6}{Inflation}{}
\subentry{1.7}{Cosmic strings}{}

\entry{2}{Strings in field theories}{}
\subentry{2.1}{Global strings}{}
\subentry{2.2}{Local or gauge strings}{}
\subentry{2.3}{Vortices and topology}{}
\subentry{2.4}{Semi-local and electroweak vortices}{}
\subentry{2.5}{Composite defects: strings and domain walls}{}
\subentry{2.6}{Composite defects: strings and monopoles}{}
\subentry{2.7}{Superconducting strings: bosonic currents}{}
\subentry{2.8}{Superconducting strings: fermionic currents}{}
\subentry{2.9}{Strings in unified theories}{}

\entry{3}{String dynamics}{}
\subentry{3.1}{Equations of motion}{}
\subentry{3.2}{Strings in Minkowski space}{}
\subentry{3.3}{Strings with damping}{}
\subentry{3.4}{Superconducting string effective action}{}
\subentry{3.5}{Global string effective action}{}
\subentry{3.6}{Intercommuting}{}

\entry{4}{String interactions}{}
\subentry{4.1}{Gravity}{}
\subentry{4.2}{Gravitational radiation from loops}{}
\subentry{4.3}{Gravitational radiation from infinite strings}{}
\subentry{4.4}{Pseudoscalar and electromagnetic radiation}{}
\subentry{4.5}{Particle emission}{}
\subentry{4.6}{Particle scattering}{}
\subentry{4.7}{Baryon number violation}{}

\entry{5}{Strings in the early Universe}{}
\subentry{5.1}{High-temperature field theory}{}
\subentry{5.2}{Formation of strings}{}
\subentry{5.3}{Early evolution}{}
\subentry{5.4}{The role of loops}{}
\subentry{5.5}{Simulations of string evolution}{}
\subentry{5.6}{Analytic treatment of string evolution}{}
\subentry{5.7}{Inclusion of small-scale structure}{}
\subentry{5.8}{Scaling configuration}{}
\subentry{5.9}{Superconducting strings}{}
\subentry{5.10}{Axion strings}{}

\entry{6}{Observational consequences}{}
\subentry{6.1}{Gravitational lensing}{}
\subentry{6.2}{Effect on microwave background}{}
\subentry{6.3}{Density perturbations}{}
\subentry{6.4}{Gravitational waves}{}
\subentry{6.5}{Cosmic rays}{}

\entry{7}{Conclusions}{}
\subentry{7.1}{GUT-scale strings}{}
\subentry{7.2}{Lighter strings}{}
\subentry{7.3}{Summary}{}

\entry{}{References}{}

\section{Introduction}

\subsection{Outline}

There have been rapid and exciting developments over the
last few years on the interface between particle physics
and cosmology.  Particle physicists pursuing the goal of
unification would like to test their theories at energy
scales far beyond those available now or in the future in
terrestrial accelerators.  An obvious place to look is to
the very early Universe, where conditions of extreme
temperature and density obtained.  Meanwhile cosmologists
have sought to understand presently observed features of
the Universe by tracing their history back to that very
early period.

The early Universe was an intensely violent environment, so
it is not easy to find direct traces of very early
events.  There are however some special events that may
have left traces still visible today --- in particular,
phase transitions.  If our present ideas about
unification of forces and about the Big Bang are correct,
then the Universe, in the first fraction of a second
after its birth, underwent a series of phase
transitions.  Like more familiar transitions in
condensed-matter systems, these may have led to the
formation of defects of one kind or another  --- domain
walls, strings or vortices, monopoles, or combinations of
these.   In many cases, such defects are stable for
topological or other reasons and may therefore survive, a
few of them even to the present day.  If defects exist,
they constitute a uniquely direct connection to the
highly energetic events of the early Universe.

Cosmic strings in particular have very intriguing
properties.  They are very massive objects and may have
played an important role in structure formation, perhaps
providing at least some of the density inhomogeneities
from which galaxies eventually grew.  Some of their observational
signatures are quite distinctive, and within
observational reach.

In this review, our aim is to explain what cosmic strings
are, how they are formed, how they move and evolve, and
what their cosmological implications might be.

In this introductory section, we review the necessary
background material, the basic ideas about phase
transitions, defects and the Big Bang.  We begin with the
concept of symmetry and spontaneous symmetry breaking,
and discuss the restoration of symmetry at
high temperatures.  Next we examine the conditions for
defect formation at a phase transition.  Then we review
briefly the standard cosmological model, the Hot Big
Bang, and explain why, in conjunction with ideas about
unification, it suggests a sequence of phase transitions
in the very early Universe.  The section ends with a
summary of the properties of cosmic strings.

In section 2, we discuss the topological conditions under
which string defects are formed in field theories.  The
dynamics of strings are reviewed in section 3 and their
interaction with other constituents of the Universe in
section 4.  Section 5 is devoted to the formation and
evolution of defects, especially strings in the early
Universe, and section 6 to their observational
implications.  The present status of the theory is
summarized in section 7.

\subsection{Symmetry, spontaneous symmetry breaking and
symmetry restoration}

The concept of symmetry in particle physics runs very deep.
In the space available we can only sketch the issues involved, and refer
readers to the reference books (for a selection see references~[\ct{QFT}]) for
more detail.  The idea that
physical laws should be invariant under groups of transformations is a
powerful one, as the theories of relativity attest.  In the context of quantum
field theories, symmetries group particles together and relate their
scattering probability amplitudes.  The importance of symmetry in the current
understanding of the fundamental structure of nature is such that particle
physics can be characterized as the search for a single underlying symmetry
behind the interactions of particles and fields.

The symmetries of quantum field theory come in two classes: space-time and
internal.  The former are those of the space-time through which the field
propagates; for example, empty space has the Poincar\'e group of symmetries,
consisting of translations, rotations and Lorentz boosts.  Internal symmetries
relate fields to one another.  To implement these symmetries, a field must
have some well-defined transformation properties under the action of the
symmetry group: that is, it must form a representation of the group.  The
labels of different representations are often called quantum numbers. For
example, the representations of the Poincar\'e group are labelled by their
mass and their spin.   Internal symmetries transform fields into one another,
and (with the exception of fundamental string theory --- not directly
related to cosmic strings) there are only finite numbers of
fields.  The corresponding groups are finite-dimensional and compact.  If
they are not discrete groups then they are Lie groups;
thus the study of Lie groups and their representations assumes fundamental
importance in particle physics.  Generally, a relativistic quantum field
theory is specified by the representations of the fields which comprise it
({\ie}, their masses, spins and internal quantum numbers), which are then
assembled into an invariant Lagrangian.

The most powerful type of symmetry is the gauge (or `local')
symmetry, where
the Lagrangian is invariant under a symmetry transformation which may be
different at every point. (If the invariance exists only for transformations
which are constant in space and time, the symmetries are distinguished by
being called `global' or `rigid'.)  For internal symmetries this requires a
spin 1 field, the gauge field, of which the electromagnetic field is the
prototypical example.  Gravity can also be formulated as a gauge theory of
local Lorentz transformations: the gauge field is the gravitational field
itself, which has spin 2.

Lastly, there is supersymmetry (see for example [\ct{SUSY}]).
This is a powerful
symmetry which relates particles of different spins and, in a sense, combines
internal and space-time aspects.  The prospect of uniting the particles of
matter (spin 1/2 quarks and leptons) with those of the known forces (spin 1
gauge bosons) is clearly an exciting one, which is so far unrealized, for none
of
the known particles fit together into representations of supersymmetries,
called supermultiplets.  Perhaps the superpartners of today's particles will
be discovered at the Large Hadron Collider currently under construction at
CERN.

The problem with invoking all these symmetries is how to hide or `break'
them.  If one proposes a symmetric Lagrangian, as well as a symmetric ground
state
for the theory, then there is a deep theorem which states that the existence
of a gauge symmetry implies that the associated spin 1 particles are
massless.  Only one such particle is known: the photon.  The resolution of
the problem is the breaking of the symmetry of the ground state by what is
known as
the Higgs mechanism.  This works by introducing spin 0 fields (denoted
$\phi$) transforming non-trivially under the symmetry group $G$, and
constructing for them an energy density which is minimized at some non-zero
value $\phi_0$.  The theory then has a ground state which is invariant only
under the subgroup of $G$ that leaves $\phi_0$ unchanged.  We say that the
symmetry $G$ is broken to $H$.

It is often useful to draw analogies with condensed matter physics, where
similar mechanisms occur.  A good example is a nematic liquid crystal
[\ct{NLC}], which
consists of rod-like or disk-like molecules. The free energy density of this
system is invariant under the group of spatial rotations, SO(3).  This is a
global symmetry: the thermodynamics should not depend on the orientation of the
sample. However, at low temperatures and at high pressures the molecules prefer
to
line up: thus in any particular sample the full rotational symmetry will not be
respected by the direction of alignment.  The rotational symmetry is still
there, for there is no preferred alignment direction, but it is broken by the
equilibrium state, which has all the molecules pointing one way.  A subgroup of
the
rotation group is left unbroken, which consists of rotations around the
alignment direction and rotations through $\pi$ about axes in the orthogonal
plane (the interactions between the molecules do not distinguish between their
ends).  This group is O(2), called $D_\infty$ in the condensed matter
literature.  This breaking is usually denoted
$$
{\rm SO(3)} \to {\rm O(2)}.
\eqqno
$$
The degree and direction of alignment of the molecules, when averaged over
many molecular spacings, can be described by a director field $\Phi({\bi
x})$.  This field must transform appropriately under SO(3) to describe a
`headless vector': it turns out to be a traceless symmetric tensor.  In the
nematic phase,
$$
\Phi_{ij} = A(n_in_j - \frac{1}{3}\delta_{ij}),
\eqqno
$$
where $\pm n_i$ is the alignment axis of the molecules, known as the
director field.

A good description of the transition between the disordered $(\Phi=0)$ and
the ordered $(\Phi\ne 0)$ phases can be obtained by postulating that the
free-energy density for constant fields takes the form
 $$
f(\Phi) = \al + \be \tr\Phi^2 + \ga\tr\Phi^3+\de(\tr\Phi^2)^2 + \ldots
\eqqno
$$
The value of $\Phi$ at the minimum of $f$ depends on the coefficients
$\al$--$\de$, which are temperature and pressure dependent.  In
particular, the transition occurs when the sign of $\be$ changes.  The form
of the free-energy function can tell us whether the
transition is first- or second-order: if $\ga\ne0$, the equilibrium free
energy changes discontinuously from one minimum
to another, so that we have a first-order transition; only if $\ga=0$ is
this transition second-order.

In the quantum field theory of scalar fields at non-zero (often called
finite) temperature the free energy density for constant fields is known as
the finite-temperature effective potential $V_T(\phi)$
(see reference [\ct{Kap89}] and section 5.1).
This is in principle
calculable from the zero-temperature classical potential in a perturbative
expansion in powers of $\hbar$, the loop expansion.  One finds that
broken symmetries in quantum field theories are almost always restored at
high enough temperatures.

\subsection{Defect formation}

In some phase transitions it happens that the order parameter does not take
up its equilibrium magnitude everywhere, and indeed may vanish inside
objects known as topological defects.  Defects may be two-, one- or
zero-dimensional, and their origin lies in the topological properties of
the set of equilibrium states, \ie\ the minima of the free energy or the
effective potential.

As an example, let us take a theory of a single real scalar
field $\ph$, with a reflection symmetry $\ph \leftrightarrow -\ph$.  The
energy functional for static fields has the form
$$
{\cal H} = \half(\nabla\ph)^2 + V(\ph),
\eqqno
$$
where the zero temperature classical potential is
$V(\ph) = \frac{1}{8}\la(\ph^2 - \eta^2)^2$.  The equilibrium states have
$\ph = \ph_\pm = \pm\eta$.  Suppose that in one region of space we find the
field at $\ph_+$, and nearby at $\ph_-$.  Then along any
path joining these two regions the field must necessarily pass through zero at
least once.  The set of points where the field vanishes form a two-dimensional
surface, or domain wall, which separates domains of $\ph=\ph_+$ from domains
of $\ph=\ph_-$.  There is energy associated with the wall, because the
potential energy density inside is higher than its equilibrium value, and
also because the field is not constant.

It is the disconnected nature of the set of equilibrium states of the theory
which  allows domain walls to exist.  Other types of defect have different
topological requirements, which we shall discuss in section 2.  Here
we will merely note that our example system, the nematic liquid crystal,
allows both line and point defects [\ct{Mer79}].
(In the condensed matter literature, such defects, arising
from broken rotational symmetries, are called {\it disclinations}.)
Around the line disclinations the molecules
change their orientation by a rotation through angle $\pi$ around an
orthogonal axis, while in
a point defect the molecules are in a `hedgehog'
configuration, directed away from a central point.  Along the central
lines and points of the disclinations, the molecules cannot have any alignment
directions and so the order parameter vanishes.

In section 5.2 we consider in detail how strings (line defects) are formed in
cosmological phase transitions [\ct{Kib76}].  However, the analogy between the
nematic
liquid crystal and a field theory makes it plausible that what happens in the
laboratory in a cooling sample of the liquid crystal also happens in the early
universe.  Figure 1.1 shows a sequence of frames taken from a video of
an isotropic-nematic phase transition in action [\ct{Bow+94}].
The transition is
first order, and we see that it proceeds by the nucleation of bubbles
of the low temperature, nematic phase.  These bubbles grow, so reducing the
total energy of the system, and then finally merge.  Out of the mess emerges
a network of strings.  What happens is that the director field takes
up random directions in each bubble when it nucleates.  When a set of
bubbles nucleate there is an appreciable probability that the field will be
so misaligned that disclinations are formed at the interstices.

\subsection{The Big Bang}

The standard model of the early Universe is based on two
observational pillars --- the recession of galaxies and
the cosmic microwave background.

Hubble observed in 1926 that recession
velocities determined from redshifts are roughly
proportional to distance [\ct{KolTur90}]:
 $$
v=Hr,
\label{\eHubLaw}
$$
where $H$ is the Hubble parameter,
 $$
H = 100h \unit{km\,s^{-1}\,Mpc^{-1}}.
$$
The dimensionless parameter $h$, in the range $0.5\lap
h\lap1$, encodes our present uncertainty.  The
characteristic expansion time is
 $$
H^{-1} = 10^{10} h^{-1}\unit{years}.
\eqqno
$$

The microwave background radiation, first observed by
Penzias and Wilson [\ct{PenWil65}], is the redshifted
relic of radiation emitted when the Universe was dense
and hot.  It has a very accurate blackbody spectrum, with
a temperature of $2.726\pm0.010$ K [\ct{CobeT}].

It is often convenient to use {\it comoving\/}
coordinates, so that (neglecting small random velocities)
each galaxy retains the same fixed coordinates.  The true
distance $r(t)$ to a galaxy is related to its coordinate
distance $x$ by a universal scale factor $a(t)$:
 $$
r(t) = a(t)x.
\eqqno
$$
Hubble's law, (\rf{\eHubLaw}), then follows, with
 $$
H = {\ad\over a}.
\eqqno
$$
The redshifts of distant objects are given by $1+z = a(t_0)/a(t)$,
where $t_0$ is the current time and $t$ is the time when the radiation
was emitted by the object.

The evolution of the scale factor $a$ is described
according to general relativity by the Friedmann equation,
 $$
{\ad^2\over a^2} = {8\pi G\over 3}\rh - {K\over a^2},
\label{\eFdmEqn}
$$
where $G$ is Newton's constant,
$\rh$ the mass density in the Universe (assumed
homogeneous on the scales of interest), and $K$ a
constant, the (uniform) curvature of the spatial sections.
If $K>0$, the Universe is spatially closed; it will
expand to a maximum radius and then contract, eventually
reaching the `big crunch'.  If $K\le0$, it is open and
will continue expanding for ever.  The condition for
closure may also be expressed in terms of the density:
the Universe is closed if $\Om>1$, where
 $$
\Om = {\rh\over\rh\rms{c}},\qquad \rh\rms{c} =
{3H^2\over8\pi G} = 1.88 h^2 \times 10^{-26}
\unit{kg\,m^{-3}};
\label{\eOmeDef}
$$
$\rh\rms{c}$ is the critical density.  Observation tells
us that $\Om$ lies in the range $0.1 < \Om <2$.
Theoretical prejudice suggests that $\Om=1$, as predicted
if the Universe underwent an early inflationary phase (see section 1.6).

The Friedmann equation (\rf{\eFdmEqn}) must be
supplemented by the energy equation,
 $$
\dot\rh + 3 {\ad\over a}(\rh + p) = 0,
\label{\eEngEqn}
$$
and also by an equation of state giving the pressure $p$ in
terms of the density $\rh$.  The most important cases are
a universe dominated by radiation, or extremely
relativistic particles, with $p = {1\over3}\rh$, and one
dominated by dust, or very non-relativistic matter, for
which $p=0$.  The energy equation (\rf{\eEngEqn}) tells us
that in these cases $\rh\pt a^{-4}$ and $\rh\pt a^{-3}$,
respectively.  In either case, $\rh$ varies much more
rapidly than the curvature term in (\rf{\eFdmEqn}), so at
early times it is always  a good approximation to neglect
$K$.  The solution is then very simple: $a \pt t^{1/2}$
in the radiation-dominated universe and $a\pt t^{2/3}$ in
the case of dust.

The observed large-scale homogeneity and isotropy of the
Universe implies that to a good approximation space-time
may be described by the Robertson--Walker metric.  When $K$
is negligible, we have the spatially flat Einstein-de
Sitter Universe, with metric
 $$
\d s^2 = \d t^2 - a^2(t) \d\bx^2.
\label{\eRobWal}
$$

The Universe today is probably matter-dominated, although it
has been suggested recently that we are entering a phase of
vacuum energy domination ($\rh\ap$ constant),
based on indirect arguments about
galaxy correlations [\ct{EfsSutMad90}].  There is
compelling evidence that much of the matter whose
gravitational effects we can detect, for example through
measuring rotation curves of galaxies,
is invisible [\ct{BahPirWei87}].
Moreover, studies of the synthesis of helium
and other light elements in the Big Bang [\ct{Oli+90}] place firm
limits on the density of ordinary baryonic matter
(protons and neutrons), of
$$
0.01 < \Om_{\rm B} h^2 < 0.015.
$$
It is therefore
probable that most of the matter in the Universe is of
some other type, known as `dark matter'.  Dark matter candidates
are generally
classified as either `hot' or `cold', depending on
whether the particles were still relativistic when they
decoupled from the rest of the matter in the Universe.

Because
$\rh\rms{rad}$ falls faster than $\rh\rms{mat}$, the Universe was
initially radiation-dominated, until the epoch
$t\rms{eq}$ of equal radiation and matter densities. If
$\Om=1$, this occurred at a redshift $\zeq \simeq 24000h^2$, when
the Universe was about $1000 h^{-4}$ years old.  Shortly
after this, the Universe became sufficiently cool for the
electrons and the nuclei to `re'-combine into neutral gas.
This happened when the  temperature was of around 0.3 eV,  at
a redshift
$\zrec \simeq 1100$.

These are important milestones, for density perturbations in
cold dark matter can
start to grow only when the Universe becomes matter-dominated, while
radiation pressure prevents the growth of perturbations in the baryons
until recombination.

The period when cosmic strings may have been formed is
very early  in the history of the Universe, when it was
certainly radiation-dominated.  The Hubble parameter
was then $H = 1/2t$.  Since $K$ was negligible, $\rh$ was
very nearly equal to $\rc$.  Thus the density was, by
(\rf{\eOmeDef}),
 $$
\rh = {3\over32\pi Gt^2}.
\label{\eDenRad}
$$
One can also find a useful relation between time and
temperature.  The density of a  relativistic gas is given
by
 $$
\rh = {\pi^2\over30}g_*T^4,
\label{\eDenRel}
$$
where $g_*$ is the effective number of distinct spin
states of relativistic particles (one for particles of
zero spin and 2 for higher spins, multiplied by
7/8 for fermions) [\ct{KolTur90}].  Equating these two
expressions for the density, we find
 $$
T^2t = \sqrt{45\over\pi g_*}{\MPl\over4\pi} = {2.42\over
g_*^{1/2}} \unit{MeV^2\,s}.
\label{\eTemTim}
$$
Here $\MPl$ is the Planck mass,
 $$
\MPl = G^{-1/2} = 1.22\times 10^{28} \unit{eV}.
$$
(We use units in which $c = \hbar = k\rms{B} = 1$.)

\subsection{Unification}

It is now very well established that two of the four
fundamental interactions of elementary particles, the
electromagnetic and weak interactions, can be described
by a unified gauge theory, the Weinberg--Salam theory,
based on the gauge group SU(2)$\times$U(1).  At low
energies, the two interactions appear very different, but
the true underlying symmetry emerges above the
characteristic energy scale of the unified theory,
around 100 GeV (the rest energy of the W and Z gauge
bosons).  The theory exhibits a phase transition at a
temperature of this order.

The strong interactions too are described by a
very successful gauge theory, quantum chromodynamics,
based on the group SU(3).  This theory may also exhibit a phase
transition, with a critical temperature of order 100 MeV,
associated with quark confinement.  (Above this
temperature, we have a dense soup of quarks and gluons;
below it we have individual hadrons, like protons,
neutrons and pions.)

The low-energy physics of elementary particles is thus
described by a gauge theory with three independent coupling
constants, $g_3$, $g_2$ and $g_1$, associated with the
three groups in the low-energy symmetry
SU(3)$\times$SU(2)$\times$U(1).  These couplings are
energy-dependent, though only logarithmically.
Extrapolating from their low-energy behaviour, it appears
that all three will become roughly equal at an energy
scale of around $10^{15}$ to $10^{16}$~GeV, only three or
four orders of magnitude below the Planck mass.  It is
therefore very plausible to suppose that there is a grand
unified theory, encompassing all three interactions.  This
grand symmetry would be manifest only above that
characteristic energy scale.

If this is correct, there is probably a grand
unification phase transition, with a critical
temperature of about $10^{15}$~GeV.  There is some reason
to believe that this is not a single transition but
involves two or more separate transitions.  In the
simplest models with only a single transition, the
extrapolated couplings do not in fact quite meet.  A
better fit can be obtained in models with more than one
transition, for example in a model with supersymmetry
broken at some intermediate scale.

Thus the picture emerges of a sequence of phase
transitions occurring in the very early evolution of the
Universe.  The corresponding times can be estimated from
(\rf{\eTemTim}).  The expected number of spin states
$g_*$ depends on the precise particle-physics model, but
during the relevant period is roughly of order 100.  Thus
we see that the electro-weak and quark-hadron transitions
occur when the age of the Universe is about $10^{-11}$~s
and $10^{-5}$~s, respectively, while a grand unification
transition would occur at between $10^{-39}$ and
$10^{-37}$~s.

Of course one can legitimately question whether we
really understand enough of the basic physics to make
reliable predictions of the behaviour of the Universe so
close to the Big Bang.  So far as electro-weak and
quark-hadron transitions are concerned, we do now have a
rather firm understanding of the basic physics.  Grand
unification does admittedly represent a very considerable
extrapolation from the region where we have reliable
tests.  On the other hand, our theories of fundamental
particle interactions work extremely well at ordinary
energies and there is no intrinsic reason to expect them
to break down until we reach the Planck energy, the energy
at which gravity also becomes strong and quantum gravity
effects necessarily enter.  So although we should be
properly cautious, we do have good reasons to think that
our basic understanding can be pushed this far.  Naturally,
if our confidence turns out to be misplaced, that
would be an even more dramatic and interesting discovery.

\subsection{Inflation}

Some discussion is in order of the relation between the concepts of inflation
and of cosmic strings.  There are rival theories of large-scale structure
formation based on these concepts, which  at first sight appear
incompatible, but can in fact be reconciled if one really wishes to do so.

The idea of inflation
(for a much more complete exposition see [\ct{KolTur90}])
was invented to solve a number of cosmological
puzzles, particularly the horizon problem, the flatness problem and the
monopole
problem.

Since the Universe is of finite age, there is at any epoch $t$ a {\it particle
horizon}; no signal from beyond the horizon can yet have reached us.  In the
radiation-dominated era of the standard Friedmann--Robertson--Walker (FRW)
Universe, the particle horizon at time $t$ is of radius $2t$.  When we
observe the cosmic microwave background radiation coming to us from opposite
directions in the sky, we are seeing light emitted from regions that were then
separated by nearly 100 times the horizon distance at the time.  It therefore
seems very surprising that the temperature of the radiation coming from
opposite
directions is the same: no causal process could have established thermal
equilibrium between such distant parts of the Universe.  This is the {\it
horizon problem}.

Another puzzle, the {\it flatness problem}, is why $\Om$ today, as
defined by (\rf{\eOmeDef}), is so close to unity.  It is easy to verify that
$\Om=1$ is an {\it unstable\/} point of the FRW evolution equation
(\rf{\eFdmEqn}).  To have $\Om$ close to 1 today, it must have been very close
indeed at early times --- seemingly requiring fine tuning by many orders of
magnitude.

Almost all grand unified theories predict the existence of ultra-heavy stable
magnetic monopoles.  Once formed, they are very hard to eliminate and so would
rapidly come to dominate the energy density of the Universe, exceeding by many
orders of magnitude the energy density in baryons.  This is the {\it monopole
problem}.

All these problems can be cured by invoking the idea of inflation --- a very
early period of rapid expansion in which the energy density is dominated by
`vacuum energy'.  This is achieved by introducing a scalar inflaton field $\si$
with an appropriately chosen potential $V(\si)$.  The energy density and
pressure due to such a field in a FRW background are given by
 $$
\eqalign{
\rh &=\half\dot\si^2 + \half a^{-2}(\nabla\si)^2 + V(\si),\cr
p &= \half\dot\si^2 - \frac{1}{6}a^{-2}(\nabla\si)^2 - V(\si).
}
 $$
The trick is to arrange that in some sufficiently large region these
expressions are dominated by the potential term.  Thus we have the strange
equation of state $p\ap-\rh$.  It follows that $\rh\ap$ constant, and hence,
by (\rf{\eFdmEqn}), that $a(t)$ increases exponentially.  Actually, $\rh$ is
not exactly constant.  The field $\si$ does evolve slowly, eventually reaching
a point where the conditions for inflation are no longer satisfied, bringing
the period of rapid expansion to an end.  During the period of accelerating
expansion $\Om$ tends {\it towards\/} 1 rather than away from it, thus solving
the flatness problem.  The horizon problem is also cured: during inflation, the
causal horizon distance increases by a huge factor, so that the entire
presently
visible Universe was originally well within a single horizon volume.  Finally,
the expansion dilutes the density of any previously existing monopoles to an
undetectably low level.

Unfortunately, inflation also dilutes any other topological defects, in
particular cosmic strings.  At first sight, therefore, the two ideas are indeed
mutually in\-com\-pat\-ible.  However, it is possible to reconcile the two
with models in which
strings or other defects are formed during the late stages of inflation. We
discuss them briefly at the end of section 5.2.

\subsection{Cosmic strings}

In the following sections, we shall deal with many
different properties of cosmic strings in detail.  To set
the scene, it may be useful, however, to review their
most important characteristics here.

Cosmic strings are linear defects, analogous to flux
tubes in type-II super\-conductors, or to vortex filaments in
super\-fluid helium.  They might have been formed at a
grand unification transition, or, conceivably, much
later, at the electro\-weak transition --- or somewhere
in between.  The standard electro\-weak model does not
predict stable strings, but some quite simple
generalizations of it, entirely compatible with available
data, do.

These strings have enormous energy.  In the simplest,
canonical type of string, the energy per unit length,
$\mu$, and the string tension are equal.  Equivalently,
the characteristic speed of waves on the string is
the speed of light.  This equality is a consequence of
local invariance of the field configuration around a
string under Lorentz boosts along its direction.  Roughly
speaking, one expects that for strings produced at a
phase transition at $\Tc$, $\mu\sim \Tc^2$.

An important number in the theory is the dimensionless
quantity $G\mu \sim (\Tc/\MPl)^2$, which characterizes the
strength of the gravitational interaction of strings.
For grand unified strings, $G\mu$ is of order $10^{-6}$ or
$10^{-7}$; it is useful to define a parameter $\mu_6$ by
 $$
G\mu = 10^{-6}\mu_6.
\label{\eDefMu6}
$$
Translated into more ordinary units, $\mu$ is very large
indeed; for example the mass per unit length is
 $$
\mu = 1.35 \times 10^{21} \mu_6 \unit{kg\,m^{-1}} = 2.09
\times 10^7 \mu_6 M_{\odot}\!\! \unit{pc^{-1}}.
\label{\eMuMag}
 $$
A grand unified string of length equal to the solar
diameter would be as massive as the sun, while such
a length of string
formed at the electroweak scale would weigh only 10 mg.
The gravitational effects of the latter
are essentially negligible, though such strings
may still be of great interest, because of other
types of interactions.  In particular, cosmic strings may
behave like thin superconducting wires, with a critical
current $\sim \Tc$.  Superconducting strings
interact strongly
with magnetic fields in the universe, and even light ones may
have significant astrophysical effects.

The parameter $G\mu$ plays a crucial role in discussions of the
observational consequences of such heavy strings, for
the size of the gravitational perturbations induced
by the strings is O($G\mu$); $G\mu$ is then the order of
magnitude
both of the seed density perturbations for galaxy formation,
and of the induced fluctuations in the Cosmic Microwave Background.
In section 6 we shall see that $\mu_6 \sim 1$ provides
fluctuations of the order of $10^{-5}$,
just the right magnitude to explain
the observed features of the CMB spectrum and the matter
distribution in the Universe today.  The fact that grand unified
strings give the right amplitude for the gravitational
perturbations is a good feature of string-based theories of
structure formation.  In theories invoking the amplification
of quantum fluctuations during inflation, the
figure $10^{-5}$ is not easily explained [\ct{KolTur90}].

\section{Strings in Field Theories}\ssf
\subsection{Global strings}
The simplest theory exhibiting string solutions is that of a complex scalar
field $\phi(x)$,  described by the Lagrangian density
 $$
\L = \pa_\mu\phi^*\pa^\mu\phi - V(\phi),\qquad
V = \half\lambda(|\phi|^2-\half\eta^2)^2
\label{\eGloLag}
$$
which has a global U(1) symmetry, under  the transformation
$\phi\to\phi e^{i\alpha}$, with $\alpha$ constant.

The Euler-Lagrange equations that follow from (\rf{\eGloLag}) are
$$
[\pa^2 +\lambda(|\phi|^2-\half\eta^2)]\phi = 0.
\label{\eGloEom}
$$
The ground state, or vacuum, solution is $\phi =
(\eta/\surd2) \exp(i\alpha_0)$ with $\alpha_0$ constant, which has zero energy.
 Since
the energy is bounded below by zero this solution (unlike $\phi=0$) is clearly
stable.    It is not invariant under the U(1) symmetry transformation:  the
symmetry is  said to be broken by the vacuum.
The mass $\ms$ of the scalar particle in the symmetry-breaking vacuum is
given by $\ms^2 = \lambda\eta^2$.  There is also a massless  particle, the
Nambu-Goldstone boson, which is associated with the broken  global symmetry.
It corresponds to space-depndent oscillations in the phase of $\phi$.

Besides the vacuum, there are also static solutions with non-zero energy
density.  Let us make the
following  cylindrically symmetric ansatz:
$$
\phi = {\eta\over\surd2} f(\ms\rho)e^{in\varphi},
\eqqno
$$
where $\{\rho,\varphi,z\}$ are cylindrical polar coordinates, and $n$ is an
integer.  The field equations
then reduce to a single non-linear ordinary differential equation
$$
f'' + {1\over\xi}f' - {n^2\over\xi^2}f - \half(f^2-1)f = 0,
\eqqno
$$
where $\xi \equiv m_{\rm s}\rho$.  As $\xi\to 0$,  continuity of $\phi$
requires that $f\to0$.  At infinity, $f\to 1$, so that the field approaches
its  ground state $|\phi|=\eta/\surd2$.  Writing  $f = 1-\delta f$, it is not
hard to show that at large $\xi$, $\delta f \sim n^2/\xi^2$. Figure
2.1 displays $f$ over the whole range of $\xi$, along  with the
energy density
 $$
{\E} = |\dot\phi|^2 + |\nabla\phi|^2 + V(\phi).
\eqqno
$$
Although $\E$ is well localized near the origin, it has a $\xi^{-2}$ tail at
large $\xi$, which comes from the angular part of the gradient term. This means
that the energy per unit length of this solution is infinite: inside a
cylinder of radius $R\gg\ms^{-1}$ it is approximately $\pi n^2\eta^2
\ln(\ms R)$.

These solutions are known as global strings or vortices
[\ct{VilEve82},\ct{ShaVil84}].  They are closely related to the vortices in
superfluid helium--4 [\ct{Hel},\ct{DavShe89a}], where the complex scalar
field represents the wave  function of the condensed
$^4$He atoms, and $-i\phi^* {\dbw}\!_k\phi/2|\phi|^2$
is proportional to the superfluid velocity, which has
a circulation around  the vortex core.  In particle
cosmology, the most commonly considered global  strings
are those associated with a spontaneously broken axial
U(1) symmetry,  the axion strings
[\ct{VilEve82},\ct{LazSha82}].

\subsection{Local or gauge strings}
Let us now consider what happens when the internal symmetry is a local one.
This requires the introduction of a vector field $A_\mu$.   The Lagrangian
density is then
$$
\L = -\frac{1}{4}F_{\mu\nu}F^{\mu\nu}+|D_\mu\phi|^2 - V(\phi),
\label{\eLocLag}
$$
where $D_\mu = \pa_\mu + ieA_\mu$ and $F_{\mu\nu}=\pa_\mu A_\nu -
\pa_\nu A_\mu$.
This is the Abelian Higgs model [\ct{QFT}], the prototypical
example of a gauge field  theory with spontaneous symmetry breaking.
The U(1) invariance is realized by the transformations
$$
\phi \to\phi e^{i\Lambda(x)}, \qquad A_\mu\to A_\mu - {1\over e}
\pa_\mu\Lambda(x), \eqqno
$$
where $\Lambda$ is a real single-valued function.  The field equations are
$$
\eqalign{ \[D^2 + \lambda(|\phi|^2 - \half\eta^2)\]\phi&= 0, \cr
\pa_\nu F^{\mu\nu} +ie\(\phi^*D^\mu\phi-D^\mu\phi^*\phi\) &= 0.\cr}
\label{\eLocEom}
$$
The particle spectrum still
has the Higgs with mass $\ms = \sqrt{\lambda}\eta$, but the Nambu-Goldstone
boson is incorporated into the vector field, which gains a mass
$\mv=e\eta$.

A vortex solution still
exists [\ct{NieOle73}], but its properties are quite different from those of
the global vortex. Now the energy per unit length is finite, because the
dangerous angular derivative of the field is replaced by a covariant
derivative, which can vanish faster than $\rho^{-1}$ at infinity.  To be
more specific, let us choose the radial  gauge $A_\rho = 0$.  Then we
may write the general cylindrically symmetric  ansatz for the gauge field as
 $$
\phi = {\eta\over\surd2} f(\mv\rho)e^{in\varphi},\qquad A^i =
{n\over e\rho}\hat{\varphi}^i a(\mv\rho).
\label{\eNOAns}
 $$
The resulting coupled ODEs do not have solutions in terms of known
functions, but it is straightforward to obtain their asymptotic behaviour:
 $$
{f\simeq\cases{f_0\xi^{|n|}, & \cr
               1-f_1\xi^{-1/2}\exp(-\sqrt{\beta}\xi), & \cr}}
\qquad
{a\simeq\cases{a_0\xi^2-{|n|f_0^2\over4(|n|+1)}\xi^{2|n|+2},
& as $\xi\to 0$;\cr
               1-a_1\xi^{1/2}\exp(-\xi), & as $\xi\to\infty$.\cr}}
\eqqno
 $$
Here, $\xi = \mv\rho$ and  $\beta=\la/e^2=(\ms/\mv)^2$.  (In the case
$\be>4$, $\xi^{-1/2}\exp(-\sqrt{\beta}\xi)$ is replaced by
$\xi^{-1}\exp(-2\xi)$.)
Note that the energy density is much more
localized than in the global string.

The local string also contains a tube
of magnetic flux, with quantized flux
 $$
\int \d^2x\, \bi{B}\cdot\bhat{z}
= \int_{\SiInf}\!\d x^iA^i= {2\pi n\over e},
\eqqno
 $$
where $\SiInf$ denotes a circle of infinite radius centred on the string.
This flux quantization is a result of the vanishing of the covariant
derivative,
which determines $A_i$ in terms of derivatives of $\phi$.  The phase of
$\phi$ must change by an integer multiple of $2\pi$, which forces the flux to
be quantized.

Local vortices are stable for any $n$ if $\beta < 1 $
[\ct{JacReb79},\ct{BogVai76}].  If $\beta>1$, a
flux $2\pi n/e$ vortex with $|n|>1$ is unstable with respect to splitting into
$n$ vortices  carrying the elementary unit of flux $2\pi/e$.  At the boundary
value,  $\beta=1$, the multiply charged vortex is neutrally stable with respect
to  dissociation: it has $n$ zero modes, or fluctuations with eigenvalue zero.
One  can interpret these results in terms of forces acting between the
vortices.   The gauge field generates a repulsive force, because lines of
magnetic flux  repel each other, while the scalar field produces an attractive
force,  essentially because it is energetically favourable to minimize the area
over  which the potential energy density is non-zero.  The range of these
forces
is  controlled by the Compton wavelength of the mediating boson, and whichever
has  longer range dominates.  For example, when $\beta>1$ the scalar boson is
heavier, the gauge force dominates and vortices repel.  If $n$ vortices are
sitting on top of one another, as in the $n$-unit vortex, any perturbation is
likely to disturb the cylindrical symmetry and enable it to break apart.

The vortices in the Abelian Higgs model have condensed matter analogues:
flux tubes in superconductors [\ct{Abr57}].  There are differences
[\ct{Dav90},\ct{Dav91}], stemming from the fact that Nielsen--Olesen
vortices exist in a vacuum background, whereas superconductor vortices
live in a background of charged bosons, the Cooper pairs.

The energy per unit length of the local string is
$\mu= \int\!\rho\,\d\rho\,\d\varphi\,\E(\rho) $.
Substituting the cylindrically symmetric ansatz for the fields into the energy
functional, it is not hard to see that it must have the form
$$
\mu = \pi\eta^2\epsilon(\beta).
\label{\eLocMu}
$$
Remarkably, it is possible to show analytically that $\epsilon(1)=1$
[\ct{Bog76}]. Numerical studies [\ct{JacReb79},\ct{HilHodTur88}]
show that $\ep$ increases monotonically with $\be$, albeit rather slowly,
going as $\log \be$ for $\be > 1$.

\subsection{Vortices and topology}
In order to decide whether a theory with a larger symmetry group than U(1)
possesses
stable vortex solutions, we need to examine the topology of the vacuum
manifold, the set of minima of the potential
[\ct{TyuFatSch75}--\ct{Ste57}].

Let us consider
a theory of a scalar field $\phi$ which transforms under some representation
of a compact Lie group $G$, with hermitean generators $T_a$ satisfying
$$
[T_a, T_b] = if_{abc}T_c.
$$
We are  interested in finding
vortex solutions, or finite-energy static solutions in  $\R^2$ to the field
equations.  The  static
energy functional is
$$
E = \int \d^2x\(\frac{1}{4}F_{ij}^aF_{ij}^a + |D_k\phi|^2 + V(\phi)\),
\label{\eStaEne}
$$
where $F_{ij}^a = \pa_i A_j^a-\pa_j
A_i^a - e f^{abc}A_i^bA_j^c$ and $D_k = \pa_k + ieA_k$, with $A_k =
A_k^aT_a$.  The potential $V$ is some $G$-invariant quartic polynomial in
the fields  $\phi$.  By the addition of a constant we can set $V=0$ at its
minima,  thus ensuring that $E$ is non-negative.

Let us denote the vacuum manifold by $\M$, and
let $\phi_0$ be one point on $\M$.  Then for any $g\in G$, $g\phi_0$ is also
on $\M$.  But many different elements $g$ may yield the same point.  It is
convenient to introduce the isotropy group (or little group) $H$ of $\phi_0$,
the set of all elements $h\in G$ such that $h\phi_0=\phi_0$.  Then clearly
$g\phi_0=g'\phi_0$ if and only if $g'=gh$, with $h\in H$.  In other words,
the points of $\M$ are in one-to-one correspondence with the left cosets of
$H$ in $G$; we write $\M=G/H$.

For a finite energy solution, each term in (\rf{\eStaEne}) must tend
sufficiently rapidly to zero as $\rho=|x|\to \infty$.  As before let us
choose the radial gauge $A_\rho=0$, in which the fields tend to definite
values as $\rho\to\infty$.  To make the potential term finite, we require
that in any direction $\vp$, $\ph(\rh,\vp)\to\bphi(\varphi)\in\M$.  Let us
choose $\bphi(0)=\ph_0$.  To make the first, magnetic term finite, we
require that $A$ tend to a pure gauge field, \ie,
$$
ieA_k(x) \to  -\pa_k{g}(\vp){g}^{-1}(\vp)
\quad {\rm as} \quad \rh\to\infty.
\label{\eGauAsy}
$$
for some $g(\vp)\in G$.  Moreover, we can always choose $g(0) = \id$, the
identity. Finally, to make the gradient term finite, we require that
$D_k\ph\to0$ as $\rh\to\infty$.  It follows at once that
$\pa_k[g^{-1}(\vp)\bphi(\vp)]=0$.  Since $g^{-1}(0)\bphi(0)=\ph_0$, this
means that
 $$
\bphi(\vp)=g(\vp)\ph_0.
\eqqno
$$

Now the fields $A^a_k$ and $\ph$ must be single-valued and continuous, but
$g(\vp)$, though continuous, need not be single-valued; $g(2\pi)$ need not
coincide with $g(0)$, though we do require $g(2\pi)\in H$.

We may regard $\bphi(\vp)$ as defining a loop in $\M$, a map from the circle
$S^1$ into $\M$, based at $\ph_0$.
Whether or not there is a vortex solution depends on the 
topological characterization of this loop.               
If it is contractible, i.e., can be smoothly shrunk to a point
within $\M$, then it is possible to find a zero-energy solution with this
asymptotic value, so we should not expect to find a stable vortex.

Non-contractible loops are classified by the elements of the fundamental
group, or first homotopy group of $\M$, denoted $\pi_1(\M,\ph_0)$.
 Two loops based at $\ph_0$ are {\it homotopic\/} if one can be smoothly
deformed into the other without leaving $\M$.  This is an equivalence
relation; the equivalence classes, or homotopy classes of loops, are the
elements of $\pi_1(\M,\ph_0)$.  These classes have a group structure:
the identity is the class of contractible loops,
homotopic to the trivial loop which remains at $\ph_0$;
the inverse is the class comprising the same loops
traversed in the reverse sense; and the product is defined by traversing two
loops in succession.
It is intuitively clear that, if $\M$ is connected, then
$\pi_1(\M,\ph_0)$ does not depend on
the base point $\ph_0$, and so the first homotopy group is
often denoted simply by $\pi_1(\M)$.
If $\pi_1(\M)$ is trivial, comprising the identity
element only, then $\M$ is said to be {\it simply connected}.  A
necessary, but not sufficient, condition for the
existence of stable vortices is that $\pi_1(\M)$ be non-trivial,
or {\it multiply} connected.  Figure 2.2 depicts a multiply-connected
manifold: the dashed loop is in the identity class, while the solid
loop is not.  Hence $\pi_1(\M)$ has more than one element, and is
non-trivial.

If $G$ is chosen to be simply connected (which can always be done by going
to the `universal covering group', for example replacing SO(3) by its
two-fold covering group SU(2)), then an equivalent condition is that $H$
contains disconnected pieces.  In fact, if $g(2\pi)$ belongs to the connected
component, $H_0$ say, of $H$, then $g(\vp)$ can be smoothly deformed,
without
changing $\bphi$, so that $g(2\pi)$ becomes $\id$.  But then since any loop
in $G$ is contractible by hypothesis, so is $\bphi$.  In this case, the group
$\pi_1(\M)$ is isomorphic to the quotient group $H/H_0$, also denoted by
$\pi_0(H)$, whose order is the number of disconnected components of $H$.
Figure 2.3 shows a path joining two
disconnected parts of $H$.  Such a path would be mapped to a non-trivial
path in $G/H$, such as the solid line represents in figure 2.2.

The Abelian case discussed earlier may easily be included in this picture.
To do so, we have to replace U(1) by its covering group, the additive group
of reals, $\R$.  Then the isotropy subgroup is the group $\Z$ of integers
(transformations with phase equal to a multiple of $2\pi$).  Here
$\M=\R/\Z=S^1$, a circle, and $\pi_1(\M) = \pi_0(\Z) = \Z$.

If $\bphi$ is non-contractible, we may suspect the existence of stable
vortices, but there is certainly no guarantee.  For example, in the Abelian
theory, all loops with non-zero winding number $n$ are non-contractible, but
for $\beta>1$ and $|n|>1$ there are no stable vortices.  In any $n$-vortex
configuration, the vortices will repel each other; the energy can always be
lowered by expanding the spatial scale.  We shall encounter a more
interesting counter-example below.

To illustrate these ideas, let us consider the simplest non-Abelian gauge
theory, with $G=$ SO(3) and a scalar field $\ph=(\ph^a)$ in the vector
representation.  We take $V =\frac{1}{8}\la(\ph^2-\et^2)^2$, so here $\M$ is
the sphere in $\ph$ space of radius $\et$.  This is consistent because the
isotropy group of a fixed vector $\ph$ is $H=$ SO(2) and so $\M\simeq{\rm
SO(3)/SO(2)}\simeq S^2$.  Since all loops on the two-sphere are contractible,
$\pi_1(\M)$ is trivial and there are no vortex solutions.

It is not difficult, however, to extend the model to accommodate vortices.
Let us suppose there are two scalar fields $\ph_1$ and $\ph_2$ each in the
vector representation, and that the potential is chosen so that at the
minima, $|\ph_1|$ and $|\ph_2|$ are fixed and $\ph_1\cdot\ph_2=0$.  Then the
group $G$ is completely broken, i.e., $H$ consists of the identity element
only, so $\M\simeq$ SO(3).  To apply the general criterion, however, we
should replace $G$ by its covering group $\tilde G=$ SU(2).  Then the
isotropy group of a fixed pair of vacuum fields $(\ph_1,\ph_2)$ is $\tilde
H=\Z_2=\{\id,-\id\}$.  So $\pi_1(\M)\simeq\pi_0(\Z_2)\simeq\Z_2$.  The
single class of non-contractible loops in $\M$ comprises paths from the
identity to a $2\pi$ rotation (paths to a $4\pi$ rotation are trivial).

More detailed study reveals that there are two distinct vortex solutions in
this theory  [\ct{HinKib85}--\ct{AryEve87}]. The solutions have
the asymptotic forms
 $$
\bphi_A(\varphi)=e^{i\varphi M}\bphi_A(0),\qquad A^k =
M\hat{\varphi}^k/e\rho,
\label{\eSOAns}
$$
where $M$ is an element of the SO(3) algebra in the adjoint
representation.  For simplicity, let us take the case where
$|\phi_1|=|\phi_2|=\eta$.  If we set $\Phi = (\ph_1 + i\ph_2)/\surd2$,
we can resolve $\Phi$ into eigenvectors of $M$:
 $$
\Phi(\rho,\varphi)=\eta\sum_{n=-1}^{1}f_{An}(\rho)e^{in\varphi}\bi{e}_n,
\label{\eSOSca}
$$
where $\bi{e}_n\cdot\bi{e}_n^*=1$.
It can be shown that the ansatz (\rf{\eSOAns}) allows two solutions with
different forms:
\item{(i)} $\Phi =  \eta f_{+1}(\rho)e^{i\varphi}\bi{e}_{+1}$;
\item{(ii)} $\Phi = \eta [f(\rho)(e^{i\varphi}\bi{e}_{+1} -
e^{-i\varphi}\bi{e}_{-1}) + if_0(\rho)\bi{e}_0]$,
\par\noindent
as well as their charge conjugates.
In the first case the equations reduce to those of the Abelian string
(\rf{\eLocEom}).   Thus we find
a solution which is an embedding of this string in the SO(3)
model:  the scalar field changes phase by $2\pi$ at infinity, and vanishes at
the origin.  In the second case, there are two independent scalar fields,
$f(\rho) =  f_{+1}(\rho)=-f_{-1}(\rho)$, and $f_0(\rho)$.  The boundary
conditions are  $f(\infty)=\case{1}{2}$ and $f_0(\infty) = 1/\sqrt{2}$, while
at the origin only  $f(\rho)$ need vanish.

We can picture the two types of solution by denoting the magnitudes and
directions in internal `isospin' space of $\phi_1$ and $\phi_2$ by arrows, as
in figure 2.4.  The solid arrows represent $\phi_1$, and dotted
ones $\phi_2$.  This makes the difference between the two solutions especially
clear:  in (i) both fields $\phi_1$ and $\phi_2$ rotate around the origin,
while only $\phi_1$ does in (ii).  This makes it plausible that
vortex (ii) has lower energy, which is indeed the case
in the regions of parameter
space that have been investigated numerically
[\ct{AryEve87}].  This is because $\phi_2$ does not have to vanish at the core.

\subsection{Semilocal and electroweak vortices}
So far we have assumed that either all or none of the symmetries of the
scalar field theory are gauged.  Let us now suppose that only a subgroup is a
gauge symmetry.  The local and global symmetries must commute,
and so the general symmetry breaking pattern $G\to H$ has the form
$$
[\Gl\times\Gg]/D_G \to [\Hl\times\Hg]/D_H,
\label{\eSLSB}
$$
where the subscripts indicate whether the symmetry is local or global, and
$D_G$ and $D_H$ are possible discrete subgroups in common between the local
and  global groups.  We can move the scalar field to any point on its vacuum
manifold $\M$ by a transformation in $G$, and factoring out the little group we
have as before $\M \simeq G/H$.  The vacuum manifold need not separate into
local and global parts, $\Ml\simeq\Gl/\Hl$ and $\Mg\simeq\Gg/\Hg$.  If
$$
\M\not\simeq [\Ml\times\Mg]/D,
\eqqno
$$
with $D$ a discrete subgroup, the theory is called
{\it semilocal}. If there are vortex solutions they show
quite surprising behaviour, at odds with the intuition gained from pure gauge
vortices.

Let us illustrate some of these points with the original semilocal theory
[\ct{VacAch91}].
This has a complex scalar doublet $\Phi$ with components $\phi_1$ and $\phi_2$.
The Lagrangian is constructed so that it has total symmetry $U(2)\simeq
[{\rm SU(2)}\times {\rm U(1)}]/\Z_2$, but only the Abelian part is gauged:
$$
\L = -\frac{1}{4}F_{\mu\nu}F^{\mu\nu} + |D_\mu\Phi|^2 - \half\lambda(|\Phi|^2-
\half\eta^2)^2,
\eqqno
$$
where $D_\mu=\pa_\mu+ieA_\mu$ and $A_\mu$ is the Abelian gauge field.  This is
essentially the bosonic sector of the electroweak theory, in the limit in
which the Weinberg angle is $\pi/2$, so that the $W$ field decouples
[\ct{Vac92}].   In the ground  state, the potential ensures that $\Phi$ gains
an expectation value, which we  can choose to be $(0,1)\eta/\surd2$.  This
leaves unbroken a global U(1) symmetry,  generated by $(\tau_3+\id)/2$.  The
symmetry breaking pattern is therefore
 $$
[{\rm SU(2)}_{\rm g}\times {\rm U(1)}_{\rm l}]/\Z_2 \to {\rm U(1)}_{\rm g}.
\eqqno
$$
The gauge symmetry is fully broken, so
the gauge orbit space $\Gl/\Hl$ is isomorphic to a circle,
as for the ordinary Abelian Higgs model.  The full vacuum
manifold is the 3-sphere, defined by $|\phi_1|^2+|\phi_2|^2=\half\eta^2$.
Now,  although $S^3$ contains circles, it is not the direct product of a
circle with  any other space and thus, by the above criterion, the theory is
semilocal.

Recalling the discussion of section 2.3, the gauge orbits are topologically
non-trivial with respect to the first homotopy group, and thus we should expect
to find finite energy vortex solutions.  From the equations of motion, we can
see immediately that one vortex solution is
 $$
\Phi ={\eta\over\surd2} {0\choose{f(\rho)e^{i\varphi}}}, \qquad A^k =
\hat{\varphi}^k{{a}(\rho)\over e\rho},
\label{\eSemLoc}
$$
where ${f}$ and ${a}$ are the functions introduced in (\rf{\eNOAns}) for
the Nielsen--Olesen vortex.  Others are obtained by global SU(2) rotations.

We issued a warning
earlier that the topological condition did not guarantee the existence of a
stable vortex solution.
In the case at hand, $\M$ contains $\Ml$ but is not identical to it,
and it  happens that non-contractible loops in $\Ml$ are contractible in
$\M$.  In  that case it is possible, by allowing the upper component
of $\Phi$ to be non-zero, to construct field configurations which
lie in $\M$  everywhere in the plane (still reaching $\Ml$ at infinity).
These configurations have
vanishing  potential energy density, resulting  in a tendency to
spread out in order to minimize the total gradient and magnetic field
energies [\ct{Pre92}].

The result of a detailed analysis of this spreading instability
[\ct{Hin92},\ct{Hin93a}] is that the embedded
Nielsen--Olesen vortex is stable only for $\beta<1$.  If $\beta>1$ the string
is
unstable and the vortex spreads out to infinity. At
the borderline, where $\beta=1$, the vortices can be any size, and there are
also stable multivortex solutions. We refer the reader to references
[\ct{Ach+92},\ct{Gib+92}] for more information.

The significance of semilocal strings is that they violate the canonical law
about gauged topological defects: that stable defects exist if the relevant
homotopy group is non-trivial.  This has implications for cosmology, for one
has to careful in predicting whether a particular theory will make strings in
the early universe.  It has also been shown that the perturbative stability of
the $\be<1$ string persists when one gauges the SU(2) symmetry, provided the
Weinberg angle remains close to $\pi/2$ [\ct{JamPerVac93a}], although
for physical values of the Weinberg angle these electroweak strings are
unstable [\ct{JamPerVac93b}].
However, they do have some features in common with the sphaleron
[\ct{HinJam94},\ct{BarVacBuc94}], another
unstable defect of the Standard Model, so electroweak strings may still have an
important role to play.

\subsection{Composite defects: strings and domain walls}
Realistic theories have several stages of symmetry breaking, each associated
with a Higgs field gaining an expectation value.  As the early Universe cooled
it would have gone through a series of phase transitions  (see section 5.1)
at which the symmetry was progressively reduced, so a symmetry breaking
sequence  arranged in order of decreasing symmetry can be
thought of as a sequence in time.  This section is concerned with studying
situations where $H$ is not the  final unbroken symmetry.

Let us first consider the sequence
$$
G \breaksto_\eta^\phi H \breaksto_v^\chi H',
\eqqno
$$
where the letter over each arrow denotes the field which breaks the symmetry,
and the letter underneath the scale of the symmetry breaking.  We
assume that $\pi_1(G/H)$ is non-trivial.  Without
loss of generality we may take $G$ to be simply connected, in
which case  $H$ must be disconnected, and $\pi_1(G/H) \simeq \pi_0(H) \simeq
H/H_0$. For simplicity we also suppose that $v\ll\eta$, so that when $\chi$
gets an  expectation value, any change in $\phi$ can be ignored.

The direction of $\chi$ in the vacuum is partly determined by its coupling
to $\phi$ in the potential $V(\phi,\chi)$.  Let us suppose that $V$ is
minimized at $(\phi_0,\chi_0)$, and normalized to zero at the minimum.  Then
$V(\phi_0,g\chi_0) \ne V(\phi_0,\chi_0)$ unless $g\in H$ or
unless there were an enlarged symmetry.
\footnote{*}{If one could make separate transformations on both fields, this
would enlarge the symmetry group to $G_{\rm l}\times G_{\rm g}$, or some
subgroup thereof.  In this case the string becomes `frustrated'
[\ct{HilKagWid88}].}
In the string background
$$
\phi(\varphi) = g(\varphi)\phi_0, \qquad
A^k = -ig \pa^k g^{-1}/e = M\hat\varphi^k/e\rho,
\eqqno
$$
and it follows that there will be a
quadratic divergence in the potential energy per unit length
unless $\chi$ `follows' $\phi$ around the string, or
$$
\chi(\varphi) = g(\varphi)\chi_0.
\eqqno
$$

Consider for example the model whose symmetry breaking pattern is
$$
{\rm SO(3)} \breaksto_\eta^{\bf 5} {\rm O(2)}
\breaksto_v^{\bf 3} {\rm SO(2)},
\eqqno
$$
where the boldface numbers indicate the dimension of the representation.  The
5-dimensional representation of SO(3) is a traceless symmetric tensor $\Phi$,
while the {\bf 3} is just the adjoint representation of SO(3) vectors.  The
sequence can be arranged with a suitable potential containing couplings
$\chi^{\rm T}\Phi\chi$ and $\chi^{\rm T}\Phi^2\chi$, such that
$$
\Phi_0 = \eta \,{\rm diag}(-1,-1,2)/\surd{6},\qquad \chi_0 = v(0,0,1).
\eqqno
$$
The intermediate O(2) then consists of matrices $\exp(i\al T_3)$ and
$\exp(i\pi T_1)\exp(i\al T_3)$.  The disconnected component of O(2) changes
the sign of $\chi_0$,  and so the final symmetry is SO(2).

After the first symmetry breaking, strings will form.  A typical
string configuration at large distances may be written
$$
\Phi = e^{i\varphi T_1/2} \Phi_0 e^{-i\varphi T_1/2}, \qquad A^k
=T_1 \hat{\varphi}^k/2e\rho.
\eqqno
$$
Thus, as $\chi$ follows $\Phi$ around the string it comes back to $\exp(i\pi
T_1)\chi_0 = -\chi_0$ [\ct{HilKagWid88}--\ct{Hin89}].  The problem
is a discontinuity in $\chi$ at $\varphi=2\pi$, where it  must change from
$-\chi_0$ back to $\chi_0$.  The best that can be done is
to allow $\chi$ to leave its vacuum manifold, and possibly vanish, on one or
more planes of constant $\varphi$, creating structures called {\it domain
walls} [\ct{ZelKobOku75},\ct{KibLazSha82}].
Domain wall solutions appear in theories with disconnected vacuum
manifolds, and can be
characterized by two quantities: their thickness, which is
the Compton wavelength of $\chi$ in the vacuum, $m_\chi^{-1}$; and their
surface energy $\sigma \sim V_{\rm w}m_\chi^{-1}$, where $V_{\rm w}$ is the
potential energy density at the centre of the wall.

Topologically, the string exists because
$\pi_1({\rm SO(3)/O(2)}) \simeq \Z_2$.  The
existence of the domain wall depends on the properties of the element
$\htil=g(2\pi)\in H$; in this case
$\htil = {\rm diag}(1,-1,-1)$.  Domain walls appear if $\htil$ is not
equivalent to the identity under the action of $H'$, and thus lies
in a non-trivial class of $H/H'$.   This condition is
satisfied here, for $\htil$ is invariant under $H'$, which in this
case is the SO(2) generated by $T_1$, and clearly
inequivalent to the identity.

Similar symmetry breaking sequences occur in SO(10) models, where there is
a discrete charge conjugation symmetry at intermediate energies,
analogous to $\htil$, which is broken at a lower scale [\ct{KibLazSha82}].

\subsection{Composite defects: strings and monopoles}
Let us now suppose that $G$ is not the full symmetry group of the theory.
There will then be a sequence
$$
G'\breaksto^\sigma_s G \breaksto^\phi_\eta H,
\eqqno
$$
and we shall assume that the scales $s$ and $\eta$ are well separated.  If we
were unaware of the existence of the larger symmetry, we would predict strings
if $\pi_1(G/H) \not\simeq \{\id\}$.  The topological stability of these strings
ultimately depends on the fundamental group of the true vacuum manifold,
$\pi_1(G'/H)$.  By analogy with the above discussion of domain
walls and strings, we might guess that it might be possible to break strings
corresponding to trivial elements of $\pi_1(G'/H)$ by creating a pair of
point defects, or monopoles [\ct{Vil82}].\footnote*{Monopoles
[\ct{GodOli78},\ct{Col88},\ct{Pre87}]
play a similar role in 3+1-dimensional Yang--Mills--Higgs theories to
vortices in 2+1 dimensions.  Gauge monopoles have
finite energy, while the energy of global monopoles is linearly divergent
[\ct{BarVil89}].}

It can be shown that if $G'$ is simply connected then monopoles exist if and
only if $G$ is not.  To see heuristically how this comes about, let us again
choose the radial gauge $\hat{x}^kA_k = 0$, in which the  scalar field
$\sigma$ has a well-defined value on the 2-sphere at infinity  $S_\infty^2$,
and maps it into the vacuum manifold $\M' \simeq G'/G$.  The  monopole field
at infinity can be written
 $$
\si(\theta,\varphi) = g'(\th,\varphi)\si(N),
\qquad A_k = -ig'\pa_kg'^{-1}/e,
\eqqno
$$
where $N$ is the north pole $\th=0$.   The image of $g'$ in
$G'$ can be thought of as a smooth set of loops $g'(\tau,\psi)$ ($\tau,\psi
\in  [0,1]$), fixed at $N$ for $\psi=0,1$.  These loops sweep
over the whole sphere as $\tau$ and $\psi$ vary over their ranges
(see figure 2.5).
Without loss of generality, we can choose
$g'(0,\psi)=\id$.   Then, in order for $\si(\th,\varphi)$ to be continuous
at $N$, the loop $\tilde g(\psi)\equiv g'(1,\psi)$ must lie in
$G$.  If $G$ is multiply connected, and $\tilde g$ is non-contractible
in $G$, $g'(\tau,\psi)$ cannot be
continuously deformed to the trivial constant configuration without
$g'(1,\psi)$
leaving $G$.  Therefore, one cannot eliminate the monopole without
introducing discontinuities in the field.

When the symmetry group $G$ is broken further to $H$ by the field $\phi$, the
potential forces $\phi$ to follow $\sigma$:
$$
\phi(\th,\varphi) = g'(\th,\varphi)\phi(N).
\eqqno
$$
However, if $\tilde{g}(\psi)$ is not in $H$, $\phi$
must fall into a string configuration near $N$ to maintain
continuity.  In this way monopoles become attached to strings in subsequent
stages of symmetry breaking.  Conversely, strings can break by the creation of
monopole-antimonopole pairs [\ct{Vil82},\ct{LazShaWal82},\ct{PreVil93}].

As an instructive example, let us take the SO(3) model of section 2.3, and
separate the scales of the fields $\phi_1$ and $\phi_2$, so
that the breaking sequence expressed in terms of the covering group is
$$
{\rm SU(2)}\breaksto_{\eta_1}^{\phi_1} {\rm
U(1)}\breaksto_{\eta_2}^{\phi_2} \Z_2,
\eqqno
$$
where $\Z_2=\{\id,-\id\}$.
After the first stage the theory supports
monopoles, since $\pi_1({\rm U(1)}) \simeq \Z$.  There
are strings of the Nielsen--Olesen type associated with the second stage,
because $\pi_1(G/H) \simeq \Z$.  The significance of the unbroken $\Z_2$ is
that strings with winding number $n=1$ are stable while those with $n=2$ (or
any even $n$) can be
broken by creating monopole-antimonopole pairs.  A  flux 2 string is in the
same topological class as two flux 1 strings: this  leads one to envisage a
configuration in which the monopole is attached to two  flux 1 strings at the
north and south poles.  This topologically stable  configuration, in which the
string `threads' the monopole, is known as a {\sl bead} [\ct{HinKib85}].
This situation is depicted in figure 2.5: the non-contractible loop in
$G$, which defines the monopole, passes through the two disconnected
components of $H$.  This path can be divided into two segments,
each of which is a closed loop in $G/H$, and thus represents a string.

Beads can also be thought of as kinks [\ct{Raj82}]
interpolating between topologically
distinct string solutions.  At first sight there appears to be only one type of
string, since there is only one non-trivial class in $\pi_1(G'/H)$.  However,
the vacuum possesses a discrete symmetry which is broken by the string
solution,
given for this model in section 2.3.
If we write the vacuum $\Phi_0 = (\eta_1,
i\eta_2,0)/\surd2$, then this discrete symmetry is $\Phi_0\to
{\rm diag}(1,1,-1)\Phi_0$.  This is violated by type (ii) strings, but not by
type (i).  This broken symmetry means that there are two distinct type (ii)
strings which can be smoothly deformed into each other via a type (i)
configuration, which has higher energy.  This is the bead.  The number of
different string solutions is clearly related to the number of discrete
vacuum symmetries broken by the string solution.  If there are more
than two, then the bead can be the junction of several strings [\ct{HinKib85},
\ct{Ary+86},\ct{VacVil87}].

\subsection{Superconducting strings: bosonic currents}

Models with extra
fields can often support currents in the core of the string [\ct{Wit84}],
which can have quite
dramatic dynamical effects, as we shall see in section 3.4.  If the
currents are electromagnetic, then the string behaves as a thin superconducting
wire, with an enormous critical current by terrestrial standards.

Consider an Abelian U(1)$\times$U(1) model with Lagrangian
$$
\L = |D_\mu\phi|^2+|D_\mu\chi|^2-V(\phi,\chi) -
\frac{1}{4}G_{\mu\nu}G^{\mu\nu}-\frac{1}{4}F_{\mu\nu}F^{\mu\nu},
\eqqno
$$
where $D_\mu\phi=(\pa_\mu+ieB_\mu)\phi$, $D_\mu\chi=(\pa_\mu+iqA_\mu)\chi$,
$F_{\mu\nu}$, $G_{\mu\nu}$ are field strengths associated with $A_\mu$ and
$B_\mu$, respectively, and the potential is given by
 $$
V(\phi,\chi) = \half\la_1(|\phi|^2-\half\eta^2)^2+\half\la_2|\chi|^4
+\la_3(|\phi|^2-\half v^2)|\chi|^2.
\eqqno
$$
Let us suppose that the parameters are chosen so that the
minimum of $V$ is at $|\phi|=\eta/\surd2$, $|\chi|=0$, so that the U(1)
symmetry associated  with $A_\mu$ remains unbroken; this we consider
to be the electromagnetic U(1).
(This minimum is ensured by having $v^2<\eta^2$ and
$\la_3^2v^4<\la_1\la_2\eta^4$).  As in the case of the semilocal
vortex, there is a solution in which $\phi$ and $B_\mu$ make up a
Nielsen--Olesen vortex, and $\chi$ and $A_\mu$ vanish.

However in large regions of parameter space, a solution is preferred in which
$\chi\ne0$ in the core [\ct{HilHodTur88},\ct{HawHinTur88},\ct{DavShe88a}].
For given couplings, there is a critical value of the ratio
$r=m_\chi^2/m_\phi^2$, where $m_\chi^2 = \half\la_3(\et^2-v^2)$ and
$m_\phi^2 = \la_1\et^2$, below which a $\chi$ condensate exists.   This
ratio tends to 1 as $\la_3/\la_1 \to \infty$, which is when the fields
are strongly coupled to each other.  Given a condensate,
there is, in fact, a family of solutions, for if $\chi=X(\rho)/\surd2$ is a
solution to the field  equations then so is $\chi=X(\rho)e^{i\al}/\surd2$.
This phase constitutes an  extra, `internal' degree of freedom for the
string.  This is the Nambu-Goldstone boson of the electromagnetic U(1)
symmetry,  broken in the core of the string.

Suppose the phase varies linearly with time and position, \ie\ that
$\al = kz -\om t $. Then the string carries an
electromagnetic current
$$
J^\mu = +iq\chi^*{\dbw}^\mu\chi - 2q^2A^\mu|\chi|^2
= qX^2(\om-qA^0;0,0,k-qA^3).
\eqqno
$$
Now let us compute $\pa_tJ^3$ in
the gauge $A_0=0$.  Providing the current is sufficiently small for us to
ignore
time-dependent back-reaction terms on $X$, we have
 $$
{\pa J^3\over\pa t} = q^2X^2E^3.
\label{\eLondon}
$$
Thus the current increases linearly in time in proportion to the electric
field: this is the London equation, characteristic of a superconductor
[\ct{SupCon}].
Borrowing from our knowledge of superconductors, we see that if the penetration
depth $\sim (qX(0))^{-1}$ is much greater than the width of the condensate, the
applied electric field changes very little over the cross-section of the
string, so we may integrate equation (\rf{\eLondon}) to get
$$
{\d I\over \d t} = 2q^2\kappa E^3,
\label{\eBosCur}
$$
where $I$ is the total current and $\kappa={1\over2}\int \d x \d yX^2$.
The current  cannot continue increasing indefinitely: just as in ordinary
superconductors  there is a critical current above which the string goes
`normal'.   Substituting $\chi = X e^{ikz}/\surd2$, which corresponds to a
current $\sim qk\ka$, the effective mass-squared
ratio is increased:
$$
r\to r^{\rm eff} = r+I^2/4q^2\ka^2m_\phi^2.
\eqqno
$$
Thus, if $I^2$ becomes too large, we leave the region of parameter space
where the $\chi$ condensate exists. Since
there is an implicit dependence on $X$ in this equation, it is not
straightforward to calculate the critical current $I_{\rm c}$ from the
parameters of the potential.  Roughly speaking, we would expect $I_{\rm c} \sim
q\ka m_\phi$.  Numerical work [\ct{HawHinTur88},\ct{DavShe88a},
\ct{BabPirSpe88}--\ct{Pet93}] bears this out,
showing that the implicit
dependence of $I/\ka$ on $X(\rho)$ causes the quenching to occur very rapidly
at $I_{\rm c}$ [\ct{DavShe88a}].

Current can also be lost by quantum tunnelling
[\ct{Wit84},\ct{Zha87},\ct{HawHinTur88}].  The rate per unit length
of this process is estimated to be [\ct{HawHinTur88}]
$$
{\d \Gamma \over \d z} \sim m_\chi^2 e^{-2\pi\ka}.
\eqqno
$$
Thus strings with long-lived currents must have large values of $\ka\sim
v^2/m_\chi^2$, meaning  that the condensate must be large and wide.
Stability against quantum tunnelling then turns out to favour
the  region of parameter
space where $\la_2$ is small [\ct{HawHinTur88}].

To summarize, a bosonic superconducting string in this
U(1)$\times$U(1) model must be constructed from fields
which are weakly self-coupled but strongly coupled to each other, and
preferably with a large difference between the masses of the string field
$\phi$ and the current-carrying field $\chi$.

More general cases can be analysed.
The underlying point is that the vortex solution need not preserve
the vacuum symmetry $H$.  If it does not, there then exist transformations,
generalizations of $\chi \to \chi e^{i\al}$, which generate
new vortex solutions [\ct{Eve88},\ct{Hin89},\ct{Alf+91}].
A supercurrent is set up on the string
when the transformations are space- and time-dependent.  The analysis
is complicated if $H$ does not commute with the generator of flux on
the string, $M$, for it is then conjugated by $g(\vp) = \exp(i\vp M)$ around
the string, giving a position-dependent isotropy group:
$$
H(\vp) = g(\vp) H(0) g^{-1}(\vp).
$$
It is certainly true that $H(2\pi)=H(0)$, and hence $\htil = g(2\pi)$ is
an inner automorphism of $H(0)$.  If it is trivial, so that
$\htil h \htil^{-1} = h$ for all $h \in H(0)$, then the isotropy group is
well-defined around the string and it is possible to show that there
are bosonic zero modes corresponding to every generator of $H(0)$
acting non-trivially on the string fields [\ct{Eve88},\ct{Alf+91}].  If
the automorphism is not trivial, then $H(0)$ is said to be globally
unrealizable, and the vortices are so-called `Alice' strings
[\ct{Sch82},\ct{SchTyu82}].

As one might expect from their name, Alice strings have curious properties.
Any particle transported around the string comes back conjugated by
$\htil$.  In many cases (the SO(3) string of section 2.5 is one of them),
$\htil$ is actually the charge conjugation operator, and so particles
come back as their own antiparticles.  Charge conservation
is maintained by the string's acquiring a balancing charge, although it
is charge of a rather peculiar nature, for it cannot be localized
anywhere (continuing the theme,
this has been dubbed `Cheshire' charge [\ct{Alf+91}]).
The reason it cannot
be pinned down is that it is actually a non-Abelian charge, and thus is
not gauge-invariant [\ct{Eve93}].
Alice strings can also carry magnetic charge [\ct{BucLoPre92}],
and a closed loop can
collapse to form a monopole (indeed, this process  can be observed
in nematic liquid crystals [\ct{Chu+91}]).
It is perhaps unfortunate for students of the bizarre that there
can be no Alice strings in the Universe today, for charge conjugation
is not a symmetry of the Standard Model of particle physics.

\subsection{Superconducting strings: fermionic currents}
It is also possible for currents to be carried by fermionic degrees of freedom
confined to the core of the string [\ct{Wit84}].  Let us extend our bosonic
U(1)$\times$U(1) symmetric Lagrangian to include a fermionic sector
$$
\L_{\rm f} = \bar\psi_{\rm l}i
\Slash{D}\psi_{\rm l} + \bar\psi_{\rm r}i
\Slash{D}\psi_{\rm r} -
g\phi\bar\psi_{\rm l}\psi_{\rm r} - g\phi^*\bar\psi_{\rm r}\psi_{\rm
l},
\label{\eFerLag}
$$
where $\psi_{\rm r}$ and $\psi_{\rm l}$ are Weyl spinors with chirality $\pm1$.
The coupling to the $\phi$ field constrains the couplings to $B_\mu$: they
must  differ by an integer multiple of $e$.  We choose $\pm e/2$, and we will
also  suppose that the electromagnetic charges are both
$q/2$.\footnote{*}{This  theory is anomalous: however, since the gauge
symmetry is
violated in an instructive  way, we will tackle the problem later.}

In the vacuum background $|\phi|=\eta/\surd2$, this Lagrangian describes a
Dirac fermion with mass $m_{\rm f}=g\eta/\surd2$. In the vortex background
it appears that the fermion mass is position dependent, and vanishes at the
core of  the string.  Thus there could be massless states confined to the
string.  There is indeed a solution to the transverse Dirac equation  with
zero eigenvalue [\ct{CarDeGMat63},\ct{JacRos81}], $$
\eqalign{
i\gamma^AD_A\psi_{\rm l} - g\phi\psi_{\rm r} &= 0, \cr
i\gamma^AD_A\psi_{\rm r} - g\phi^*\psi_{\rm l} &= 0. \cr}
\eqqno
$$
It has the form
$\psi_{\rm r} = \ze_{\rm r} P[\bar\phi,\bar B]$,
where $\zeta_{\rm r}$ is a constant right handed spinor satisfying
$i\ga^1\ga^2\zeta_{\rm r} = \zeta_{\rm r}$, and $P$ is a function of
the background vortex fields
$\bar{\phi}$ and $\bar{B}$.  At large $\rho$, $P$ goes as
$\exp(- m_{\rm f}\rho)$, showing the
localization of the state.

Now, we can generate new space- and time-dependent solutions $\psi_{\rm
r,l}\exp{i\beta(t,z)}$, provided
$$
i[\ga^0\pa_t + \ga^3\pa_z]\psi_{\rm r,l}\exp{i\beta(t,z)} = 0.
\eqqno
$$
Using $\ga^0\ga^3\zeta_{\rm r} = \zeta_{\rm r}$, we see that $\beta = \beta(t-
z)$ solves this equation.  Thus the string supports modes moving at the speed
of light in the $+z$ direction only, which we call right-moving.
If we want a left-moving fermion, we should couple $\bar\psi_{\rm
l}\psi_{\rm r}$ to $\phi^*$ instead of $\phi$.

These massless fermionic solutions are often called zero modes.
Their existence is actually a consequence of the topology of the background
field: it is possible to show by quite general methods [\ct{Wei81}] that in
the winding number $n$ sector of the field configuration space there are at
least $2|n|-1$ zero modes.

Let us now consider the behaviour of our string-fermion system under an applied
electric field.  In the  ground state,  all
levels of the Dirac sea up to zero energy are occupied,
much like a 1D metal with zero Fermi momentum $k_{\rm F}$.
In the presence of an electric field $E^3$ applied along
the string, the occupied states move under the Coulomb force,
so that after time
$\Delta t$ the Fermi surface is at $\half q \int_{\Delta t}E^3(t) \d t$.
Depending on which way the surface moves, fermions or antifermions are created.
The 1D density of states is $1/2\pi$, so the result is the appearance of a
current $I$ at a rate
$$
{\d I\over \d t} = {1\over 2\pi}\({q\over 2}\)^2 E^3.
\eqqno
$$
This is identical in form to (\rf{\eBosCur}), the equation for the current on a
bosonic superconducting string in an electric field.  However, it is not
strictly fair to call this a superconductor:  it is more precisely a perfect
conductor.  A metal would also behave in this way if there were no impurities
to scatter fermions from one side of the Fermi surface to the other.

It may have been noticed that we are
creating a charge density on the string at the same rate as the current
increase.  This type of violation of charge conservation at the quantum
level is known as an {\it anomaly} [\ct{Anom}].
The classical field theory with Lagrangian (\rf{\eFerLag})
has two conserved currents coupled to the two gauge
fields, but adding the fermion gives anomalies to both
[\ct{Wit84},\ct{Wid88},\ct{HilLee88}].
We may restore charge conservation to the theory by adding another fermion
with charges $(-e/2,q/2)$, a procedure known as `cancelling' the anomaly.
This fermion
couples to $\phi^*$ where the first coupled to $\phi$, and so produces a set of
left-moving zero modes.  When the electric field is applied along the string,
we create (assuming both $q$ and  $E^3$ positive) left-moving {\it
anti\/}particles of charge $-q/2$ at the same rate as the right moving
particles, and thus charge is conserved.  The antiparticles contribute to the
current with the same sign as the particles, so the rate of current increase is
doubled:
${\d I/ \d t} = {q^2E^3/ 4\pi^2}$.

Another way to ensure a consistent theory is to set $e=0$.  This means that we
are considering fermions in the background of a global string (section 2.1).
There still seems to be a violation
of charge conservation, but a careful calculation of the anomaly in
the global theory shows that the scalar field contributes an extra
piece to the current [\ct{CalHar85}]:
$$
j^\mu = \half q\langle\bar\psi\ga^\mu\psi\rangle -i{q\over 32\pi^2}
\ep^{\mu\nu\rho\si}F_{\nu\rho} {\phi^*{\dbw}\!_\si\phi\over|\phi|^2}.
\eqqno
$$
In the string background, ${-i\phi^*{\dbw}\!_\si\phi/2|\phi|^2}$ is in the
azimuthal direction, so an electric field along the string induces an inward
radial current of just the right magnitude to keep the charge conserved on the
string.  This is an important case, for it models the interaction of an axion
string with an ordinary fermion of the Standard Model [\ct{LazSha85}].

It was mentioned above that fermionic superconducting strings should really be
called perfect conductors.  If there were a way for right-moving fermions to
scatter into left-moving states, then the string would behave like an ordinary
conductor and exhibit resistance: the current would stop growing when the
scattering rate matches the accelerating effect of the electric field.  In our
fully gauged model, charge conservation ensures that this cannot happen.
The only way that current can be lost is if the zero modes scatter off each
other
with enough energy to produce a fermion with sufficient energy to escape the
string.  The threshold is when the sum of the Fermi momenta equals the sum of
the fermion masses.  This is the analogue of the critical current in the
bosonic superconductor.
However, models can be constructed which do exhibit
resistance [\ct{BarMat87}],
although they are somewhat complicated.  Whether or not this
complexity is in some sense generic to grand unified theories, and fermionic
superconducting strings are therefore rare, is another question.

In general, one might expect to see both bosonic and fermionic
superconductivity present at once.  The possibility then arises that the
bosonic condensate $\chi$ couples left and right-movers together
[\ct{HilWid87},\ct{Dav87b},\ct{Hin88}].
Remarkably, in the light of the above argument, this does not
destroy the superconductivity, unless $\chi$ is electrically neutral and the
sum of the electric charges of the zero modes is zero [\ct{Hin88}].

\subsection{Strings in unified theories}
As mentioned above, the electroweak theory does not possess any topologically
stable string solutions, because its gauge orbit space is isomorphic to $S^3$,
which is simply connected.
Thus if cosmic strings
exist in nature, they must arise from the breaking of as yet unknown
symmetries, perhaps those of a grand unified theory or GUT [\ct{GUT}].
In a GUT, where
the gauge interactions are unified in a simple compact Lie group, stable gauge
strings require an unbroken discrete subgroup at low energies.

Minimal SU(5) grand unification [\ct{GeoGla74}]
does not incorporate any such discrete symmetries:  the smallest
group to do so is SO(10), or rather its simply connected covering group
Spin(10) [\ct{SO10GUT}].
In this scheme, each family of fermions is supplemented by a
left-handed antineutrino and assembled into a spinorial {\bf 16}.
A possible discrete symmetry $D$ is then
just $-\id$ in Spin(10).  In order to leave $D$ unbroken, subsequent symmetry
breakings must be performed by scalar fields invariant under this element.
Such representations can be constructed from the tensor product of an even
number of spinor {\bf 16}s and $\overline{\bf 16}$s.  The natural
representation to  use is the {\bf 126}, which is in the symmetrized product
{\bf 16}$\times${\bf  16}, for it alone
gives the left-handed antineutrinos
Majorana masses at tree level via the `see-saw' mechanism [\ct{Seesaw}].

The symmetry breaking scheme Spin(10) $\breaksto\limits^{\bf 126}$ SU(5)
$\times$ $\Z_2$ is closely analogous to the breaking
SU(2)$\breaksto\limits^{{\bf
3}\times{\bf 3}} \Z_2$ discussed earlier.  The Spin(10)
string is indeed very similar to the SO(3) string of section 2.3
[\ct{AryEve87},
\ct{Yaj87}].

{}From a phenomenological point of view, this Spin(10) theory is similar to
SU(5)  unification, and so supersymmetry seems to be necessary for the
consistency  of the scheme [\ct{AmaDeBFur91}].
There are many other types of SO(10) unification [\ct{SO10GUT}]:
non-supersymmetric ones tend to go via S[O(6)$\times$O(4)] at the grand
unification scale, with SU(2)$_{\rm R}$ surviving to lower energies.
The strings produced at the first symmetry breaking
do not survive to low energies [\ct{KibLazSha82}]
and would therefore not be important for
producing density perturbations (see section 6), while stable strings produced
much below $10^{15}$ GeV could only be astrophysically important if they were
superconducting [\ct{Chu+86}--\ct{MalBut89}].

There may also be spontaneously broken global symmetries in nature.  The most
commonly considered is an axial U(1)$_{\rm A}$ or Peccei-Quinn symmetry
[\ct{PecQui77}--\ct{Wil78}],
which rotates the phases of the left- and right-handed fermions in the opposite
sense.  This can only be accommodated in a model with at least two Higgs
doublets.  Its breaking allows global strings of the type discussed in
section 2.1.  However, axion strings are in a category of their own,
because the axial symmetry is anomalous.  Under normal circumstances,
the axion field $a$, which is essentially the phase of a complex scalar
field, would be a massless Goldstone boson. At the quantum level however,
it gains a temperature-dependent potential,
via its anomalous coupling to instantons, which
are topologically non-trivial configurations of the gluon field
[\ct{GroPisYaf81}].  This potential has the form
$$
V(a) = \Om(T)[1-\cos(Na/\fa)]
\eqqno
$$
where $N$ is a model-dependent integer, and $\fa$ is the expectation value
of the field breaking the Peccei-Quinn symmetry, also known as the
axion decay constant.  $\Om(T)$ increases with decreasing temperature,
giving the axion a mass-squared $\ma^2 = N^2\Om(T)/\fa^2$.
At temperatures which are low in comparison to the QCD scale of
$\sim 100 $ MeV, a current algebra calculation gives the mass
directly: $\ma = m_\pi f_\pi/\fa$, where $m_\pi$ and $f_\pi$ are the
pion mass and decay constant respectively.  The axion decay constant
also controls the strength of the coupling of the axion to
ordinary fermions.  This coupling is axial, of the form
$i(m_{\rm f}/\fa)a\bar\psi\ga_5\psi$.  Thus, the larger the scale $\fa$
the smaller the couplings to fermions.
There are very stringent astrophysical and cosmological bounds
on $\fa$.  If $\fa$ is too low, the couplings to ordinary fermions become
sufficiently large to affect calculations of stellar models
[\ct{Dic+78}].  The strongest astrophysical constraint comes
from SN1987a [\ct{EliOli87}], which puts $\fa \gap 10^{9-10}$ GeV.
Radiation from axion strings gives a cosmological {\it upper\/}
bound of about $10^{11}$ GeV [\ct{Dav86},\ct{HarSik87},\ct{DavShe89b}].
We shall say more about this in section 5.

Larger global symmetries, such as family symmetry [\ct{Wil82}],
have been proposed.
If there are three families of fermions, the
symmetry can only be SU(2) or
SU(3), with a possible U(1) factor for the overall phase of the fermions.
There are many possibilities for topological defects in
models with family unification [\ct{JoyTur94}], and several models with
strings have been proposed [\ct{BibDva90}].  Perhaps the most intriguing
proposal is for quaternionic strings [\ct{DvaSen94}], where the first
homotopy group is the non-Abelian group of quaternions.  This results
in strings with rather unusual proporties: for example,
they can come together in 3-way junctions.  Although such strings
are seen in biaxial nematic liquid crystals, very little is
known about their cosmological effects.

\section{String dynamics}\ssf
\subsection{Equations of motion}

So far we have found a rather limited class of solutions to the field
equations: straight, static strings.  Here, we seek an effective action whose
extrema give moving string solutions.  The idea is to start from the field
theory
action and to reduce  the number of degrees of freedom from those of a
four-dimensional field theory  down to the coordinates of the two-dimensional
string worldsheet.   We consider only the case in which the curvature of the
worldsheet is small when compared to the width of the string.

We begin by defining the position of the string as the coordinates of the
zeroes of the Higgs field $\phi(x)$, which we denote $X^\mu(\si^a)$
($a=0,1$).  (We ignore technical complications posed by non-Abelian theories
whose strings do not necessarily have a zero of the Higgs field.)  There is a
metric $\ga_{ab}$ induced on this surface by the embedding in the background
space-time, given by
 $$
\ga_{ab} = \pa_aX^\mu\pa_bX^\nu g_{\mu\nu}.
\eqqno
 $$
The Lagrangian density for a static solution in a flat background space-time
[\ct{NieOle73},\ct{For74}] is just the negative of the energy density.  In
terms of the energy per unit length, $\mu$, the action for a straight string
on the $z$ axis is therefore the Nambu-Goto action
[\ct{GodGolRebTho73},\ct{NamGot70}]
 $$
S_0 = -\mu\int\d t\,\d z = -\mu \int\d^2\si\sqrt{-\ga}.
\label{\eNamGot}
 $$
This last expression is generally covariant both in two and
in four dimensions, so it holds for any background metric $g_{\mu\nu}$ and for
any  embedding $X^\mu(\si)$, providing the string remains close to being
straight.  This is the first term in the expansion of the effective action for
a string in powers of the curvature; we shall consider the next term
briefly below.

Let us first study the equations of motion of the pure Nambu-Goto string.
Two of the three  degrees of freedom of the worldsheet
metric may be removed by using 2D reparametrization (general coordinate)
invariance [\ct{GreSchWit87}]. That is, we can always make a coordinate
transformation $\si^a
\to \tilde\si^a(\si)$ such that the metric takes the form
 $$
\ga_{ab} = \Om^2(\si)\eta_{ab},
\eqqno
 $$
where $\eta_{ab}={\rm diag}(1,-1)$.  This is known as the conformal gauge.
Let us call the timelike worldsheet coordinate $\tau$ and the spacelike one
$\si$, and denote differentiations of $X$ with respect to $\tau$ and $\si$
by $\Xd$ and $\Xp$ respectively.
Then the conformal gauge imposes the constraints
 $$
\Xd^2+\Xp^2 =0, \qquad \Xd\cdot\Xp=0,
\eqqno
 $$
and the equations of motion obtained by varying $S_0$ are
 $$
\ddot X^\mu - \dprime{X}^\mu + \Ga_{\nu\rho}^\mu(\Xd^\nu\Xd^\rho-
\Xp^\nu\Xp^\rho)=0.
\label{\eConEqu}
 $$

The string stress tensor $T^{\mu\nu}(x)$ is obtained by variation of the
action with respect to $g_{\mu\nu}$, giving
 $$
\sqrt{-g}T^{\mu\nu}(x) = \mu\int\d^2\si\sqrt{-\ga}\ga^{ab}\pa_aX^\mu\pa_bX^\nu
\de_4(x-X).
\label{\eNGStTen}
 $$

If the string curvature is small but no longer negligible, we may consider
an expansion in powers of the curvature.  The relevant measure of curvature
is given by the extrinsic curvature tensor $K_{ab}^A$ ($A=1,2$), which is
computed in terms of an orthogonal pair of spacelike unit normals $n_A^\mu$,
satisfying
 $$
n_A\cdot n_B = -\de_{AB}, \qquad \pa_aX\cdot n_A = 0.
\eqqno
 $$
Then
 $$
K_{ab}^A = -\pa_an^A_\mu \pa_bX^\mu = n_\mu^A  \pa_a\pa_b X^\mu.
\eqqno
 $$
In two dimensions, the Ricci curvature  scalar $R$ is a function of the
extrinsic curvature, namely,
 $$
R = K^{abA}K_{ab}^A - K^AK^A,
\eqqno
 $$
where $K^A = \ga^{ab}K_{ab}^A$.

It can be shown that to second order in $K$, the effective action takes
the form
 $$
S=-\int\d^2\si\sqrt{-\ga}(\mu-\al K^AK^A + \be R)
 \label{\eCorAct}
 $$
where $\al$ and $\be$ are dimensionless numbers.  By the Gauss-Bonnet
theorem [\ct{DifGeo}], the integral of $R$ over the worldsheet is $4\pi$
times its Euler characteristic, which is a topological invariant.
Thus $\beta$ does not affect the equations of motion.

Both the sign and the magnitude of the constant $\al$ have been subjects
of some disagreement in the literature [\ct{MaeTur88}--\ct{Lar93}],
due in part to some ambiguity in the definition of the problem.  A
simple energy argument appears to show that $\al$ is in fact positive
\ct{MaeTur88}].  However, the correction term proportional to
$\al$ is proportional to
the extrinsic curvature $K^A$, which vanishes identically when the
lowest-order Nambu-Goto equation (\rf{\eConEqu}) is satisfied.  Thus, any
solution of the Nambu-Goto equation is actually also a solution of the
corrected equation.  To this order, $\al$ too does not affect the equations
of motion.

\subsection{Strings in Minkowski space}

The conformal gauge condition still leaves considerable freedom.  To
analyse the solutions of the equations of motion, it is convenient to fix
the gauge further.  In Minkowski space, with metric
$g_{\mu\nu}=\eta_{\mu\nu}\equiv{\rm diag}(1,-1,-1,-1)$, the usual choice is
the temporal gauge [\ct{GodGolRebTho73}],  in which one identifies the
worldsheet time $\tau$ with  the Minkowski time $X^0$.  The constraints and
the equations of motion become
 $$
\eqalign{
\bXd^2 + \bXp^2 = 1, \qquad \bXd&\cdot\bXp  =0,\cr
\ddot{\bX} - \dprime{\bX} = 0. & \cr
}
\label{\eTemEqu}
 $$

This choice has two happy results: the velocity is orthogonal to the string
tangent vector; and $\si$ measures equal energy intervals along the string.
{}From (\rf{\eNGStTen}), we have
 $$
T^{0}_{\,0}(x) = \mu\int\d\si\de_3(\bi{x}-\bX).
 \eqqno
 $$
The general solution to the wave equation and constraints (\rf{\eTemEqu}) is
 $$
\bX = \half[\ba(\si-t)+\bb(\si+t)], \qquad \bap^2 = 1 = \bbp^2.
\label{\eGenSol}
 $$
Thus we solve the Nambu-Goto equations by specifying two curves on the unit
sphere [\ct{KibTur82}].

If we require a length $L$ of string to form a closed loop, there is an
additional constraint on the curves:
 $$
\int_0^L\d\si\bXp = \half\int_0^L\d\si(\bap+\bbp) =0.
\eqqno
 $$
Furthermore, in the centre of momentum frame
 $$
\int_0^L\d\si\bXd =
\half\int_0^L\d\si(-\bap+\bbp)=0.
\eqqno
 $$
Thus the curves $\bap$ and $\bbp$ are both centred on the origin for
stationary loops.  From the
periodicity of $\bap$ and $\bbp$ it follows that:
 $$
\bX(t+L/2,\si+L/2)=\half\[\ba(\si-t)+\bb(\si+t+L)\] = \bX(t,\si).
\eqqno
 $$
Thus although the defining curves have period $L$ the loop itself has period
$L/2$.

Another property of loops in Minkowski space is that their mean square
velocity is 1/2  in their rest frame [\ct{Kib85}], as may easily be
verified from (\rf{\eTemEqu}).

There can be points on the worldsheet where the string moves at the speed of
light.   Through the constraints we see that if $|\bXd|=1$, the string tangent
vector vanishes, and so at that point
 $$
\bap(\si-t)+\bbp(\si+t) = 0.
\eqqno
 $$
Providing $\dprime{\bX}$ is not zero at that point, the string takes the
appearance of a cusp.  These cusps are in a loose sense generic, for $\bbp$
and $-\bap$ are both closed curves centered on the origin, so it is
intuitively `likely' that there will be points of intersection.  However,
this need not happen: we could, for example, wrap the curves rather like seams
on a tennis ball [\ct{GarVac87}].   There is also no formal need for $\bap$
and $\bbp$ to be  continuous [\ct{GarVac87}].
The string then has discontinuities in the tangent vector, or
`kinks', moving in one or other direction at the speed of light.  It is
relatively easy to construct kinky loops with no cusps.  A simple example is a
loop made up entirely of kinks and straight segments:
 $$
{\bap(u) = \cases{\bi{e}_1, & $nl\le u < (n+\half)L$, \cr
                          -\bi{e}_1, & $(n+\half)L \le u < nL$; \cr}}\quad
{\bbp(v) = \cases{\bi{e}_2, & $nl\le v < (n+\half)L$, \cr
                          -\bi{e}_2, & $(n+\half)L \le v < nL$;\cr}}
\eqqno
 $$
where $\bi{e}_1\cdot\bi{e}_2 = 0$.
Over one period, this solution changes from a square to a doubled line in
the $\bi{e}_1+\bi{e}_2$ direction, back to a square, and then to a doubled
line again, but in the $\bi{e}_1-\bi{e}_2$ direction.

In the simplest class of continuous solutions, $\bap$ and $\bbp$ are great
circles at an angle $\psi$ [\ct{Bur85}]:
 $$
\eqalign{
\bap(u) &= \cos (2\pi Mu/L) \bi{e}_1 + \sin (2\pi Mu/L) \bi{e}_2 \cr
\bbp(v) &= \cos(2\pi Nv/L)\bi{e}_1 + \sin(2\pi
Nv/L)(\cos\psi\bi{e}_2+\sin\psi\bi{e}_3)\cr}
\eqqno
 $$
with $M$ and $N$ relatively prime.  In figure 3.1 we display a
sequence of projections of the string over one period for $M=1$ and $N=2$.
This trajectory has two cusps, at $t=0$ and $t=8$, on opposite sides
of the loop.

In an infinite universe we may also consider strings which do not join up on
themselves and do not therefore satisfy $\int\d\si\bXp=0$.  The simplest such
solution is of course just a stationary straight string $\bXd=0$,
$\bXp=\hat{\bi{n}}$.  There is also a simple solution describing helical
standing waves [\ct{Tur83b}]
 $$
\eqalign{
\bap(u) &= \om A[\cos(\om u)\bi{e}_1+\sin(\om u)\bi{e}_2] + (1-
\om^2A^2)^{1/2}\bi{e}_3 \cr
\bbp(v) &= \om A[\cos(\om v)\bi{e}_1+\sin(\om v)\bi{e}_2] + (1-
\om^2A^2)^{1/2}\bi{e}_3 \cr}
\eqqno
 $$
where $A$ is the amplitude, and the pitch of the helix is $(1-
\om^2A^2)^{1/2}/2\pi\om$.

The temporal gauge has the advantage of being easy to interpret, although
solutions in closed form are hard to find because of the non-linear
constraints.  There is another gauge which does solve the constraints: the
light-front  gauge [\ct{GreSchWit87},\ct{Tho88}].   Instead of the condition
$\tau=X^0$, we identify $\tau$ with a  parameter labelling a family of null
3-surfaces, or
  $$
\tau = X^+ \equiv X^0+X^3.
\eqqno
 $$
Thus a constant-$\tau$ slice of the worldsheet is simply its intersection
with a family of null geodesics with 4-velocity $(1;0,0,-1)$.  It is exactly
as if we  were looking at the shadow of the string illuminated by a distant
source and projected on an imaginary screen.  The independent degrees of
freedom
are the  two transverse coordinates $X^A$, which satisfy a wave equation
 $$
\ddot{X}^A - \dprime{X}^A = 0.
\eqqno
 $$
The constraints are now solvable in terms of these coordinates,
for
 $$
\dot{X}^- = (\Xd^A)^2+(\Xp^A)^2, \qquad \Xp^- = 2 \Xd^A\Xp^A.
\eqqno
 $$
The general solution in the light-front gauge is thus given by any two curves
$\Xp_{\rm L}^A(\si-\ta)$ and $\Xp_{\rm R}^A(\si+\ta)$.  There is in fact
one further constraint if one wishes to consider only closed loops of string,
for not  merely do we have to make the transverse coordinates periodic, but
$X^3$ must  be periodic too.  Hence $\int\d\si\,\Xd^A\Xp^A = 0$.

We shall see in section 5.2 that when strings are formed in a phase
transition, they are random walks.  The mean square distance is given by
 $$
\langle(\bX(\si)-\bX(0))^2\rangle \simeq \xi^2(\si/\xi)^{2\nu},
\eqqno
 $$
where  $\xi$ is the persistence length.  The exponent $\nu$ is $1/2$ for
$\si/\xi \gg 1$, and $\sim 1$ for  small $\si/\xi$.
Numerical and analytic studies, discussed in section 5.5, indicate
that as the string network evolves, $\xi$ settles down to a `scaling'
solution where it grows in proportion to the time $t$.  On small scales, $\nu$
approaches 1 rather slowly as $\si/\xi\to 0$.  This slow approach towards
the value for a straight string indicates that there is a significant amount
of `wiggliness' on the string network on scales well below its persistence
length.

We would like to have some description of the string dynamics which
ignores the details of the wiggles while keeping track of their effects.
Consider for example a  string along the $z$-axis, which is straight
apart from small
wiggles [\ct{Vil90}].  For the two functions in the general solution
(\rf{\eGenSol}) we take
  $$
\ba(u) = ku\bhat{z}+\ba_\perp(u),\qquad
\bb(v) = kv\bhat{z}+\bb_\perp(v),
\eqqno
 $$
where $k$ is a constant in the range $0<k<1$ and $\ba_\perp$ and $\bb_\perp$
lie in the $x$-$y$ plane.  The gauge constraints require
 $$
\ba'{}^2 = \bb'{}^2 = 1 - k^2.
 $$

We may define an effective stress-energy tensor for the string,
by averaging over a length scale
$\De$ which is large compared to the wiggles:
  $$
T^{\mu\nu}\rms{eff}(t,\bi{x}_\perp,z) =
\th^{\mu\nu}(t,z)\de_2(\bi{x}_\perp),
\eqqno
 $$
where
 $$
\th^{\mu\nu} = {1\over\De^2}\int_{t-\De/2}^{t+\De/2}\d t'
\int_{z-\De/2}^{z+\De/2}\d z' \int \d^2\bi{x}_\perp\,
T^{\mu\nu}(t',\bi{x}_\perp,z').
 $$
We then find that the only two non-vanishing components of $\th^{\mu\nu}$ are
the coarse-grained energy per unit length and the coarse-grained tension,
  $$
\muf \equiv \th^{00} = \mu_0/k, \qquad \tf \equiv -\th^{33} = k\mu_0.
 $$
Note that the product of the effective energy density and tension remains
constant:
 $$
\muf\tf = \mu^2,
\eqqno
 $$
a result first derived for wiggly or `noisy' strings by Carter [\ct{Car90}],
and then
demonstrated for straight strings with wiggles by Vilenkin [\ct{Vil90}].

\subsection{Strings with damping}

The motion of strings in the early Universe is damped for two reasons ---
the Hubble expansion and the friction due to interaction with other
particles.  Let us first consider the effect of expansion.

We shall limit ourselves to a spatially flat
Friedmann--Robertson--Walker cos\-mology (a very good approximation in the
early
universe) for which
  $$
\d s^2 = \d t^2 - a^2(t)\d x^i\d x^i = a^2(\eta)(\d\eta^2 - \d x^i\d x^i),
\eqqno
 $$
where $\eta$ is the `conformal time'. Then the temporal and conformal
gauge conditions are no longer compatible; one must be sacrificed.  It is
very convenient for numerical  purposes to keep the identification of
worldsheet time with the background (conformal) time, so instead we drop the
condition $\Xd^2+\Xp^2=0$.  We define $\ep$ by
$\ep^2=- \Xp^2/\Xd^2=\bXp^2/(1-\bXd^2)$,
where dots now denote $\d/\d\eta$, and find that the equations of motion
become  [\ct{TurBha84}]
  $$
\eqalign{
\dot\ep + 2h\dot{\bX}^2\ep &= 0, \cr
\ddot{\bX} + 2h(1-\bXd^2)\bXd -
{1\over\ep}\pa_\si\({1\over\ep}\bXp\) &= 0, \cr
}
\label{\eEomExp}
 $$
with $h=\dot{a}/a = aH$ (not to be confused with the dimensionless parameter
$h$ defined in section 1.4).  The effect of the  expansion of the universe is
to provide a damping term.  The significance of  $\ep$ is seen if we compute
the
total energy of the string:
 $$
E = \mu a^{-2}\int\d\si\,\ep.
\eqqno
 $$
We see that $\ep$ is simply a dimensionless measure of the linear mass density
of the string, which in Minkowski space we could set to be unity for all time.

A rough idea of string dynamics in an expanding background is obtained by
considering two opposite limits for waves on infinite string, set by
whether the curvature radius is large or small compared with the
comoving Hubble length $h^{-1}$  [\ct{Vil81b},\ct{TurBha84}].
For very long wavelength modes we can neglect the $\si$ derivatives, to obtain
 $$
\ddot{\bX} + 2h(1-\bXd^2)\bXd = 0,
\label{\eAveMot}
 $$
which has an attractor at $\bXd = 0$.  Thus the comoving velocity of the
string is zero, and we infer that long wavelength configurations are
conformally stretched as the universe expands.  Conversely, we can neglect the
Hubble damping for short wavelengths, to obtain approximately Minkowski
equtions of motion
 $$
\ddot{\bX} - \pa_s^2\bX = 0
\label{\eShDMot}
 $$
where $\d s = \ep\d\si$.  This is also true of small loops of string.

We now turn to the effect of friction.
In the early universe the strings are
moving through a hot medium of particles  with which they interact, resulting
in a frictional force  [\ct{Kib76},\ct{Eve81}].  In the rest
frame of the string, the force per unit length $\bi{f}$ exerted by a fluid
of density $\rho$ moving with velocity $\bi{v}$ is, in the conformal temporal
gauge,
 $$
\bi{f} = \si\rho[\bi{v} - (\bXp\cdot\bi{v})\bXp/\bXp^2],
\label{\eFriFor}
 $$
where $\si$ is the scattering cross-section per unit length (the calculation
of $\si$ will be discussed in section 4.6).  A covariant
generalization of (\rf{\eFriFor}) is [\ct{Vil91}]
 $$
f^\mu = \si\rho(v^\mu - \ga^{ab}\pa_aX^\mu\pa_bX^\nu v_\nu).
\eqqno
 $$
Thus the frictional effect of a fluid with velocity $v^\mu(x)$  is to
change the equations of motion to
 $$
\Box X^\mu + \Ga_{\nu\rho}^\mu\ga^{ab}\pa_aX^\nu\pa_b X^\rho =
\({\si\rho\over\mu}\)[v^\mu(X) - \ga^{ab}\pa_aX^\mu\pa_bX^\nu v_\nu(X)].
\eqqno
 $$
In a radiation dominated FRW universe, the fluid velocity is $v^\mu = (a^{-
1};0,0,0)$, and it is found that the temporal gauge equations of motion
become [\ct{Vil91}]
 $$
\eqalign{ \dot\ep + (2h +a\si\rho/\mu)\bXd^2\ep &= 0,\cr
\ddot{\bX} + \(2h+{a\si\rho\over\mu}\)(1-\bXd^2)\bXd -
{1\over\ep}\pa_\si\({1\over\ep}  \bXp\) &= 0. \cr}
 \eqqno
 $$
Thus friction acts in exactly  the same way as the damping due to the
expansion of the universe.  Associated with the friction there is a length
scale $l_{\rm f} = \mu/\si\rho$ which acts in the same way as the Hubble
length.  If $l_{\rm f} <2h/a$, frictional damping dominates over expansion;
the network is conformally stretched on scales  greater than $l_{\rm f}$,
while short wavelengths evolve as for Minkowski  space.  We will see (in
section 5.3) that the cross section per unit length $\si$ goes as the  inverse
momentum of the scattering particles, so
$\si\simeq T^{-1}\propto t^{1/2}$.
The density $\rho$ goes as $t^{-2}$.  Thus, although friction is important at
early times, it eventually becomes subdominant to expansion damping at
(physical) time $t_*$ given by [\ct{Kib76}]
 $$
t_* = \si(t_*)\rho(t_*)/\mu\sim (G\mu)^{-1}t_{\rm c},
 \eqqno
 $$
where $t_{\rm c}$ is the time at which the Universe was at the string critical
temperature $T_{\rm c} \simeq \mu^{1/2}$.

\subsection{Superconducting string effective action}
Superconducting strings carry currents in the worldsheet, which have a
similar effect to small amplitude waves in that they change the effective
tension and linear mass density [\ct{VilVac87a}--\ct{SpePirGoo87}].
If these currents are sources of a long-range gauge field, then there are
additional complications due to the self-interaction of the string.  Let us
first neglect the gauge field, and consider a straight static string with
a bosonic condensate $\chi = Xe^{i\al}/\surd2$.  This condensate contributes an
extra piece to the  action
 $$
S_\chi = \int\d t\d z\d x\d y\,\half\{
X^2[(\pa_t\al)^2-(\pa_z\al)^2]-(\pa_x X)^2-(\pa_y X)^2 \}.
\eqqno
 $$
To zeroth order in the curvature, the first two terms are constants,
unimportant for the string dynamics.  Introducing a pair of coordinates $\rh^A$
in the plane orthogonal to the worldsheet, we may write
 $$
S_\chi \simeq \int\d^2\si\sqrt{-\ga}\ka \ga^{ab}\pa_a\al\pa_b\al, \eqqno
 $$
where, as in section 2.7, $\ka = \half\int \d^2\rho\,X^2$.  The phase of the
scalar field is therefore a massless bosonic degree of freedom on the
worldsheet.  In principle, $\ka$ is a function of $(\pa\al)^2$, since the
current $j_a=\pa_a\al$ reacts back on the condensate; we shall return to this
point below.  So, for small currents and curvatures, the equations of motion
for
the superconducting string are
 $$
\eqalign{
\pa_a[\sqrt{-\ga}(\mu\ga^{ab}+\th^{ab})\pa_bX^\mu] &= 0,\cr
\pa_a(\sqrt{-\ga}\ga^{ab}\ka j_b) &= 0,\cr }
\label{\eScsEom}
 $$
where $X^\mu$ now representes the string worldsheet position and is
unrelated to the field $X$ above.  Here $\th_{ab}$ is the worldsheet
energy-momentum tensor, given by
  $$
\th_{ab} = \ka(2j_aj_b - \ga_{ab}j^2).
\eqqno
 $$
By virtue of the equations both the current and the stress tensor are
covariantly conserved: moreover, $\th_{ab}$ is traceless.

Let us examine the dynamical effect of a constant current $j_a = (\om,k)$ on
a string in a Minkowski background.  The stress tensor is
 $$
\th_{ab} = \ka \pmatrix{\om^2+k^2 & 2\om k \cr
                        2\om k & \om^2 + k^2 \cr},
\eqqno
 $$
In principle, it would
seem to be possible to construct a stationary loop
solution by arranging that $\mu-\ka(\om^2+k^2)=0$
[\ct{OstThoWit86},\ct{CopHinTur87},\ct{SpePirGoo87}].
(It is easy to see that (\rf{\eScsEom}) reduces to this single
condition for a static loop with constant current.)
However, one has to be careful in pushing the equations of motion this far,
because  the stabilization of the loop
would require a current of order
$(\mu/\ka)^{1/2}$.  One would not then expect backreaction effects on
$\ka$ to be negligible.  Furthermore, if $\mu-\ka(\om^2+k^2)$
were to become negative, the loop would become unstable to small
transverse perturbations [\ct{Car89}].
The existence of classically stable loops has
cosmological relevance, for such objects would have a relatively long lifetime
and could contribute to the mass density of the universe, even to the extent
of ruling the theory out
[\ct{CopHinTur87},\ct{HawHinTur88},\ct{DavShe89c}].   The crucial issue is
whether or not the critical current, at which the condensate is `quenched' and
the superconductivity of the string is lost, is higher than the current which
stabilizes the loop.

{}From numerical investigations of superconducting cosmic strings
[\ct{HawHinTur88},\ct{DavShe88a},\ct{BabPirSpe88}--\ct{Pet93}],
it has emerged that null currents  $(j^2 =
0)$ have no back reaction, while timelike currents actually  increase the
size of the condensate.  Loops with null currents (`vortons'
[\ct{DavShe88a}])  seem to be the best bet for forming stable stationary
congurations, although it should be emphasized that all stability studies have
been carried out for straight strings. A general framework for the analysis of
the
stability of current-carrying strings has been developed by Carter
[\ct{Car89}].  The variation of the linear mass density $\mu$ and the tension
$T$ with the conserved current $J_a = \ka j_a$ gives the string an equation of
state $T=T(\mu)$, which can be computed numerically for the model field theory
of section 2.7 [\ct{Pet92}].
This equation of state allows one to calculate the
speeds of propagation of transverse and longitudinal waves, which are
respectively given by
 $$
c_{\rm T}^2 = T/\mu, \qquad c_{\rm L}^2 = -\d T/\d\mu.
\eqqno
 $$
The longitudinal waves are disturbances of varying current density,
which involve the magnitude of the condensate.  If $\d T/\d\mu$ becomes
positive an instability develops in the condensate, and so the string cannot
support currents beyond the threshold $\d T/\d\mu =0$.  We display the
generic behaviour of $\mu$ and $T$ for spacelike currents $I = \surd(-J^2)$
in figure 3.2.  Classically stable loop configurations can exist
only if the tension can vanish before the critical current $I_{\rm c}$ is
reached. Numerical studies indicate that this is not possible for spacelike
currents without the extra help of a gauge field [\ct{DavShe88a},\ct{Pet93}].

So, let us now consider the dynamics when a massless gauge field $A_\mu$, which
couples to the worldsheet current, is included.   The condensate field $\chi$
now contributes
 $$
S_\chi = \half\int \d t\d z \d x \d y\,
X^2[(\pa_t\al + qA_t)^2 - (\pa_z\al +
qA_z)^2]
\eqqno
 $$
to the action.  Provided we are interested only in gauge field
configurations with wavelengths much greater than the string width, we may
approximate $A_\mu(x)$ across the string by its value at the centre,
$A_\mu(X)$, and write
 $$
S_\chi = \int\d^2\si\sqrt{-\ga}\ka\ga^{ab}(\pa_a\al+qA_a)(\pa_b\al+qA_b),
\eqqno
 $$ where $A_a = \pa_aX^\mu A_\mu(X)$ is the (unnormalized) projection of the
gauge field into the worldsheet.
The string carries a gauge current $J^\mu = -\de S_\chi/\de A_\mu$, which in
turn acts as a source for the gauge field.   The equations of motion for the
superconducting string interacting with the gauge field are then
 $$
\eqalign{
\pa_a[\sqrt{-\ga}(\ga^{ab}+\th^{ab})\pa_bX^\mu] &=
-2q\ka \sqrt{-\ga}F^{\mu\nu}(X)\pa^aX_\nu j_a, \cr
\pa_a(\sqrt{-\ga}\ga^{ab}\ka j_b) &=
0,\cr  \pa_\nu F^{\mu\nu} &= 2q\ka\int\d^2\si\sqrt{-\ga}\,\pa^aX^\mu j_a
\de_4\(x-X(\si)\),\cr}  \eqqno
 $$
with $j_a = \pa_a\al+qA_a$.

These equations first appeared in a dimensionally reduced classical string
theory [\ct{Nie80},\ct{NieOle87}].  They arise naturally from a string
propagating in a five-dimensional space-time, where one of the spatial
dimensions is compactified to a circle.  This is the original Kaluza-Klein
construction  [\ct{KalKle}], which results in a theory of gravity and
electromagnetism.  When strings propagate in this background, the periodic
coordinate in  the compactified direction behaves in the same way as $\al$,
the phase of  the scalar field.  There is also a parameter playing the
same role as  $\ka$, which is essentially the radius squared of the
compactified dimension in units of the inverse string tension.

The
self-interaction of the current-carrying string now makes true solutions hard
to find, although perturbative solutions exist for small $q$.  The principal
effects of the gauge field are, firstly, to contribute a logarithmic term to
the stress tensor and, secondly, to cause the string to lose energy through
radiation. We consider radiation in section 4.4: here we find the
logarithmic term.  Formally, the gauge field from the string source is
 $$
\eqalign{
A_\mu(x) &= \int\d^4x'G_{\rm ret}(x-x')J_\mu(x'),\cr
         &= {q\over\pi}\int\d^2\si\sqrt{-\ga}\,\pa^aX_\mu
j_a\de\((x-X)^2\).\cr}
\eqqno
 $$
The leading logarithmic divergence can be extracted [\ct{CopHawHinTur88}],
and we find that the
gauge field on the string is
 $$
A_a(X) = {q\over \pi}\ln(L/w)j_a(X),
\eqqno
 $$
where the upper and lower cut-offs $L$ and $w$ are interpreted as a local
curvature radius and a string width.  This is quite familiar for a
wire carrying current and charge.  Thus if $j_a = (\om,k)$ as before, the
worldsheet stress tensor  becomes
 $$
\th_{ab} = \ka(2j_aj_b -
\ga_{ab}j^2)[1+(q^2/\pi)\ln(L/w)]^2.
\eqqno
 $$
The effect of the gauge field is to reduce the effective tension and increase
the effective mass density still further.  It was found in [\ct{HawHinTur88}]
that zero
effective tension (including the gauge field) could be reached with purely
spacelike currents in a straight string for some parameter values in the
simple model of section 2.7.  However, relatively large values of $L/w$
were necessary, and so it is the pressure in the magnetic field which
resists the collapse of a loop.   The local tension in the string
remains positive [\ct{Pet93}].

\subsection{Global string effective action}
The global string differs markedly from the gauge string in that it is
surrounded by a long-range Nambu-Goldstone field $\pa_\mu\al$, which falls off
as $r^{-1}$ away from the string.  The string is a source for this field and
interacts with it, making a general solution to the equations of motion
impossible.  Even perturbative solutions are not without problems, for there
is no small coupling constant in which to expand.

Let us consider once again the action of a straight string on the $z$ axis,
this time separating out the Nambu-Goldstone mode $\al$.  By writing
$\phi=fe^{i\al}/\surd2$, we have
 $$
S = \int \d t\d z\int\d x \d y[\half(\pa f)^2
-\frac{1}{8}\la(f^2-\eta^2)^2] +  \int\d^4x\,\half f^2(\pa\al)^2.
\label{\eScaAct}
 $$
We know from section 2.1 that the last term is logarithmically divergent,
but the first term is finite, and gives $-\mu_0\int\d t \d z$. The action of
the massless field $\al$ generally has two contributions:  a multivalued part
from the string itself, and single-valued propagating Nambu-Goldstone modes.
Both can be described by a single-valued antisymmetric tensor field
$B_{\mu\nu}$ and its accompanying field strength $H_{\mu\nu\rho}= \pa_\mu
B_{\nu\rho}+\pa_\nu B_{\rho\mu} + \pa_\rho B_{\mu\nu}$
[\ct{KalRam74}--\ct{VilVac87b}].  Locally, the fields
are related by $H_{\mu\nu\rho} =\eta\ep_{\mu\nu\rho\si}\pa^\si\al$, (where
$\eta$ is the expectation value of the field away from the string), but in the
presence of the string there are extra terms in the action describing the
coupling between the string and the antisymmetric tensor field. Davis and
Shellard [\ct{DavShe88b}] have found the  canonical transformation which takes
the action  (\rf{\eScaAct}) into the dual form
 $$
S^* =  -\mu_0\int\d^2\si\sqrt{-\ga} +
2\pi\eta\int\d^2\si\,\ep^{ab}\pa_aX^\mu\pa_bX^\nu B_{\mu\nu} + {1\over 6}
\int\d^4x\,H^2.
\eqqno
 $$
The equations of motion that follow from this action are
 $$
\eqalign{
\mu_0 \Box X^\mu &= 2\pi\eta H^{\mu\nu\rho}\pa_a
X_\nu\pa_bX_\rho\ep^{ab}, \cr
\pa_\rho H^{\mu\nu\rho} &=
2\pi\eta\int\d^2\si\ep^{ab}\pa_aX^\mu\pa_bX^\nu\de^{(4)}(x-X(\si)).\cr}
\eqqno
 $$
In the temporal gauge it is not hard to check that the straight string on the
$z$ axis gives rise to a field
 $$
H_{\mu\nu\rho} = \eta\ep_{\mu\nu\rho i} \hat{\varphi}^i/\rho,
\eqqno
 $$
which is consistent with $\al = \varphi$.  The strength of the coupling
between the string and the antisymmetric tensor field is in fact fixed by
this equality [\ct{VilVac87b}],
which amounts to a consistency condition:  we must always
have $\oint\d x^\mu\pa_\mu\al = 2\pi$ when taken on some curve $C$ around
the string.  By Stokes' theorem, this implies that $\eta^{-
1}\int\d\Si^{\mu\nu}\pa^\rho H_{\mu\nu\rho} = 2\pi$ over a surface whose
boundary is $C$, and hence the value for this coupling of $2\pi\eta$.  We can
think of the field generated by the string as a Coulomb-type field arising
from a `charge' per unit length of $2\pi\eta/\mu_0$.  This is not small
compared with the string scale $1/\eta$, for $\mu_0$ is also fixed in terms of
$\eta^2$ (and is independent of $\la$).
Thus the string coupling is always
strong. We shall amplify the discussion of global string self-interactions
when we come to discuss radiation in the next chapter.

\subsection{Intercommuting}
The Nambu action described in section 3.1 is a good approximation
for string segments which are well-separated.  However, a crucial dynamical
question lies outside the framework: what happens when strings cross.
To answer this question fully, one has to go to numerical solutions of
the underlying classical field theory [\ct{She87}--\ct{LagMat90}].
However, some qualitative
and heuristic remarks serve to set the scene [\ct{She88}].

In figure 3.3, two segments of string, assumed Abelian, have just crossed.
For our purposes it is irrelevant whether they are global or local.
If we examine the planes $A$ and $B$, we see that the total winding
number in $A$ is 2, while in $B$ it is zero.  Thus there is no topological
reason for plane $B$ to be pierced by any string at all.  It is
almost as if plane $B$ contains a vortex and an antivortex which,
if they get sufficiently close, can annihilate into radiation.
By the same token, plane $A$ contains two vortices, whose total
flux is conserved.  Thus the evolution of the fields subsequent
to figure 3.3 seems likely to eliminate the strings passing through
$B$, so that end 1 joins end $2'$ and vice versa.  The strings
are said to `intercommute'.

Numerical simulations of this process for both global [\ct{She87}] and
local [\ct{MorMyeReb88a}--\ct{MatMcC88}]
strings, as well as for superconducting ones [\ct{LagMat90}], confirm that this
picture is correct.  The string ends exchange partners, with the
creation of sharp bends in each string.  Each bend is resolved
into two kinks moving in opposite directions at the speed of light
(figure 3.4).  Intercommuting can even be observed in the
laboratory, between line disclinations in a liquid crystal
[\ct{Chu+91}], although in this system the strings are heavily
damped and so we do not see kinks.

One possible complication is if the strings are carrying currents.
Suppose string $11'$ carries current $I_1$ and $22'$ carries $I_2$.
Then, if reconnection takes place, charge must build up on the
bridging segment on string $12'$, at a rate $(I_2-I_1)$.  Since
each kink is moving at the speed of light, the length of that
segment is increasing at a rate equal to $2c$, and so
the charge density between the kinks is $\rho = (I_2-I_1)/2$
in natural units, where $c=1$.  However, it might be argued that if
$I_1$ and $I_2$ are oppositely oriented, the momentum in the current
will act to preserve the connectivity $11'$, $22'$.  Nonetheless,
numerical simulations by Laguna and Matzner [\ct{LagMat90}] find that
reconnection nearly always takes place, despite the build-up
of charge on the reconnected strings.

Finally, we note that this discussion has been entirely about
Abelian strings.  The original argument in favour of
intercommutation clearly does not apply to $\Z_2$ strings,
where neither plane has any topologically conserved flux.
Given that strings cannot break, reconnection must occur,
but it can happen
between 1 and 2 as well.  This may be energetically disfavoured if,
for example, it leads to the creation of a pair of beads
(section 2.6).  If $\pi_1$ is non-Abelian, then there is an
additional possibility: the two strings may correspond to
elements $\htil_1$, $\htil_2$, of the unbroken subgroup $H$
which do not commute. In that case
they pass through each other and create a third which
joins the two intersection points, associated with the element
$\htil_1\htil_2\htil_1^{-1}\htil_2^{-1}$ [\ct{Mer79}].

\section{String interactions}
The observational effects of strings (see section 6) are the results of their
interactions with the various particles and fields of nature.  To some
extent, the interactions are model-dependent: if the strings are global
they radiate Goldstone bosons; if superconducting, electromagnetic
waves.  They also interact with particles through scattering off the
strong fields in the core of the string.  However, all strings are the
source of a gravitational field, through which they become a candidate
for the origin of primordial density perturbations.

\subsection{Gravity}
The energy density of a gauge string is concentrated in its core, where
we therefore expect a region of high curvature. Rather than working
with the energy-momentum tensor of the field directly, it is often more
convenient to approximate the string as a $\de$-function source.  For a
gauge string obeying the Nambu action (\rf{\eNamGot}),
the energy-momentum tensor for a string moving through a general
background space-time was given in (\rf{\eNGStTen}).
In Minkowski space, where we can use
the conformal temporal gauge (\rf{\eTemEqu}),
a straight string on the $z$ axis has
$$
T^{\mu\nu} = \mu \, \diag(1,0,0,-1) \de(x)\de(y).
\label{\eStrEMT}
$$
The string is a line source with equal tension and
mass density, a result which can be derived purely from Lorentz
invariance along the string, together with energy-momentum conservation
[\ct{Vil81a}].  Although the space-time curvature on the string is formally
infinite in this approach, outside the string we may use the linearized
Einstein
equations for the perturbation $h_{\mu\nu} = g_{\mu\nu} -
\eta_{\mu\nu}$. In the harmonic gauge these are
$$
\pa^2 h_{\mu\nu} = -16\pi G(T_{\mu\nu} - \half \eta_{\mu\nu} T).
\eqqno
$$
With the source (\rf{\eStrEMT}) the inhomogeneous part of the solution
is
$$
h_{\mu\nu} = 8G\mu \ln(\rho/\rho_0) \, \diag(0,1,1,0)
\eqqno
$$
where $\rho^2 = x^2 + y^2$, and $\rho_0$ is an arbitrary scale.  The
linearized solution cannot be correct everywhere, since the logarithm
blows up for large and small $\rho$.  However, this solution can be
matched onto an exact solution by the coordinate transformation
$ [1-8\pi G\mu\ln(\rho/\rho_0)]\rho^2 = (1-4G\mu)^2R^2.
$ To order $G^2\mu^2$ the metric in the new coordinates is [\ct{Vil81a}]
$$
\d s^2 = \d t^2 - \d z^2 - \d R^2 - (1-4G\mu)^2 R^2 \d \vp^2
\label{\eConMet}
$$
With a new angular coordinate $\bar\vp = (1-4G\mu)\vp$ the space-time is
seen to be flat everywhere (except at $R=0$), but with the angular
coordinate running from 0 to $2\pi - \de$, with $\de = 8\pi G\mu$. Thus
it has the form of a cone in the plane transverse to the string, with
an angle deficit $\de$.   Being flat, it clearly satisfies the full
Einstein equations everywhere where $T^{\mu\nu}=0$.  It can be shown that
the solution inside the string can be matched on to the asymptotic conical
metric [\ct{Got85}--\ct{Lin85}].  In the resulting metric,
the angle
deficit is no longer quite proportional to the mass per unit length,
there being O$(G^2\mu^2)$ corrections [\ct{FutGar88},\ct{HinWra90}].
Strings may also be embedded in other space-times, including
FRW [\ct{Vic87},\ct{Gre89}].  All that is required is that the space-time
have an axis of symmetry:  then the string space-time can be created by
removing a wedge and then gluing together the resulting edges.

As mentioned above, the form of the string metric was dictated by
Lorentz invariance along the string.  As we have seen, there are string
solutions which break this symmetry: superconducting strings and
wiggly strings (see sections 2 and 3).  Their energy-momentum tensors
have the approximate form
$$
T^{\mu\nu} = \diag(\mu,0,0,-T)\de(x)\de(y),
\eqqno
$$
where $T$ is the string tension.  (There are also radial and azimuthal
stresses, dictated by energy-momentum conservation, but they are
O$(G\mu(\mu-T))$ and therefore negligible to first order
[\ct{MosPol87},\ct{HinWra90}].)
The linearized metric becomes, after a redefinition
of the radial coordinate,
$$
\d s^2 = [1+2\psi(R)]\d t^2 - [1-2\psi(R)]\d z^2 - \d R^2 -
[1-2G(\mu+T)]^2R^2\d \vp^2,
\eqqno
$$
where $\psi(R) = 4G(\mu-T)\ln(R/R_0)$.  This can also be matched to a
static, cylindrically symmetric exact solution
[\ct{MosPol87},\ct{HinWra90}].  It is conical perpendicular to the
string, with angle deficit $4\pi G(\mu+T)$, but there is also a
non-zero Newtonian potential due to the $h_{00}$ term, which results in
an attractive force
$$
F = {2G(\mu-T) \over R}
\label{\eAttFor}
$$
toward the string.

Around a global string, the energy-momentum tensor does not vanish.
As we saw in section 2.1, there is a contribution from the Goldstone
mode (the phase of the field) which decreases as $r^{-2}$ away from the
string core, and leads to a logarithmically divergent mass per unit
length.  Near the string, however, a weak field solution to the metric
exists, which is [\ct{HarSik88}]
$$
\d s^2 = [1-4G\mu\ln(R/R_0)](\d t^2 - \d z^2) - \d R^2 +
[1-8G\mu\ln(R/R_0)]R^2\d \vp^2.
\eqqno
$$
Remarkably, the straight global string has a repulsive gravitational
force $2G\mu/R$.  The angle deficit $\de$ increases with distance, which
causes problems at $\de=2\pi$, or $R=R_0\exp(1/8G\mu)$.  Unlike the
gauge string, this is not just a problem with the coordinates: there
is a genuine singularity at large $R$ [\ct{CohKap88},\ct{Gre88b}].
However, this is at such a large distance that the singularity will not
occur in physically realistic string configurations --- we do not
expect to find a completely isolated, straight string in the early universe.

\subsection{Gravitational radiation from loops}
An oscillating string radiates
gravity waves.  The power in gravitational radiation produced by an
isolated loop of length $L$ can be estimated using the quadrupole
formula [\ct{Wei72}]
$$
P \propto G({\dddot{I}}_{ij})^2,
\eqqno
$$
where $I_{ij}$ is the quadrupole moment.  On
dimensional grounds, if $L$ is the only length scale on the loop,
$I_{ij} \sim \mu L^3$, while each time derivative brings in a factor of
$L^{-1}$. Thus we find that
$$
P=\Ga G\mu^2,
\eqqno
$$
where $\Ga$ is a constant for each loop.  That is, the power is
independent of its size. In order to calculate $\Ga$, we have to go to
a full weak-field calculation.  The average flux in radiation of
frequency $\om$ and momentum $\bk$ is [\ct{Wei72}]
$$
{\d P \over \d \Om} = {G\om_n^2 \over \pi} [T_{\mu\nu}^*(\om_n,\bk)
T^{\mu\nu}(\om_n,\bk) - \half |T^\nu_{\,\nu}(\om_n,\bk)|^2 ],
\label{\eWFPow}
$$
where $T^{\mu\nu}(\om_n,\bk)$ is the Fourier transform of the energy
momentum tensor.  The string has a period $L/2$, so the frequencies
$\om$ are multiples of this fundamental: $\om_n = 4\pi n/L$.  In the
conformal temporal gauge
$$
T^{\mu\nu}(\om_n,\bk) = -\mu{2\over L}\int \d u \d v (\pa_uX^\mu\pa_vX^\nu +
\pa_vX^\mu\pa_uX^\nu) e^{ik\cdot X}
\eqqno
$$
where $u = \si - t$ and
$v = \si + t$.  Resolving the string into left and right-movers, $X^\mu
= X^\mu_R(u) + X^\mu_L(v)$, we find that the  Fourier transform can be
written
$$
T^{\mu\nu} = -2\mu{ L} (U^\mu V^\nu + V^\mu U^\nu),
\eqqno     
$$
where
$$
U^\mu(\om_n,\bk) = \int_0^L {\d u\over L}\,\pa_u X^\mu(u) e^{ik\cdot X_R},
\qquad
V^\mu(\om_n,\bk) = \int_0^L {\d v\over L}\,\pa_v X^\mu(v) e^{ik\cdot X_L}.
\eqqno
$$
Thus the source of the gravitational radiation is the interaction of
left and right-movers on the string.  A loop has to be constructed
from both, so a loop must always radiate.

Besides energy, the gravity waves also take away momentum and angular
momentum [\ct{HogRee84}--\ct{Dur89}].  Again, dimensional
arguments indicate that the rate of loss of each must go as
$\Ga_P G\mu^2$ and $\Ga_L G\mu^2 L$ respectively, where $\Ga_P$ and $\Ga_L$
are dimensionless constants.  The radiation of momentum is particularly
interesting, since it means that the loop will be accelerated in the opposite
direction --- the so-called `rocket' effect [\ct{HogRee84}]. Although each
impulse is small, of order  $G\mu^2 L$, the net effect over many oscillations
can build up so that the loop is moving with speeds comparable to  $c$ towards
the end of its life.

The constants $\Ga$, $\Ga_P$ and $\Ga_L$ can be worked out exactly for
simple trajectories [\ct{VacVil85},\ct{Bur85},\ct{Dur89}], for which it
is found that $\Ga$ is in the range 50--100, while $\Ga_P$ and
$\Ga_L$ are somewhat smaller, around 5 [\ct{Dur89}].  For some special
cases, the total power actually diverges, due to the presence
of persistent cusps.

Cusps, we recall, are points on the string where the tangent vector
vanishes and the string formally reaches the velocity of light,
usually just for an instant.  In the weak field approximation this
results in the metric diverging along a null line originating at the
cusp, and pointing in the direction of the 3-velocity of the string at
the cusp [\ct{Tur83b}].  This divergence is not very strong
[\ct{Vac87}], and for loops with isolated cusps the power is still
finite.  The only sign of it is that the power falls very slowly with
the mode number $n$:   $P_n \propto n^{-4/3}$
[\ct{VacVil85}].  However, if the cusp persists throughout the loop's
period, the power becomes infinite.

One should
take into account the back-reaction of the gravitational field on the
string
[\ct{CopHawHin90},\ct{QuaSpe90}].  What happens is entirely analogous to
Dirac's computation of the self-force of the electron.  There is a
logarithmically divergent self-energy which can be absorbed into a
renormalization of the string tension, while the finite parts
contribute corrections to the Nambu-Goto equations of motion at third
and higher order in the derivatives.  For trajectories with isolated cusps,
the weak-field self-force equations can be solved numerically.  This involves
computing the impulse on each point of the string due to the force exerted
over one period by all the others, which is finite and O$(G\mu)$.  Quashnock
and Spergel [\ct{QuaSpe90}] have followed the trajectories of several cuspy
and kinky loops, finding that the kinks seem to be rounded off, while
cusps change their shape and arrive later, but do not disappear, in
accordance with the arguments of Thompson [\ct{Tho88}].

An intriguingly different tack was taken
by Mitchell et al.~[\ct{Mit+89}],
who went to the first quantized string theory in order to compute the decay
rate of highly excited string states, which are in some sense
quasi-classical.  They found that in four dimensions, string states with level
number $N$, which in the classical limit corresponded to rotating rods of
length $\surd N$, decayed primarily into states at level $N-1$ and $1$, with a
finite rate proportional to $\surd N$.  This is a nice result for three
reasons: the level 1 states correspond to the massless graviton, dilaton and
antisymmetric tensor fields; the energy in the emitted
massless states goes as $1/\surd N$ so, as in the classical process,
the total power is independent of the
length; and the classical power from the rotating
rod is infinite, whereas Mitchell et al.~obtain  $$
P \simeq 360G\mu^2.
\eqqno
$$
It is not entirely clear, however, that this can be regarded as the
solution to the back-reaction problem for a gauge string,
since the connection between the semi-classical gauge string and the
quantized fundamental string is obscure.

\subsection{Gravitational radiation from infinite strings}

Small amplitude waves on infinite strings are also the source of
gravitational waves [\ct{Sak90},\ct{Hin90a}].  Cosmic string simulations
indicate that there is a substantial amount of this small-scale
structure on the string network (see section 5), and it cannot be
neglected as a source of radiation.
For a string on the $z$ axis with small perturbations we can write
$$
X^\mu = (t,\bXperp,X^3).
\eqqno
$$
The flux per unit length in waves with frequency $\om$ and $z$
component of momentum $k$ through a large cylinder enclosing the string
is then [\ct{Hin90a},\ct{BatShe94}]
$$
{\d P \over \d z}(\om,k) = 32\pi G\mu^2 \om (|\bUp|^2|\bVp|^2 + |\bUp
\cdot \bVp|^2 - |{\bUp}^* \cdot \bVp|^2),
\label{\eSStRad}
$$
where $\bUp$ and $\bVp$ are ordinary Fourier
transforms of the right and left-moving derivatives of the
perturbations:
$$
\bUp(\om,k) = \int_0^L {\d u \over L} \pa_u \bXperp e^{-i(\om+k)u/2}, \qquad
 \bVp(\om,k) = \int_0^L {\d v \over L} \pa_v \bXperp e^{i(\om-k)v/2}.
\eqqno
$$
Here, $L$ is a large repeat length, needed to render integrals finite.
Again we see that it is the interaction of right and left-movers which
provides the radiation.  Note that, in this small-amplitude approximation,
perturbations of frequency $\om$ also produce waves of frequency $\om$.  In
loops, the interactions also produce higher harmonics, which decrease only as a
power law.  On the straight string the higher  harmonics are exponentially
damped and are usually negligible, unless some special symmetry forbids the
lowest harmonic [\ct{BatShe94}].  A further difference from loops is that there
can be waves travelling in one direction only, which do not radiate, for the
gravitational wave moves along with the perturbation [\ct{Vac86}].  In fact,
there are exact travelling wave solutions [\ct{FroGar90}], which can be thought
of as ordinary gravitational plane wave solutions   with a wedge cut out of the
space around the direction of propagation.

The gravity waves remove energy from the string, which must result in
the decrease in the amplitude of the perturbations.   We see from
(\rf{\eSStRad}) that the power radiated is proportional to the frequency: thus
high frequency modes decay more quickly.  This has implications for the small
scale structure, which is made up of many small-angle kinks, for the
high frequencies in the kinks will disappear first, rounding them
off (see section 5).

\subsection{Pseudoscalar and electromagnetic radiation}
In section 2 we found that strings may be the sources of other massless
fields.  If the string is a global one, such as an axion string, it is
surrounded by the field of the pseudoscalar Goldstone boson associated
with the symmetry breaking.  If it is superconducting, it can produce
a long-range electromagnetic field.  When in motion the string then
radiates coherent Goldstone bosons [\ct{Dav85}] and electromagnetic
waves as well as gravitational radiation.  As we saw in section 3.6, we
can treat global strings as sources for an antisymmetric tensor field
$B_{\mu\nu}$, which is equivalent to the Goldstone boson.  In this
representation one can derive the equations for the emitted power per
unit solid angle in much the same way as for gravitational radiation,
to obtain [\ct{VilVac87b}]
$$
{\d P \over \d \Om}(\om,\bk) = 2 \om^2 J_{\mu\nu}^*(\om,\bk)
J^{\mu\nu}(\om,\bk)
\eqqno
$$
where $\om$ is quantized in units of $4\pi/L$, and $J^{\mu\nu}$ is the
Fourier transform of the source distribution for the antisymmetric
tensor field, or
$$
J^{\mu\nu}(\om,\bk) = {\eta \over L} \int  \d u \d v \,\(\pa_u X^\mu
\pa_v X^\nu - pa_v X^\mu \pa_u  X^\nu\) e^{ik\cdot X}.
\eqqno
$$
An analogous formula exists for infinite strings [\ct{Sak91}].  On
dimensional grounds, the total power from a loop is independent of
size and proportional to $\eta^2$, where the constant of
proportionality tends to be about 50--100 for the trajectories for which
it can be computed (which are the same as those for which the
gravitational power is known) [\ct{VilVac87b}].

Compared to the gravitational case, the coupling between the string and
the field is very strong, so it is legitimate to ask whether the power
calculated in this way really represents the true radiation.  Indeed, it
has been argued that the string is overdamped, and should dissipate its
energy without oscillating at all [\ct{HarSik87},\ct{HagSik91}].
However, a careful investigation by Battye and Shellard [\ct{BatShe94}]
has compared the antisymmetric tensor formulation with the radiation in
a full numerical simulation of the complex scalar field theory
represented by the Lagrangian (\rf{\eGloLag}), and found them to be in
good agreement.  The string is heavily damped --- for example, kinks are
very quickly smoothed out --- but not overdamped, for trajectories
oscillate several times before losing their energy.  Settling this
question is extremely important for axion strings, for a very strong
bound on the symmetry-breaking scale comes from cosmological limits on
the energy density in coherent axion modes (see section 5.10).

A string carrying a current will also produce electromagnetic
radiation, with luminosity
$$
{\d P \over \d \Om}(\om,\bk) =  - {2\om^2}
J_\mu^*(\om,\bk)J^\mu(\om,\bk),
\eqqno
$$
where
$$
J^\mu(\om,\bk) = e {2\over L} \int \d u \d v \,( j_u\pa_v X^\mu +
j_v \pa_u X^\nu) e^{ik\cdot X},
\eqqno
$$
which is the Fourier transform of the current distribution on the
string.  The power of a string with constant current $I$ can be
estimated with the formula for dipole radiation:   the average power
emitted by an oscillating magnetic dipole $\bi{m}$ is roughly
$\vev{\ddot{\bi{m}}^2} $.  If the string has length $L$, the dipole
moment is $\sim eIL^2$, and since the period is $2/L$, we find
$$
P = \Ga_{\rm em} e^2 I^2.
\eqqno
$$
$\Ga_{\rm em}$ is a constant for each trajectory, and as for the
gravitational radiation can be computed in special cases
[\ct{GarVac87},\ct{CopHawHinTur88}].  Again, typical values are
50--100.

\subsection{Particle emission}
A vibrating string can also be the source of other fields besides the
massless ones considered above.  For example, it is likely that the
scalar field which makes the string will also couple to the conventional
Higgs of the Standard Model.  However, it has been shown by Srednicki and
Theisen [\ct{SreThe87}] that the power radiated in massive particles is
generally not significant compared to gravitational radiation.  The essential
point is that particles of total energy $E$ are produced only in regions of the
string where the fields are changing very rapidly compared with the
scale $E$, which turn out to be small regions near cusps.

For simplicity let us consider only a single real field $h$, which is coupled
to the string field $\phi$ through terms in the Lagrangian such as
$\half\la'|\phi|^2h^2$. Let $\phi$ have expectation value $\eta/\surd2$ in the
vacuum. It can be shown that, if back-reaction on the string state $\ket{S}$
is neglected, then the amplitude for the emission of two particles with
momenta $k_1$ and $k_2$ is
$$
\vev{S';k_1k_2|S} = \la' \int
\d^4x\, e^{i(k_1+k_2)\cdot x} \bra{S} (|\phi|^2 - \half\eta^2)\ket{S}.
\eqqno
$$
The expectation value of $|\phi|^2 - \half\eta^2$ in the string state
$\ket{S}$ can be evaluated by changing to string-centred coordinates,
much as we did in section 3 to calculate the string effective action,
and we find (in the conformal temporal gauge)
$$
\bra{S}(|\phi|^2 - \half\eta^2)\ket{S} \simeq {1\over \la}\int \d t\,\d \si\,
{\bXp^2}\de^{(3)}(\bx-\bX(\si,t)),
\eqqno
$$
where $\la$ is the self-coupling of the field $\phi$, so that the
string width is $M^{-1} \simeq (\sqrt{\la}\eta)^{-1}$.

Using this approach, Srednicki and Theisen [\ct{SreThe87}] found that
ultrarelativistic
particles are emitted in a narrow cone around the cusp velocity of angle $(\om
L)^{-1/3}$, where $L$ is the length of the loop.  The sum over modes is
apparently divergent, but in fact the neglect of back-reaction must fail when
$\om\sim M$, so $M$ acts as a cutoff. The emitted power is found to be
 $$
P \sim {\la'{}^2\mu\over\la ML}.
\eqqno
$$
Comparing with the gravitational power $P_G\sim G\mu^2$, we see that (for
$\la'=\la$) loops larger than about $(G\mu)^{-1}M^{-1}$ emit an
insignificant fraction of the energy in 2-particle decays.  Equally, loops
smaller than this critical value quickly end their lives in a burst of
particle emission.  The reason for graviton emission being much more
likely is the dimensionful coupling and the 1-particle final state, which
yields a convergent mode sum
 $$
P_G \sim  G\mu^2\sum_n n^{-4/3}.
\eqqno
$$
Note that the fall-off with $n$ is just that advertized for loops with cusps
in section 4.2.

These methods are unable to give much insight into the emission of
particles with the same mass scale as the string, especially those
which actually make up the string.  For particles of mass
$\sim M$ the back-reaction is not negligible, and the replacement of
$\ket{S'}$ by $\ket{S}$ is no longer valid.  The
prime site for particle emission is the cusp, because the fields are
changing very rapidly there.  The cusp is also a region where the
string changes direction, and oppositely oriented segments get very
close.  In fact, they overlap in a region of length $\si\rms{c} \sim
(M|\tprime{\bX}_0|)^{-1/3}$, where the subscript 0 indicates the value at
the cusp. Thus there is the possibility of a classical annihilation of this
region, with the energy being transferred into propagating modes of the scalar
and gauge fields which make up the string [\ct{Bra87}].  This `cusp
annihilation' can result in the emission of energy $\mu
\si\rms{c} \sim \mu L(ML)^{-1/3}$.  If this happened every period, the
power would be $(ML)^{-1/3}\mu$, which dominates light particle emission and
also gravitational radiation for loops smaller than $\sim
(G\mu)^{-3}M^{-1}$.  However, it is not clear that the cusp does
recur:  the replacement of a segment of string of length $\Delta \si
\sim L(ML)^{-1/3}$ with one of length $\Delta \si \sim M^{-1}$ could
also result in $|\dprime{\bX}_0|$ becoming very large. The centre of
the string now takes a short cut of physical distance $M^{-1}$, so
$|\dprime{\bX}_0|$ is now of order $M$, and the new overlap region
contains energy $M$ --- much less than the original  $\mu L(ML)^{-1/3}$.

Small amounts of energy are also available for conversion into
particles when strings self-intersect and reconnect, especially at
high relative velocities, where the scalar field can be seen to be
highly excited in the intersection region after the strings have moved
away [\ct{She87}--\ct{LagMat90}].  However, none of these particle emission
processes challenge gravitational radiation as the dominant decay mode
of the gauge string; or coherent Goldstone boson radiation for the
global string, except when the loops are smaller than $(G\mu)^{-1}$
times their width.

\subsection{Particle scattering}
A perturbative calculation of the scattering cross-section per unit
length for a scalar particle coupled as in the last section reveals
that it takes the form
$$
\({\d \si\over\d l}\)\rms{b} \sim \({\la'\over\la}\)^2{1\over E},
\eqqno
$$
where $E$ is the energy of the incident particle. The cross-section is
therefore determined by the Compton wavelength of the incident boson. A nice
physical argument for this behaviour has been given by Brandenberger, Davis
and Matheson [\ct{BraDavMat88}]. The string effectively consists of a chain of
scatterers of size $M^{-1}$, with geometric cross-sections $M^{-2}$. A
boson of energy $E$ will scatter coherently off $M/E$ of them, thus
amplifying the cross-section of a length $E^{-1}$ by a factor $(M/E)^2$.
The final cross-section per unit length is therefore $M^{-2}(M/E)^2 E
\sim E^{-1}$.

Fermions have different couplings and a different phase space.  With a
coupling $g\phi^*\bar{\psi^{\rm c}}\psi$, the cross-section per unit
length goes as [\ct{BraDavMat88}]
$$
\({\d\si\over\d l}\)\rms{f} \sim \({g\eta^2\over M^2}\)^2 E \sim
\({g\over\la}\)^2{E\over M^2}.
\eqqno
$$
This is again explicable in terms of our chain of scatterers: fermion
wavefunctions do not add coherently, so in the length $E^{-1}$ the
fermion scatters from only one point with cross-section $M^{-2}$.
Thus $(\d\si/\d l)\rms{f} \sim M^{-2}E$.

These estimates are perturbative estimates, and are only good if
they represent small corrections to the probability of the particle
being unaffected.  Hence the coupling of the scattering particle to
the string must be small compared with the string self-coupling.
Otherwise, one must take into account the distortion of the wave
functions of the scattering particles by the string core.  This in
general leads to amplification of the scattering amplitude by a factor
${\cal A} \sim |\psi(M^{-1})|/|\psi_0(M^{-1})|$, where $\psi_0$ is the
free wave function [\ct{BraDavMat88}].  The amplification occurs because
the expansion in angular momentum eigenstates of the exterior part of
the true wave function can contain modes which diverge at small
radius.  In the free case, the condition of regularity at the origin
means that they cannot be present, but when the string is there they
can match onto the solution in the core.  Precisely which divergent
modes are excited determines the amplification.

A crucial factor in the calculation of the cross-section for gauge
strings is the charge of the scattering particle under the generator of
the broken $U(1)$, say $q$.  If the charge of the string field is $e$,
then the cross-section depends very strongly on the ratio $q/e$.  If it
is an integer, then it is found (for both bosons [\ct{Eve81}] and
fermions [\ct{Per+91}]) that
$$
\({\d\si\over\d l}\) = {4\over k\ln^2(k/M)},
\eqqno
$$
where $k$ is the transverse momentum of the incident particle.  If,
however, $q/e$ is fractional, the particle experiences an
Aharonov-Bohm effect [\ct{AlfWil89}].  The wave function of a particle
passing the string picks up a relative phase factor $e^{i2\pi(q/e)}$
between the parts which pass on either side of the string, and so they
can interfere very strongly.  In that case it is found that the
cross-section is just the usual Aharonov-Bohm form
$$
 \({\d\si\over\d\th\d l}\) = {1\over 2\pi k}
{\sin^2(\pi q/e)\over\cos^2\th/2}.
\eqqno
$$
The cross-section does in fact have some dependence on the details of
the fields in the core of the string [\ct{FewKay93}].  It can depart
from the Aharonov-Bohm form if, for example, the flux in the core
changes direction with radius (which might happen in strings in
theories with several stages of symmetry breaking).

There is another effect, gravitational in origin, which contributes to
particle scattering.  We recall that the space-time around a straight
string has an angle deficit $\de = 8\pi G\mu$, so if we place a charge
in this background, we must satisfy the boundary condition that the
electric field is the same at $\th=\pi+\de/2$ and $\th=\pi-\de/2$. By
solving the Poisson equation with  these modified periodicity
requirements, it can be shown [\ct{Lin86}] that there is a repulsive
force
$$
\bi{F} = {G\mu q^2\over 16r^2} \hat{\bi{r}}
\eqqno
$$
where $\hat{\bi{r}}$ is the unit vector from the nearest point on the
string to the particle.  This results in a Coulomb-like scattering
cross-section [\ct{ConSca}].

\subsection{Baryon number violation}
In the interior of a string formed after the breaking of a Grand Unified
symmetry are fields which carry both baryon and lepton number.  Thus if
a quark gets to the core of the string it can interact with the
background core fields and emerge a lepton, and vice versa.  Strings
therefore have the potential for mediating baryon number violating
processes.  This is very reminiscent of the Rubakov-Callan effect
[\ct{RubCal}],
where quarks and leptons change into each other in the interior of a
monopole.  The cross-section for this process is not the geometric
cross-section: rather, it depends on the Compton wavelength of the
scattering particle, going as $E^{-2}$.  For strings, we have seen how
the elastic scattering cross-section depends crucially on the ratio
$q/e$, since it controls the amplification of the scattering
wave-function at the core of the string.  Since any $B$-violating
processes happen in the string core, the same is also true of the
$B$-violating cross-section [\ct{AlfRusWil89},\ct{Per+91}].
The cross-section also depends on the type of interaction: that is,
whether it is mediated by a scalar or by a vector field.  A vector field
produces its greatest effect for integer $q/e$, when the $B$-violating
cross-section per unit length can reach the scattering cross-section
$\sim E^{-1}$ [\ct{Per+91}]. The $B$-violating cross-section from a core
scalar field is greatest for half-integer $q/e$, when it can also reach
$\sim E^{-1}$ [\ct{AlfRusWil89},\ct{Per+91}].

There have also been suggestions that there can be $B$-violation
produced by a kind of Aharonov-Bohm effect [\ct{Per+90},\ct{Ma93}].  If
the generator of the gauge field of the string does not commute with
the generator of $B$\ or $L$, then the adiabatic transport of a particle
around the string can result in its returning with different $B$\ or $L$\
quantum numbers: this is then apparently a $B$- or $L$-violating
process.  In a scattering experiment, the outgoing wavefunction
resulting from an incident beam of pure quarks could contain a mixture
of quarks and leptons, in different proportions depending on the
scattering angle $\th$. However, it has been shown that this mixture is
an illusion [\ct{BucGol94}]. While it is true that the quark and lepton
fields in the original basis at $\th=0$ are apparently mixed as we move
around the string, the definition of what mixture actually constitutes
a quark or lepton field also changes.  The worst that can happen is
that the definition at $\th=2\pi$ differs by a phase from that at
$\th=0$: but this is just the original Aharonov-Bohm effect, without
any peculiar baryon number violation.

The physical consequences of $B$-violation are greatest when the string
density is greatest, soon after the phase transition which forms them.
Any existing $B$-asymmetry in the universe can relax to zero via
scattering off strings [\ct{BraDavMat89}].  If $\xi(t)$ is the
correlation length of the string network, and
$\si_{\Slash{B}}$ is the
$B$-violating cross-section per unit length, then
$$
{\d n_B \over \d t} \simeq -\bar{v} \si_{\Slash{B}}
{1\over \xi^2} n_B,
\eqqno
$$
where $n_B$ is the baryon number density and $\bar v$ is the thermally
averaged relative velocity of the particles and the strings, which is
of order 1.  The cross-section per unit length goes as $c/T(t)$, where
$T$ is the temperature and $c$ a dimensionless constant, while the
correlation length increases at $t^p$.  When the network is scaling (see
section 5.3) $p=\half$, but just after the phase transition it has the
value $5/4$.  Using the relation $T\sim (t\MPl)^{-1/2}$, we find
$$
n_B(t) \simeq n_B(t_0) \exp \(C\[(t_0/t)^{2p+1/2} -1\]\),
\eqqno
$$
where
$$
C \sim {c\over (t_0\MPl)^\half}\({t_0\over\xi(t_0)}\).  \eqqno $$
Thus if the baryon number is generated when the string is scaling, that
is $\xi \sim t$, the reduction factor $\exp(-C)$ is insignificant.
However, if it is generated at or before the string-forming phase transition,
when $\xi(t_0)/t_0 \sim T_{\rm c}/\la\MPl$, there is potentially a very
large suppression.

Where there is scattering there is also emission, so strings may also
be a source of baryon number.  As Sakharov pointed out [\ct{Sak67}],
generating a baryon asymmetry in the early universe requires not only
$B$-violation but also $CP$-violation, as well as a departure from thermal
equilibrium.  If quarks and leptons have $CP$-violating couplings to
the strings, or to a string decay product, then emission from strings
constitutes an injection of baryon number into the equilibrium
distribution of particles in the early universe
[\ct{BhaKibTur82}--\ct{DavEar93}].  This requires a high
string density, in order to produce a sufficient asymmetry.
However, the tendency
of the strings to reduce the baryon number density by scattering has not
been taken into account in existing estimates.

\section{Strings in the Early Universe}

If cosmic strings exist, they were formed in the first
fraction of a second after the Big Bang.  Any effects
they have on observable cosmological features occur at
much later epochs. To follow the causal chain from one to
the other, we have to study how they were formed and how
the network of strings evolves over this immense span of
time.

\subsection{High-temperature field theory}

To study the formation of strings, since the early
Universe was very hot, we need to consider the effects of
high temperature on a field theory.

The most important effect is to replace the potential
$V(\ph)$ in the Lagrangian by a high-temperature {\it
effective potential}, $V_T(\ph)$
[\ct{DolJac74},\ct{Wei74},\ct{Kap89}].
This is the {\it free
energy\/} density in a state with the prescribed value of
$\ph$.  When $T$ is much larger than all the masses
involved (i.e., $T\gg m(T)$, the temperature-dependent
mass), the leading correction terms are easily
identifiable with the free energy density of an ideal
relativistic gas:
 $$
V_T(\ph) = V(\ph) - {g_*\pi^2\over90}T^4 +
{M^2(\ph)\over24} T^2 + {\rm O}(T).
\label{\eEffPot}
$$
Here $g_*$ is the effective number of spin states of
relativistic particles (with a factor of 7/8 for
fermions) and $M^2(\ph)$ is the sum of squared masses of
particle excitations about the chosen state (with a factor
of 1/2 for fermions).  The $T^4$ term is independent
of $\ph$, and hence does not affect the symmetry breaking.
The important correction is the $T^2$ term, which yields
symmetry restoration at high temperature.

Consider for example the model with Lagrangian
(\rf{\eLocLag}) and $V(\ph) = \half\la(|\ph|^2 -
\half\et^2)^2$.  The zero-temperature symmetry-breaking is
signalled by the fact that $V$ has a {\it maximum}
at $\ph=0$.  Here we find $g_* = 4$ and $M^2(\ph) =
3e^2|\ph|^2 + \la(2|\ph|^2 - \et^2) +
{1\ov12}(2\la+3e^2)T^2$.  Thus at high
temperature, the coefficient of $|\ph|^2$ in $V_T$
becomes positive.  Then $V_T$ takes its {\it minimum\/}
value at $\ph=0$.
In this approximation, we have a second-order phase
transition.  The critical temperature above which the
symmetry is restored is given by
 $$
\Tc^2 = {6\la\et^2\over2\la+3e^2}.
\eqqno
$$
The ${\rm O}(T)$ term in (\rf{\eEffPot}) can change this
analysis: formally, it is $-T\sum_j m_j^3(\phi)/12\pi$,
where $j$ labels the particle species.
However,  it turns out that this term can only be trusted
if  $\la \ll e^2$.
Its effect is to give the potential a shallow
minimum at $\ph=0$, which suggests that for heavy vector
particles there should be a
first-order phase transition.

For the electroweak transition, the high-temperature
approximation appears to be reasonable provided that
$\la$, or equivalently the Higgs mass $m_H$, is not
too large [\ct{ArnYaf94}].
In practice, values of the Higgs
mass below about 65 GeV can probably now be ruled out.
For reasonable
values of $m_H$, the critical temperature is around
$200$ GeV, substantially larger than $m_W$ and $m_Z$.
However, the conclusion concerning the order of the
transition is still far from certain.  There are infrared
divergences in higher-order correction terms which
invalidate the perturbation calculation in the
immediate neighbourhood of the critical point, the
effects of which are not wholly understood [\ct{Kap89}].  It has
been argued that non-perturbative effects may yield a
first-order transition even in the case of large Higgs mass
[\ct{Sha93}].

In grand unified theories, there is a phase transition
(or transitions) at a critical temperature of $10^{15}$
or $10^{16}$ GeV.  In this case, the estimated critical
temperature is roughly comparable with the induced
particle masses, so a first-order transition is rather
more likely.  This is mainly because the number of
contributing light particles is much larger.

There is still much work to be done before it can be said
that we have a completely reliable understanding of the
high-temperature phase transitions in field theory.

It should be mentioned that there is probably another
important transition in the early Universe, considerably
later than those mentioned so far.  This is the
quark-hadron transition, at which the dense gas of free
quarks and gluons condenses into individual hadrons.
It occurs at a temperature of perhaps 200 MeV, and may be
associated with the breaking of chiral symmetry.
This transition, while not generating cosmic strings itself,
is an important stage in the evolution of axion strings,
as we shall describe in section 5.10.

\subsection{Formation of Strings}

Because the scales of interest are mostly very large
compared to the thickness of a cosmic string, the
long-term evolution of a string network is almost
independent of the details of the field-theory model in
which the strings appear --- though we have to
distinguish local from global strings and superconducting
from non-superconducting ones, and of course we have to
know the symmetry-breaking scale.  So, for simplicity, let
us consider the simplest model, the Abelian Higgs model
described by the Lagrangian (\rf{\eLocLag}), which predicts
non-superconducting, local strings, with tension
$\mu\ap\et^2$.  We shall also concentrate on thermal
phase transitions, which happen as a result of the cooling
of the Universe in a conventional radiation-dominated
FRW Universe, reserving a brief discussion of the formation
of defects during inflation for the end of this section.

Initially, when the Universe is very hot, it is in
the symmetric phase and no strings are present.  When
it cools to below the critical temperature $\Tc$, the
Higgs field $\phi(x)$ will tend to settle down into the
valley containing the circle of minima of the potential
$V_T$.  Eventually, as $T \to 0$, the system will
tend towards one of the vacuum states described by the
points on this circle.

Of course, this is a quantum system, so it might also be
in a superposition of these states.  If we were dealing
with a single particle then, as every student of
quantum mechanics knows, the true ground state would be a
symmetric superposition; there would be no symmetry
breaking.  However, a field theory is different, because
it involves an {\it infinite\/} number of degrees of
freedom.  To get from one to another of the degenerate
vacua, one must rotate the phase of the Higgs field at an
infinite number of points.  Thus a quantum superposition
of states with different values of the phase is for all
practical purposes equivalent to a classical
superposition.  No local observable has matrix elements
between the degenerate states.  Because of gauge
invariance, the different vacuum states are physically
indistinguishable; the overall phase of $\phi$ is not an
observable.

In fact, even the {\it relative\/} phase of
$\phi$ between different points is not an observable,
unless we have chosen a gauge.  For example, we might work
in the Coulomb gauge.  Then, provided the remaining gauge
ambiguity is removed by imposing a suitable boundary
condition, the relative phase between the $\phi$ values at
different points is an observable.

Now what happens when the temperature in the expanding
Universe falls below the critical temperature?  Once it
has fallen well below $\Tc$, the Higgs field in most
regions will have acquired a non-zero average value,
but for causality reasons the phases in widely separated
regions will be uncorrelated.  This must be true even in
a gauge theory, once the gauge ambiguity has been
removed.  We should of course also consider the coupling
to the gauge field.  However, the currents that generate
such fields arise from phase fluctuations only once the
magnitude of $\ph$ has become significantly non-zero.  It
therefore seems unlikely that this effect would make a
major difference to the probability of string formation.
(However, for a different view, see [\ct{RudSri93}]).

The phase transition may be first-order, in which case it
will proceed by bubble nucleation
[\ct{Col77}--\ct{Lin83}] or by spinodal decomposition
[\ct{Lan92}];
alternatively it may be a
continuous, second-order process.
The order depends, as we have
seen, on details such as the ratios of coupling
constants.  The net effect is much the same in both cases,
though one may perhaps expect a rather different initial
scale for the resulting string network.

In the bubble-nucleation case, we should expect the phase in
each bubble to be a more or less independent random
variable, although if the phase is determined by the
random values of other fields, one might expect the phases
in neighbouring bubbles to have some correlation.  When
two bubbles meet, the values of $\phi$ across the
boundary will tend to interpolate between those in the
two bubbles [\ct{HinBraDav94}].
When three bubbles meet, a string may or
may not be trapped along their mutual boundary, depending
on whether or not the net phase change around it is
$2\pi$ or zero [\ct{Kib76}].  The scale size of the
resulting network is related to the separation of
nucleation centres, and to the probability of trapping a
string at a boundary, which is typically of order 0.3
[\ct{VacVil84},\ct{LeePro91}].  The scale
would be larger if there were
indeed a correlation between the phases of neighbouring
bubbles ---  the separation of nucleation centres would be
replaced by the overall phase correlation length.

In the alternative case of a second-order transition, or of a
first-order transition proceeding by spinodal decomposition,
the Higgs field
everywhere will tend to move away from zero --- the peak
of the potential hump --- at about the same time.
It is of course characteristic of a
second-order transition that the values of $\phi$ will be
correlated over long distances, though the correlation
length will never actually become infinite if the rate of
change of temperature is finite.  Immediately after the
transition, when the hump is still small, fluctuations
taking $\phi$ over it will be quite probable, so the phase
is not really fixed.  However, as the temperature falls
further, such fluctuations become increasingly
improbable.  When one reaches the {\it Ginzburg
temperature\/} $T_{\rm G}$ [\ct{Gin60}], at which the
energy of a fluctuation over the hump on the scale of a
correlation volume is equal to $T$, these fluctuations
effectively cease and the phase is `frozen in'.
(Remember that in our units $k_{\rm B}=1$.)  In this case
too, string defects will be trapped in places where the
phase change around a loop is $2\pi$.  The initial length
scale of the resulting network will be determined, as
before, by the probability of string trapping and by the
correlation length at the Ginzburg temperature.

In either case the net result is a random network of
strings with some characteristic scale, which we shall
call $\xi$;  $\xi$ is certainly $<t$ and probably $\ll
t$.  The strings cannot have free ends and must either form
closed loops or be infinitely long.  Numerical
simulations of string formation on a cubic lattice
performed by Vachaspati
and Vilenkin [\ct{VacVil84}] suggest that
initially about 20\% would be in the form of loops
with the remaining 80\% in `long' strings,
defined as strings which  crossed the periodic box of the
simulation.  This figure, however, depends on the choice
of lattice.  A tetrahedral lattice (which is preferable
because it avoids the ambiguity associated with having
more than one string passing through a cubic cell) gives
37\% in loops [\ct{HinStr94}]. It is interesting that
for $Z_2$ strings on a cubic lattice
the proportion in loops appears to be much less, about
6\% [\ct{Kib86a}].

The precise details of this structure are not
important, because the subsequent evolution leads, as we
shall see, to a final state which is more or less
independent of this early structure.

As we mentioned in the introduction, there are models in which
strings or other defects can be formed during the late stages of inflation
[\ct{ShaVil84},\ct{LazSha84},\ct{VisOliSec87}].
This can be achieved by using an appropriate type of
coupling between the inflaton field $\si$ and the field $\ph$ responsible for
string formation.  For example [\ct{VisOliSec87}], if there is a term in the
potential of the form $\la(|\ph|^2 + v^2 - \al\si^2)^2$, where $\la$, $v$ and
$\al$ are constants, then when $\si$ is small, $\ph$ will remain near zero, but
as it evolves it will eventually reach the point where $\si>\sqrt{\al}v$, and
it
is then energetically favourable for $\ph$ to acquire a non-zero average value.
Thus the string-forming phase transition occurs {\it during\/} inflation.  Many
variations on this theme are possible.

A notable example is that of Yokoyama [\ct{Yok88}], who
pointed out that there would in
general be a coupling between $\ph$ and the scalar curvature $R$,
of the form $\half \xi R |\ph|^2$.  Thus $R$ behaves exactly like
the square of the temperature in the effective potential (\rf{\eEffPot}),
and as it decreases towards
the end of inflation, a phase transition in the $\ph$ field can
happen, if $\xi$ is positive and $V(\ph)$ possesses the usual
symmetry-breaking form of equation (\rf{\eGloLag}).

\subsection{Early Evolution}
After a thermal phase transition,
the string tension $\mu$ acquires its zero-temperature
value of order $\eta^2$ once the temperature becomes
small compared to $\Tc$.  Immediately after their
formation, however, the strings have smaller tension (in
the simple approximation used in section 5.1, it
is reduced by a factor $[1-(T^2/\Tc^2)]$).  At that time,
the strings are moving in a very dense environment, causing
their motion to be heavily damped.

As we saw in section 3.4, there is a length scale
$l\rms{f} = \mu\si/\rh$ associated with friction,  where
$\rho$ is the energy density of the surrounding matter
and $\sigma$ the average linear cross-section for particle
scattering (strictly speaking, the momentum-transfer cross
section).  According to (\rf{\eDenRel}), $\rh\sim T^4$
while, as we saw in section 4.6, $\sigma$ is (in very
rough terms, neglecting logarithmic factors) of order
$1/k$ for particles of momentum $k$.  Since typically
$k\sim T$, we may set $\sigma\sim 1/T$.   Thus
$l\rms{f}\sim \mu/T^3$.  Recall also that
$\mu\sim\Tc^2$.  Initially, therefore, when $T$ is close
to $\Tc$, $l\rms{f}\sim 1/T$.  We then expect $l\rms{f} <
\xi < t$.  The question is: how fast does this length
scale $\xi$ of the string network grow in response to the
effect of friction?

Think of a segment of string with radius of curvature
$r$, initially at rest.  Because of the string tension
$\mu$, it experiences a transverse accelerating force
$\mu/r$ per unit length.  When moving with velocity $v$ it
suffers a damping force equal to $\rho\sigma v$ per unit
length, so the expected limiting velocity of
our string segment is $\mu/r\rho\sigma\sim l\rms{f}/r$.
Typically, $r\sim\xi$, so the expected time scale for
straightening kinks is $\td\sim\xi^2/l\rms{f}$.

This is also the time scale on which $\xi$ increases,
i.e., $\dot\xi/\xi \sim 1/\td \sim l\rms{f}/\xi^2$.  Since
we are in the radiation-dominated era, $l\rms{f} \sim
\mu/T^3 \pt t^{3/2}$. Thus (apart from an initial
transient) the correlation length  behaves like $\xi\pt
t^{5/4}$, increasing faster than the horizon distance.
This process continues [\ct{Kib87}] until all three
lengths, $l\rms{f}, \xi$ and $t$, are of the same order.
This happens when the temperature of the Universe is
$T=T_*\sim\Tc^2/\MPl$.  For example, if $\Tc$ corresponds
to the temperature of the supposed grand unification
transition, namely $\Tc\sim10^{15}$ GeV, then
$T_*\sim10^{12}$ GeV, which still of course occurs at a
very early time, in fact when $t\sim10^{-31}$ s.

As $T$ falls further, the damping becomes less and less
important.  Once $T\ll T_*$, it is negligible, and the
motion of the strings can be considered essentially
independently of anything else in the Universe.  They
soon acquire relativistic speeds.

What happens thereafter to the characteristic length
scale $\xi$ of the string network?  As a matter of fact
there may be several different length scales, so in order to
answer the question we have to define it more precisely.
One way of defining $\xi$ is simply in terms of the
overall string density (excluding small loops, for reasons
to be explained later) --- essentially, $\xi$ is the length
such that within any volume $\xi^3$ we expect to find on
average a length $\xi$ of string.  This is equivalent to
saying that the string density $\rs$ is
 $$
\rs = {\mu\over\xi^2}.
\eqqno
$$

For causality reasons, $\xi$ can never grow larger than
$t$, so there are two logical possibilities: either the
network approaches a `scaling' regime in which $\xi$
remains a fixed fraction of $t$, or it grows more
slowly, in which case the ratio $t/\xi$ increases.

Now
in the radiation-dominated era, the total density $\rt$ of
the Universe, given by (\rf{\eDenRad}), scales like
$1/t^2$.  Thus in the scaling
regime, the strings constitute a fixed fraction of the
total energy; in fact, if we define $\ga$ to be the ratio
of the Hubble distance to $\xi$, then in the radiation era
 $$
\ga = {2t\over \xi} \to  {\rm constant}.
\eqqno
$$
The fraction of the total density in the form of strings then
becomes
 $$
{\rs\over\rt} = {8\pi\over3}\ga^2 G\mu,
\eqqno
$$
so the strings constitute
an interesting though small fraction.

On the other hand, if $\xi$ grows more slowly than $t$,
it is inevitable that the strings will eventually come to
dominate the total energy density.  Such a Universe would
be very different from ours, at any rate if the strings
are heavy, with a characteristic scale anywhere near that
associated with grand unification.  It is just possible
that the Universe might be string-dominated if the
strings were very much lighter and so formed at a much
later stage in the history of the Universe
[\ct{Vil84a},\ct{Kib86b}]; however, this does not seem very
likely.

\subsection{The role of loops}

There are, then, these two quite different final states.
What determines the choice?  Fundamentally, the answer
is: the efficiency of the energy-loss mechanism.

Suppose for a moment that the string configuration simply
stretches with the Hubble expansion.  Then, of course, the
total length of string in any comoving volume would grow
in proportion to the cosmic scale factor $a$, so the
string density would go like $1/a^2$.  In other words,
$\xi\propto a\propto t^{1/2}$.  Then the strings
would quite rapidly come to dominate the energy density.

So clearly, if this is not to happen, there must be some
efficient mechanism for transferring energy from the
strings to other forms of matter and radiation.  Strings
of course lose energy by radiating particles of all
kinds.  But non-superconducting strings generally do not
carry non-zero charges of any type, so their interaction
with most fields is rather weak (see section 4).  On
the other hand, for grand unified strings, their
characteristic energy scale is not many orders of
magnitude below the Planck scale, so their gravitational
interactions, at least over a long time, are
non-negligible.

Strings do lose a significant amount of energy in
gravitational radiation, but that in itself would not be
enough.  An important feature of the energy loss
mechanism is the formation of small closed loops
[\ct{Vil81}].  As we discussed in section 3.6, whenever
two strings intersect, they reconnect or intercommute
(see figures 3.3, 3.4).
When a string intersects {\it itself}, it breaks off a
closed loop. Once formed, such a loop will
oscillate quasi-periodically, gradually losing energy by
gravitational radiation.  This is a very slow process;
a loop of length $l$ has a lifetime of $l/\Ga G\mu$, which
for grand unified strings is about $10^4l$.  But this is
irrelevant to the fate of the long-string network; once
formed, the loop is effectively lost --- {\it unless\/} it
rejoins the long string network by a second
intercommuting event.

Large loops quite often rejoin the main network.  Smaller
loops, however, usually survive without again encountering
any long strings.  It is not too difficult to estimate
what `long' and `short' must mean here.  The
probability that any particular segment of string, of
length $l$ say, intersects another piece of string in a
short time interval $\de t$ is clearly proportional to
$l$, to $\de t$, and to the overall density of long
strings, i.e., to $1/\xi^2$.  Hence it is $\ch l \de
t/\xi^2$, where $\ch$ is a dimensionless constant
[\ct{CopKibAus92}], now estimated to be about $0.1$
[\ct{AusCopKib93}].  The probability of intercommuting is
a rapidly decreasing function of time, because the string
density is falling.  Assuming that $\xi$ at least
approximately scales, i.e., that $\xi\propto t$, we find
that the probability that a loop of length $l$ formed at
time $t$ will survive without reconnection is
 $$
\exp\left(-\int_t^\infty \d t'{\ch l\over \xi^2}\right)
\sim \exp\left(-{\ch lt\over \xi^2}\right).
\eqqno
$$
Note that the assumption about scaling is a very weak
one: if $\xi$ decreases like some other power of $t$, the
only difference will be an additional numerical factor
in the exponent.  We may conclude therefore that strings
longer than $\xi^2/\ch t$ are very likely to reconnect,
while much shorter strings are unlikely to do so.

\subsection{Simulations of string evolution}

Given that strings do have this energy loss mechanism,
how can we establish whether it is adequate to ensure
that the string network does indeed approach a scaling
regime?

Much of the evidence we have bearing on this question
comes from computer simulations of an evolving string
network (in a spatially flat
Friedmann--Robertson--Walker universe)
performed by three different groups, referred to here as
AT [\ct{AlbTur85},\ct{AlbTur89}], BB
[\ct{BenBou88}--\ct{BenBou90}] and AS
[\ct{AllShe90},\ct{SheAll90}].  There are some
disagreements of detail between the groups, but all three
find evidence for scaling of the long string network; in
other words, they find that the density of long strings
eventually behaves like $1/t^2$, as scaling predicts.
However, this is very far from a conclusive answer
to the question, for several reasons.

Firstly, the simulations can only be run for relatively
limited times, perhaps twenty Hubble times.  So if there
were, for example, logarithmic departures from scaling,
they would certainly not show up.

Secondly, there certainly are features of the network that
have {\it not\/} yet reached a scaling limit.  One feature
that has shown up very clearly in the work of BB and AS,
as well as in the more recent simulations of AT, is that on
small scales the strings are extremely wiggly.  This comes
about because of the intercommuting process: every
intercommuting event introduces four new kinks, two on
each of the strings, propagating away from each other at
the speed of light.  The kink angles do decrease, due to
the Universal expansion, but only very slowly, typically
like $t^{-0.1}$ [\ct{BenBou90},\ct{CopKibAus92}], and the
kinks are slowly rounded by the back-reaction of
gravitational radiation.  But because these processes are
so slow, the number of kinks on the strings grows
throughout the period covered by any of the simulations.
Whether this small-scale structure does eventually scale
is a crucial unsolved problem, to which we shall return
below.

Finally, there is an unresolved disagreement about the
actual density at which strings scale, i.e., the value of
the constant $\ga$ introduced above.  In the
radiation-dominated era, AT find $\ga \ap 14$, while both
BB and AS predict a value of $\ga \ap 7$.  This indicates
a difference in the predicted scaling density of a factor
of 4.

The origin of this disagreement appears to lie in the
treatment of the very small-scale structure.
Sharp-angled kinks are very hard to treat in numerical
simulations.  All the simulations use cut-offs or
smoothing of one kind or another and the way in which this
is done affects the results.  In particular, there is
also a significant disagreement about the typical size
of the loops that are formed.  The initial naive guess
which formed the basis of early work on the cosmic
string scenario, was that loops should typically have a
size comparable with the scale length $\xi$.  This is the
prediction of what has been described as a `one-scale
model' [\ct{Kib85}].  It is now agreed by all authors that
that initial guess was quite wrong: typically loops are much
smaller; the length satisfies  $l\ll\xi$.  But the various
simulations do not agree about {\it how much\/} smaller
they are: typical loops in the simulations of BB
and AS are smaller than those of AT (even though AT's
value of $\xi$ is less).

It is obviously important to resolve these issues,
and one alternative approach to studying the dynamics of string
networks is to do the numerical simulations in Minkowski
space [\ct{SmiVil87},\ct{SakVil90}].  This has the advantage that there is
an {\it exact\ } algorithm for evolving the string [\ct{SmiVil87}],
and so there
is no problem with kinks --- they remain sharp, being evolved
in accordance with the true equations of motion.
However, the damping from the expansion of
the Universe is
neglected, and it is unclear how good an approximation that is.
Ultimately, though, what is needed to complement numerical
simulations is an analytic treatment.

\subsection{Analytic treatment of string evolution}

The first attempts at an analytic treatment, by Kibble
[\ct{Kib85}] and by Bennett [\ct{Ben86a},\ct{Ben86b}], were
founded, as mentioned, on a `one-scale model' of string
evolution, based on assumptions about how the rates of
various processes depended on the scale length $\xi$.
These processes are stretching (due to the Hubble expansion),
intercommuting, and loop production.  (Gravitational
radiation was assumed to have a negligible effect on the
evolution of the long-string network itself, though being
crucial in the decay of loops.)

It soon became apparent, however, that this was a very
crude description of the behaviour revealed by the
simulations.  In particular, the recognition that the
typical loop size is much smaller than $\xi$ called in
question the whole idea of a one-scale model.

Kibble and Copeland, later with Austin, [\ct{KibCop91},
\ct{CopKibAus92}] then developed a `two-scale model'
intended to accommodate this extra structure.  The second
scale, $\xb$, was defined essentially as the persistence
length along a string --- more precisely, along the {\it
left-moving\/} string.  Recall that the motion of a string
can be resolved into `left-moving' and `right-moving'
modes (see equation (\rf{\eGenSol})),
with tangent vectors $\bp = \ba'$ and
$\bq = \bb'$ respectively.
In an expanding Universe (see equation (\rf{\eEomExp})),
there is a small
coupling between $\bp$ and $\bq$  --- negligibly small for
wavelengths much shorter than the Hubble radius.  The
definition of $\xb$ is
 $$
\xb = \int_0^\infty \d s \langle \bp(0)\cd\bp(s) \rangle,
\label{\eChiBar}
$$
where the angle brackets denote averaging over an
ensemble of string distributions and $s$ is the length
along the string.  (Thus {\it if\/} the angular
correlation function $\langle \bp(0)\cd\bp(s) \rangle$ is
assumed to have an exponential dependence on $s$, then it
would be $\e^{-s/\xb}$.)  Of course, one could equally use $\bq$ rather than
$\bp$ in the definition.

The approach adopted was to try to derive equations for
the rates of change of the length scales $\xi$ and $\xb$
due to the processes of stretching, intercommuting and
of loop formation.  A rather different, but in some ways
complementary, treatment has been developed by Embacher
[\ct{Emb}].

As an illustration of the approach let us consider  for
example the effect of intercommuting.  The probability
that a small segment of string of length $l$ undergoes an
intercommuting event during a time interval $\de t$ is
$\ch l \de t/\xi^2$, where $\ch$ is the dimensionless
parameter introduced above.

Now, what is the effect of intercommuting on $\xi$ and
$\xb$?  In this particular case, the answer is rather
simple.  Intercommuting does not change the total length
of string, so $\xi$ is unaffected.  On the other hand,
intercommuting introduces new kinks onto the string, so
its effect would be to {\it decrease\/} $\xb$.  If one
thinks of $\xb$ very roughly as representing the mean
distance between large-angle kinks on the string, one can
see that the result should be to increase $1/\xb$ by an
amount $\ch\de t/\xi^2$.  Thus the expected contribution
to $\d\xb/\d t$ due to this effect is in fact
$-\ch\xb^2/\xi^2$.  (A more detailed calculation suggests
that there should be a numerical factor, of order unity
[\ct{AusCopKib93}].)

Using similar arguments, one can derive expressions for
the rates of change of $\xi$ and $\xb$ due to stretching
and loop formation.  To look for scaling, it is
convenient to introduce new variables, $\ga$ (already
defined above) and $\gb$:
 $$
\ga = {2t\over \xi},\qquad \gb = {2t\over\xb}.
\label{\eGamDef}
$$
(We consider only the radiation-dominated era for
simplicity.)

The equations obtained [\ct{KibCop91}] (written in a
slightly modified notation) were
 $$
\eqalign{
t{\d\ga\over \d t} &= {1+\al\over4}\ga
 - {c\over4}\ga\gb \cr
t{\d\gb\over \d t} &= \(1-{3\al\over2}\)\gb -
{q-1\over2}c\gb^2 + {\ch\over2}\ga^2. }
\label{\eScaEqn}
$$
There are three new dimensionless parameters here: $\al$ is
defined by $\al = - \langle \bp(0)\cd\bq(0) \rangle$,
or equivalently in terms of the mean square transverse
velocity of the strings, $\al = 1 - 2\langle\bv^2
\rangle$; $c$ is a measure of the efficiency of loop
formation; while $q$ corresponds to a more subtle effect,
relating to the expected `kinkiness' of excised loops.
(Regions of string with more than the usual number of
kinks are more likely to produce loops than smoother
regions; consequently loops will tend to have more than
the average number of kinks on them.)

`Scaling' is achieved if $\ga$ and $\gb$ approach
definite finite limits as $t\to\infty$, i.e., if the
equations (\rf{\eScaEqn}) have a stable fixed point.  In
the most naive version of the model, the three
dimensionless parameters were taken to be constants.  Then
it is easy to see that scaling will be achieved provided
that $q$ exceeds a critical value, in fact
 $$
q > {3-2\al\over1+\al}.
$$
According to BB [\ct{BenBou90}], $\al\approx 0.14$, so
the required condition is $q\gap 2.4$.  In general
terms, this conclusion has been confirmed in a more
sophisticated version of the model [\ct{CopKibAus92}], in
which the fact that $\al$ and $c$ actually depend on the
ratio of the length scales is taken into account.

\subsection{Inclusion of small-scale structure}

There are two major problems, however, with the model just
described.  Firstly, there seems to be no obvious
way of reliably estimating the magnitude of the most
important parameter, $q$.  Even more critically, though
the model was originally intended to provide some
understanding of the small-scale structure seen in the
simulations, it does nothing of the sort.  For reasonable
values of the parameters, the two length scales $\xi$ and
$\xb$ obstinately turn out to be of the same order of
magnitude.

The trouble here is that the definition
(\rf{\eChiBar}) of $\xb$ is dominated by long-range
correlations and has rather little to do with the
small-scale structure.  What the simulations reveal is a
very complex structure.  Viewed on a sufficiently coarse
scale, the strings are roughly straight on a scale ($\xb$)
quite comparable with the typical inter-string distance
($\xi$); but looking at the finer detail, one sees many
wiggles on each of these `straight' segments.  The
typical inter-kink distance ($\ze$, say) is very much less
than $\xb$.

To deal with this problem, the model has been further
refined [\ct{AusCopKib93}], now becoming a {\it
three\/}-scale model.  The definition of the third scale
must involve the short-distance behaviour of the angular
correlation function.  It can be taken to be
 $$
{1\over\ze} = -{\pa\over\pa s} \langle \bp(0)\cd\bp(s)
\rangle \bigg|_{s=0}.
\eqqno
$$

The full set of equations describing the evolution of all
three length scales is too complicated to quote here.
The conclusions of the analysis, however, are fairly easy
to describe.

If we include only the effects discussed so far, the
equations {\it do not\/} scale (unless one of the
parameters is larger than seems at all probable).
Initially, if all the length scales start out roughly
comparable, $\xi$ and $\xb$ start to grow, approximately
proportionally to $t$ as scaling would require.  However,
$\ze$ does not grow, but remains roughly constant.  So as
their overall dimensions grow, the strings would become
more and more kinky.  To put this another way, let us
define another scaling parameter $\ep$, by analogy with
(\rf{\eGamDef}), by
 $$
\ep = {2t\over \ze}.
$$
We then find that while $\ga$ and $\gb$ approach constant
scaling values, $\ep$ continues to grow throughout the
evolution.

But there {is} another effect which we must include,
namely gravitational radiation.  It is actually
relatively straightforward to do so, using results
derived by Hindmarsh [\ct{Hin90a}].  Several authors
[\ct{AllCal91}--\ct{Aus93}] have already
argued that it is the back-reaction of gravitational
radiation that will eventually cause the small-scale
structure on strings to scale.

This hypothesis is confirmed by the analysis of the
evolution equations [\ct{AusCopKib93}].  Under reasonable
conditions, the string network does eventually reach a
full scaling regime in which all three length scales grow
in proportion to the time, but with a large ratio between
$\xi$ and $\xb$ on the one hand and $\ze$ on the other:
in fact, one expects $\ze/\xb \sim \Ga\rms{ls} G\mu$.  Here,
$\Ga\rms{ls}$ is a numerical constant, analogous to our
earlier constant $\Ga$, which
characterizes the rate of energy loss by gravitational
radiation from long strings,
rather than from loops.  We expect $\Ga\rms{ls}$ to be of
order 10.

This conclusion still rests on the values of some
undetermined constants, and in particular on a parameter $\hat
C$ that governs the effectiveness of gravitational
back-reaction in smoothing small-scale kinkiness on the
strings.  This must exceed a critical value if scaling is
to be achieved.  Various physical arguments suggest that
it is likely that this condition is fulfilled (if it were
not, the amount of small-scale structure would continue to
grow, apparently without limit).

One important conclusion from this work is that when
the gravitational-radiation effect switches on and the
full scaling regime is reached, the value of $\gb$ will
drop, as will that of $\ga$, though by a smaller factor.
This means that the final scaling values of these parameters
are likely to be smaller than those found in the
simulations, where gravitational radiation has not yet
been taken into account.

\subsection{Scaling configuration}

In terms of observational consequences, the most
important questions concern the likely scales of
the string structures.

Assuming that scaling has been achieved, the mean
inter-string distance, $\xi$, at the present time is
likely to be a sizable fraction of the horizon distance,
determined by the parameter $\ga$ above.  The simulation
of Bennett and Bouchet, for example, [\ct{BenBou89}]
predicts that in the matter-dominated era, $\ga\ap
2.8\pm 0.7$. As remarked above, it is likely that the
simulations have if anything over-estimated $\ga$, so a
reasonable guess is that $\ga$ is today of order 2.
This would mean that the typical interstring distance now
is perhaps 1500 $h^{-1}$ Mpc; the nearest long string to
us might have a redshift of perhaps 0.3.

The persistence length $\xb$ along
the strings is of similar order of magnitude, but the
scale $\ze$ characteristic of the small-scale structure is
much less, perhaps a few hundred kiloparsecs.

For many purposes, it is not the
strings present today that are important, but those that
were present around the time of decoupling of matter and
radiation.  Decoupling occurs not very long after the
time of equal matter and radiation densities, so $\ga$
was probably still somewhat above its final
matter-era scaling value.  (The string density in the
radiation-dominated era was proportionately much larger
than it is now: a reasonable estimate of $\ga$ in the
radiation-dominated universe, based on the value of 7 from
the simulations, might be around 5.)  Allowing for
subsequent expansion by a factor of $10^3$, the comoving
size of the inter-string distance at decoupling could be
about 25 $h^{-1}$ Mpc, comparable with the scales of the
large-scale structures.

String loops also play a very significant role.
At one time, it was thought that loops might act as seeds
on which galaxies could grow.  However, the relatively
small size of the loops now envisaged means that that is
no longer a viable scenario.  Nevertheless, loops may
still have very important effects.

Typically, the
lengths of the loops produced at any given time are
probably a few times the small-scale length $\ze$, which is
itself of order $\Ga\rms{ls} G\mu t$.  We write
the mean size of a loop born at $t\rms{b}$ as $(\ka-1)\Ga
G\mu t\rms{b}$.  This
parametrization is convenient because the size of
this loop at a later time $t$ is then given by
 $$
l = \Ga G\mu (\ka t\rms{b} - t).
$$
The loop finally disappears when $t=\ka t\rms{b}$.
The parameter $\ka$ is not at present well
determined, but is expected to lie between 2 and 10.   So
any loops we see today were probably born in the fairly
recent past.

The total number
density of loops in the matter dominated era is
[\ct{AusCopKib94}]
 $$
n\rms{loops}(t) = {\nu\over\Ga G\mu t^3},
\label{\eLooNum}
$$
where $\nu$ can be expressed in terms of the various
scaling parameters, and is probably of order 0.1.  (In
the radiation era, the expression is slightly more
complicated, involving also a function of $\ka$.)

The typical separation between loops would then be
 $$
n\rms{loops}^{-1/3} \ap \({\Ga\over100}
{\mu_6\over\nu}\)^{1/3} 90h^{-1}\unit{Mpc},
$$
where as before $\mu_6 = 10^6 G\mu$.  The nearest loop to
us is probably at about half this distance.  Thus the
expected number of loops with redshift less than 0.3,
say, would be a few hundred, so it is quite likely that
some loops could be detectable.  A typical size for a loop
might be hundreds of kiloparsecs; if so by (\rf{\eMuMag}),
their masses would be comparable with those of large
galaxies.

\subsection{Superconducting strings}

In several important ways superconducting strings behave
quite differently.  In particular, as we saw in section
3.4, the string tension, $T$ say, and the energy per unit
length, $\mu$, are no longer constants --- they depend on
the current in the string.  Even more significantly, they
are no longer equal.

The fact that $T\ne\mu$ means of course that the
invariance under local Lorentz transformations along the
direction of the string is broken.  The string can have a
longitudinal momentum; a distinguished rest frame is
defined, in which the longitudinal momentum vanishes.  By
definition, $\mu$ is the energy per unit length, and $T$
the tension, in that frame.

In terms of the effect on the evolution of a string
network the most significant effect is that loops of
string may be dynamically stabilized.  Even nonrotating
loops might be stable; they have been called {\it
springs} [\ct{CopHinTur87}].  However, there is some doubt about
whether this can occur at currents less than the critical
current, beyond which superconductivity disappears.  On
the other hand, when the angular momentum is non-zero,
there are definitely configurations, termed {\it
vortons\/} [\ct{DavShe88a}], which are classically stable,
although susceptible
to decay by quantum tunnelling.  A stable
vorton is a circular loop rotating in its own plane,
a configuration which would not radiate gravitationally.
Therefore, if vortons can be formed, they are likely to be
very long-lived; they will then come to dominate the
energy density of the Universe.

In the case of grand unified strings, where the
characteristic energy scale is large, the result would be
catastrophic: at a relatively early stage in its history
the Universe becomes dominated by stable vorton loops.
This effectively rules out theories in which high-energy
superconducting strings are formed, unless
quantum tunnelling is rapid [\ct{HawHinTur88}].

On the other hand, if the energy scale is much lower, say
the electroweak scale, things are very different. Very
light non-superconducting strings are not of much
cosmological interest; their contribution to the energy
density would be so small as to be negligible.  However,
light superconducting strings would have several
interesting effects.  The vortons they
generate could eventually come to dominate the energy
density, but only at a rather late epoch.  They could
indeed constitute the dark matter [\ct{Car90b}].
Superconducting strings interact strongly with
magnetic fields, which exist both within and between galaxies.
A segment of string moving through a galactic magnetic field
gains an induced current, and could
become a strong source of synchrotron radiation [\ct{Chu+86}].
Very light strings would still be evolving towards scaling, and
so their density could be quite high.  For example, electroweak
scale strings would have a correlation length $\xi \sim
(t_0/t_{\rm ew})^{5/4} \xi(T_{\rm G})$.  Depending on the self-coupling
of the relevant scalar field, this could be as low as 10 kpc.
In fact, it has been argued [\ct{ChuVil88}] that the density of these
lightest strings could be even higher, because of their strong
coupling to the cosmic plasma.  The coupling might even
be so strong that the strings are frozen in to the plasma.
Turbulence in the plasma would then actually stretch them and increase
the total length of string.  In this way the interstellar medium could be
filled with very light ($\mu_6 \sim 10^{-28})$ strings.  Even
the centres of stars could become sites for the generation of string
through turbulence.

A theory of structure formation in the Universe involving
GUT scale super\-con\-ducting strings was proposed some years ago
by Ostriker, Thomson and Witten [\ct{OstThoWit86}].  It required
a primordial magnetic field to induce very large currents
on the strings, which could then move matter around with the
pressure of the resulting low-frequency electromagnetic radiation.
This is essentially a variant of the explosive galaxy formation
scenario [\ct{Explo}], using strings instead of supernovae as a
power source.
However, it is thought that this class of theories would induce
large microwave background fluctuations, and so they have fallen
from favour [\ct{LevFreSpe92}].   Massive superconducting strings
were also proposed
as power sources for quasars [\ct{FieVil87}], although once again
a primordial magnetic field is required to induce sufficiently
large currents.

\subsection{Axion strings}
Axion strings are formed at the Peccei-Quinn phase transition,
which, if it happens at all,  is quite tightly constrained to
take place at around $10^{10-11}$ GeV.  Axion strings are global
strings, which are thought to evolve towards a scaling solution
in much the same way as gauge strings.  The efficiency of
the energy-loss mechanism for global strings is if anything
higher because of the strong coupling between
the strings and the axion field.
Early on in the evolution of the axion string network the axions
are essentially massless, and redshift as radiation.  However,
axions are in fact {\it pseudo}-Goldstone bosons:  they have a
small temperature-dependent mass $\ma(t)$,
supplied by QCD instanton effects,
which ensures that they become non-relativistic
shortly before the QCD phase transition.  The energy
density in axions, $\rho\rms{a}$, then redshifts more slowly,
and may eventually come to dominate the total energy density.
The critical density forms an approximate upper bound on
the total energy density in the Universe, and so from this
an upper bound can be derived on the PQ symmetry-breaking scale $\fa$,
through the dependence of the mass and the density of the axions on
$\fa$ [\ct{Dav86}].

Let us outline the history of the axion field.  The axion string
network is formed at around $t\rms{a} \sim (G\fa^2)^{-1} t\rms{Pl}$,
when the temperature is $\sim \fa$.  As for gauge strings,
the network is thought initially to lose energy mostly by friction,
until $t_* \sim (G\fa^2)^{-1}t\rms{a}$, when the strings can all
move freely and radiate axions.  Thus the energy density in the string
network is gradually transferred into axion radiation.  However,
although the axion is very light, its mass is increasing as the
temperature decreases, while the characteristic radiation frequency
$\xb^{-1}$ decreases as the network coarsens.  Eventually, the
Compton wavelength becomes smaller than $\xb$, and the network
can no longer radiate.  This happens at a time
$\tilde t$ defined by $\ma(\tilde t) = 1/\xb(\tilde t)$.
If $\xb \sim t$, this turns out to be at a temperature of around
1 GeV [\ct{Dav86}].

We recall from section 2.9 that the QCD instanton effects introduce
a potential into the Lagrangian for the axion field $a$, of the
form $\Om(T)[1-\cos(Na/\fa)]$, where $\Om(T) \propto
T^{-8}\ln^6(\pi T/\La)$ [\ct{GroPisYaf81},\ct{Kim87}] and
$\La\sim200$ MeV is the QCD scale.
One can think of the growth of $\Om(T)$ as a
`buckling' of the  circle of minima of the potential of the underlying complex
scalar  field.  At temperatures high compared to the QCD scale,
all points on the circle have equal free energy.  However, as QCD
effects become important, only the points at $a=2\pi\fa/N$ remain
unaffected [\ct{Sik82}].
If $N=1$ there is only one minimum left, at $a=0$,
and we can imagine the potential `tilting'.  Since $a/\fa$ is
identified with the CP-violating $\th$-parameter [\ct{QFT}]
of QCD, we have an explanation (originally due to Peccei and Quinn
[\ct{PecQui77}]) of why that parameter is so small.

The tilt or buckling of the axion potential greatly changes the
evolution of the string network, for each string becomes attached to
$N$ domain walls. The domain walls appear as a result of the energy
density at $a/\fa = \pi/N$, for it is energetically favourable to
concentrate the change in the axion field $a$ in $N$ sheets rather
than having is spread around the string (this is essentially the
same argument as in section 2.5).  If $N=1$, then every string
forms the boundary of a wall, and the walls' surface tension acts
to draw the strings together so that they can annihilate
[\ct{VilEve82},\ct{She90}].  If $N>1$,
then disaster ensues, for annihilation cannot take place.  Thus,
if the Universe was ever as hot as $\fa$,
any viable axion model must have $N=1$ [\ct{LazSha82}].

Following the disappearance of the string/wall system, its energy lives
on in the form of axions, most of which are non-relativistic.
They are very weakly interacting, for their self-coupling is
O($\La^4/\fa^4$), and their couplings to fermions are O($m\rms{f}/\fa$).
Hence they cannot annihilate, and the comoving axion number density
$a^3 \na $ is conserved.  The energy density in axions is therefore
$a^3\ma\na$, which grows relative to the energy density in the rest of
the particle species in the Universe.  This is required to be
not much greater than the critical density today.  Both $\ma$ and
$\na$ depend on $\fa$, and as a result one can derive an upper bound
on $\fa$: $\fa \lap 10^{11} \; {\rm GeV}$
[\ct{Dav86},\ct{DavShe89b},\ct{BatShe94}].

It should be mentioned that Sikivie and coworkers
[\ct{HarSik87},\ct{HagSik91}] arrive at a higher bound, $\fa \lap
10^{12}$ GeV. This is essentially the same as that obtained
if the axion field is homogeneous [\ct{AbbSik83}--\ct{PreWisWil83}],
which would be the case after a period of inflation.
The difference arises through an argument that an axion string
releases its energy very quickly, and without any oscillation at all.
This results in the energy density of the network being transferred
into fewer, higher momentum axions, and hence in a decrease in
the axion energy density at late times.
The careful numerical simulations of Battye and Shellard
[\ct{BatShe94}], however, do not bear out this contention.

\section{Observational consequences}

The concept of cosmic strings, like any scientific
hypothesis, must ultimately be tested by examining its
observational implications.  In this section we shall
examine the various way in which the notion of cosmic
strings might be tested.  We confine our attention to the
case of non-superconducting gauge strings with a large
characteristic energy scale, around $10^{15}$ or $10^{16}$
GeV, so that the dimensionless parameter $G\mu$ is roughly
of order $10^{-6}$ or $10^{-7}$.

\subsection{Gravitational lensing}

The most important interactions of ordinary strings are the
gravitational ones.  The gravitational field of a string
is quite remarkable: around a straight, static string,
the gravitational acceleration vanishes.  This is a
consequence of the exact equality of the energy per unit
length, $\mu$, and the tension, $T$.  In linearized
gravitation theory, tension acts as a negative source of
the gravitational field and, when $T=\mu$, the two effects
exactly cancel.

As we saw in section 4,
the space-time near a string is locally, but not globally, flat.  The space
around the string is cone-shaped, as though a wedge of
angle $\de$ had been removed and the faces stuck
together [\ct{Vil81a}].  The curvature is
entirely localized within the string.  The deficit angle
$\de$ is related to the string tension:
 $$
\de = 8\pi G\mu = 5.18\mu_6\,'',
\label{\eDelDef}
$$
where $\mu_6 =
10^6 G\mu$.

The most immediate observational prediction from the
conical space-time around a string is of gravitational
lensing
[\ct{Vil81a},\ct{Vil84b},\ct{Got85},\ct{Gar85}].  If we
look at a distant quasar, for example, (see figure
6.1) we may see two images, one on either side of
the string.

Suppose the comoving distances to the string and the
galaxy are $\xs$ and $\xg$ respectively.  Let
$\bn$ be the unit vector along the line of sight from us
to the string, and  suppose the string is moving with
transverse peculiar velocity $\bv$.

It is best to transform to the rest frame of the string,
in which both the observer and the galaxy are moving with
velocity $-\bv$.  If we work in terms of the angular
coordinate $\bar\vp$ defined after (\ref{\eConMet}), then light will
propagate along straight lines, but there is a missing
wedge of space of angle $\de$ around the string (see
figure 6.1).  It is then easy to show [\ct{Vil86}]
that the angular separation between the two images of the
galaxy, to first order in $\de$, is
 $$
\ps = {\de\sin\th\over\ga_v(1-\bv\cd\bn)}
{\xg-\xs\over\xg},
\label{\eLenAng}
$$
where of course
$\ga_v=(1-\bv^2)^{-1/2}$.

The probability that a randomly chosen galaxy with
redshift $\zg$ will be lensed by a long string is then
found to be
 $$
\eqalign{
{\rm probability} &= {64\pi\over 3}\ga^2 G\mu
\({\sqrt{1+\zg}+1\over\sqrt{1+\zg}-1}\ln\sqrt{1+\zg}-2\)
\cr
&= {8\pi\over9}\ga^2 G\mu (\zg^2 - \zg^3 + \dots).}
\eqqno
$$
Including the effects of loops is straightforward; this
does not change the result very much, because the total
energy in loops is almost certainly substantially less
than that in long strings in the matter era.

As we have seen, a reasonable value for $\ga$ in the
matter-dominated era is around 2, so for say $\zg=0.4$
and $G\mu=10^{-6}$, only about one galaxy in $10^6$ is
lensed.

There are of course many other cosmological objects that
act as gravitational lenses, such as giant galaxies and
rich clusters (see e.g.~[\ct{GLens}]).  What is special about
strings is the linear nature of the lens.  A long string
stretched across our field of view would be likely to
induce many lensed pairs [\ct{Vil84b}] arranged
in a roughly linear array.  A loop would give
several lensed pairs within a small region of
the sky.
There has in fact
been one report of a small field (less than one
arcminute across) containing seven apparent
lensed pairs [\ct{CowHu87},\ct{Hu90}], but it is unclear
whether all these are in fact genuine double images,
rather than merely accidental pairs of similar galaxies.
Neither is the signal of a line of lensed pairs likely
to be obvious, because the density of string is too low
to make them stand out clearly above the `noise' consisting
of line-of-sight galaxy pairs [\ct{Hin90b}].  The alignment of the
position angles of the pairs could help in filtering out the
noise, but the small-scale structure on the string tends to reduce
the correlations between them.  It remains an open question as to
whether it is possible to search for string this way.

An important feature is that because strings are so
narrow and fast moving, their effects can change rapidly
with time.  Patterns of magnification produced by
gravitational lenses generically exhibit caustics along
which the magnification is formally infinite.  Pairs of
images coalesce and disappear as one crosses a caustic.
(In reality the magnification never becomes actually
infinite because of the finite size of the objects being
imaged.)  Hogan and Narayan [\ct{HogNar84}] pointed out
that in the case of lensing by very small loops, objects
may cross these caustics in a relatively short space of
time, of the order of months or less.  Unfortunately, most
loops are probably not small enough to produce significant
magnification of images.  Nevertheless, pairs of images
may appear or disappear within a time scale comparable
with the light travel time across the source.

Another special feature that, if found, would be a very
clear signal of cosmic strings is the appearance of
sharp edges on images [\ct{Pac86}].  One of the images of
an extended object seen behind a cosmic string might well
exhibit such sharp edges.  Even if the object is too far
away for its size to be resolved, this effect would lead
to a pair of images of different magnitude, and even of
different colour if, for example, only the arms of a
spiral galaxy were lensed.

\subsection{Effect on microwave background}

Possibly the most definitive test of the idea of cosmic
strings with GUT energy scale will come from observations
of the predicted discontinuities in the temperature of
the cosmic microwave background [\ct{KaiSte84}].

Returning to figure 6.1, let us suppose that in the
absence of the string, the light would have an observed
angular frequency $\om$.  The effect of the moving string
is that the light is redshifted in the beam ahead of the
string and blue-shifted behind it.  The difference in
frequency is given by [\ct{Vil86},\ct{Vac86}]
 $$
{\De \om\over\om} = \bn\cd\bt\wedge\bv\de =
v\de\sin\th\sin\al,
\eqqno
$$
where $\bt$ is the unit tangent vector on the string,
and $\al$ is the azimuth angle of $v$.  Clearly there
will be a discontinuity in the observed microwave
background temperature, with $\De T/T = \De\om/\om$.

To compare with observation, one needs to calculate the
expected pattern of temperature fluctuations generated by
the string distribution.  Typically, the observations
compare the temperature in two
directions in the sky separated by a given angle.  To
distinguish cosmic strings from other sources of
perturbation, observations on small angular scales are
particularly important, because it is only on small
scales that the stringy nature of the perturbing source
will be visible.

Fortunately, calculating the temperature fluctuations
on angular scales that are small compared to the horizon
at decoupling is simpler than one might expect,
because it turns out that a Minkowski-space treatment is sufficient
[\ct{Ste88},\ct{Hin94}].  Consider an initially parallel
beam of photons, propagating towards us with unperturbed
momentum $p^\mu = (E,0,0,E)$.  In the presence of a weak
gravitational field the perturbation in momentum is given
to first order by
 $$
\De p_\mu(x) = -{1\over2}\int_{\la_1}^0 \d\la\,
h_{\nu\rh,\mu}(x(\la))p^\nu p^\rh,
\eqqno
$$
where the unperturbed trajectory of the photon arriving
at our position $x^\mu$ is $x^\mu(\la) = x^\mu + \la
p^\mu$.  It is convenient to choose the harmonic gauge
for the perturbation $h_{\mu\nu}$ around the
Minkowski metric, i.e., $h_\mu{}^\nu{}_{,\nu} -
{1\over2}h_\nu{}^\nu{}_{,\mu} = 0$.  Then by
applying a transverse two-dimensional Laplacian
${\bnabla}_\perp^2$ and dropping an unimportant surface
term, one finds a direct relation between $\De p_\mu$ and
the perturbing stress-energy-momentum tensor, namely
 $$
{\bnabla}_\perp^2 \De p_\mu = -8\pi G \pa_\mu
\int_{\la_1}^0 \d\la\,T_{\nu\rh}(x(\la))p^\nu p^\rh.
\eqqno
$$

This formula can readily be applied to the case where
$T_{\mu\nu}$ is the stress tensor of a cosmic string,
(\rf{\eEMTen}).  The calculation is simplified by
using the conformal light-front gauge, in which
 $$
X^+(\ta,\si) \equiv X^0 + X^3 = \ta,
\eqqno
$$
and as usual
 $$
\Xd\cd\Xp = 0, \qquad \Xd^2+\Xp^2 = 0.
\eqqno
$$
Then, using the fact that $\De T/T = \De E/E$ one easily
finds
 $$
{\bnabla}_\perp^2{\De T\over T} = -8\pi G\mu \int \d\si\,
\bXd\cd{\bnabla}_\perp\de^{(2)}
(\bx_\perp-\bX_\perp(x^+,\si)).
\label{\eMHForm}
$$

This result means that, in Minkowski space, the temperature
distortion depends only on the apparent positions of the
strings and their apparent velocities, not on the entire
history of the world sheet.
{}From here one can go on to calculate the power spectrum
$\<|\De_\bk|^2\>$ of the temperature perturbation,
where $\De_\bk$ is the two-dimensional
Fourier transform of $\De T/T$.

This same problem must
be tackled numerically in an expanding Universe, using
the string distributions obtained from numerical
simulations.  Com\-parison with observation yields
information on the magnitude of the string parameter
$G\mu$.  Bouchet \etal [\ct{BouBenSte88}] have
used equation (\rf{\eMHForm}) (expressed in the temporal gauge)
to compute the fluctuations from a set of numerically generated
string configurations.  From this they obtained an upper bound
on $G\mu$ from the small angular scale ($\sim 10'$) OVRO NCP data
[\ct{Rea+88}].  However, it turns out that the temporal gauge
formula was incorrect [\ct{SteVee94}]. A recent reanalysis [\ct{Hin94}]
yields the 95\% confidence limit
$$
\ga G\mu < 6\times10^{-6}.
\eqqno
$$
The bound is expressed on $\ga G\mu$ because of the systematic and
statistical uncertainties in the string density parameter $\ga$, which we
defined in section 5.3.

Deriving bounds on larger angular scales is more problematic, because
one must take the rest of the matter in the Universe into account.
Scales of around $1^\circ$ are particularly complicated: one must
incorporate gravitational potential fluctuations at recombination,
as well as the motions of the coupled electron-photon fluid.  On
larger angular scales one only has to worry about the general
relativistic equations for the gravitational field and the density
field of the dark matter.

Bennett \etal [\ct{BenSteBou92}] argued that the dark matter
perturbations
could  be neglected on large angular scales, and found that the COBE
1st year results [\ct{Smo+92},\ct{Ben+92}] gave
 $$
G\mu = (1.5\pm0.5)\times 10^{-6}.
\label{\eBSBMuCOBE}
 $$
However, this used the old, incorrect, formula, which
underestimates the string-induced CMB perturbations.
Coulson \etal [\ct{Cou+94}], using a Minkowski space approxi\-mation
for the string network, but otherwise incorporating
general relativistic effects in their code, quote
$$
G\mu = (2.0\pm0.5)\times 10^{-6}.
\label{\eCFGTMuCOBE}
$$
They were also able to incorporate the effect of the
electron motion at last scattering, by assuming that
the electrons were subject to the same perturbations as the
dark matter.  This is appropriate for a reionized Universe
in which last scattering was at $z\sim10^2$.

Useful information can be obtained from other statistical
measures besides the power spectrum.  On small scales,
strings are expected to generate distinctly non-Gaussian
perturbations.  In particular, because of the localized
nature of the source, the phases of the various Fourier
components $\De_\bk$ are not independent random variables.
It has been argued [\ct{MoePerBra94}] that a sensitive test
of the cosmic string picture will be to measure the
kurtosis of the distribution of the gradient of $\De T/T$
(the ratio of the fourth moment to the square of the second
moment).  The distribution predicted by strings has higher
peaks and a longer tail than a purely random distribution.

\subsection{Density perturbations}

One of the great unresolved problems of cosmology is the
determination of the origin of the initial density perturbations
from which galaxies eventually condensed.  There are two main contenders at
present: quantum fluctuations
during an early epoch of inflation, and perturbations due
to topological defects of one kind or another, such as
strings or textures.

The cosmic string scenario in its original form placed
the main emphasis on the role of loops.  It was supposed
that loops would be formed with a characteristic size
equal to a substantial fraction of the horizon size.
Such large loops would live for a long time, so that
there would always be a large population of loops
present.  The idea was that these loops would form seeds
on which galaxies would start to condense.  Much of the
early excitement generated by the idea of cosmic strings
was due to the apparent success of the scenario in predicting
 the correlation function of Abell clusters
[\ct{Tur85}].

More recent results from the simulations have shown that
loops are born with much smaller size and therefore live
a much shorter time than was originally supposed.  It now
seems that they are too small and do not survive long
enough to perform the function of seeding condensations.
Strings may still have a very important role
to play in generating density perturbations, but the centre
of attention has shifted from the loops to the long
strings.  This is not to say that the loops are
unimportant; although they probably represent a smaller
fraction of the total density of the Universe, they are
far more uniformly distributed than the long strings, so
their effects may well be significant.

The first indication of the role of long strings in
creating density perturbations was due to Silk and
Vilenkin [\ct{SilVil84}], who pointed out that moving
strings will leave behind them an over-dense wake.  This
is another consequence of the conical structure of the
space around a cosmic string.  Consider a straight string
moving through the matter in the Universe with transverse
velocity $\bv$.  In the rest-frame of the string, other
matter is streaming past it with velocity $-\bv$.
Let us suppose the matter is collision-free, so that it
will follow geodesics around the string.  In terms of the
coordinate $\bar\vp$, these are straight lines, so in
terms of the original coordinate $\vp$, every particle is
deviated inwards as it passes the string by an angle
$\de/2 = 4\pi G\mu$.  In other words, every particle
acquires an inward velocity component, of magnitude
$v\ga_v\de/2$.  Note that this is independent of
the impact parameter.  The result is (see figure
6.2) that in the region behind the string matter
from both sides converges, leaving an over-dense wake.

In fact, however, as we have seen, the strings are by no
means straight.  There is a great deal of small-scale
structure, so that viewed from a large scale the
effective energy per unit length, $\mu$, is increased,
while the effective tension $T$ is decreased by the same
factor.  This means that it is no longer true that
space-time around a string is locally flat.  In linear
gravitation theory, there is no longer a cancellation
between the effects of $T^{00}$ and $T^{33}$.  A kinky
string, therefore, acts as an ordinary gravitational
attractor (equation \rf{\eAttFor}),
so that even a string at rest will generate a
density perturbation.  In fact, a particle moving past a
kinky string will be deviated from its path by two separate
effects, the conical defect angle and the normal
gravitational attraction.  The resulting velocity
perturbation is given by [\ct{VacVil90}]
 $$
\De v = 4\pi G\mu v\ga_v + {2\pi G(\mu-T)\over v\ga_v}.
\label{\eVelPer}
$$

Clearly, the first of these two terms is more important
for fast-moving strings, while the second is relevant for
slower-moving ones.  Fast-moving string will generate
planar structures in the surrounding matter, whereas with
slower strings the shape will be more nearly linear.  We
might expect to see a combination of these planar and
linear features in the Universe, but there is an
important question, to which we shall return shortly,
about the extent to which such `stringy' features will
be visible above the general stochastic background.

The simulations suggest [\ct{SheAll90}] that the r.m.s.\
velocity of strings on a small scale is large (about
$0.6c$), but that the coherent velocity of long, roughly
straight, sections of string is much less, no more than
$0.1$ or $0.2c$.  For this reason, the second term in
(\rf{\eVelPer}) is usually more important than the
first.  On the whole, linear features should predominate
over planar ones.

Quantitative predictions are hard to come by in the cosmic string
scenario.  In particular, we would like to know the power spectrum
of the density perturbations in the dominant matter component,
$$
P(\bk) = |\de_\bk|^2,
\eqqno
$$
where $\de_\bk = (8\pi^3V)^{-1}\int\d^3\bk\,\de(\bx)e^{i\bk\cdot\bx}$,
the (comoving) Fourier transform of the fractional density
perturbation $\de(\bx) \equiv \de\rho(\bx)/\rho$.

A comprehensive account of cosmological perturbations in the presence of
strings and other `stiff' sources has been given by Veeraraghavan
and Stebbins [\ct{VeeSte90}], who derived the linearized
Einstein equations in a flat FRW model for the perturbations in
the metric and in the matter, given a string energy-momentum tensor
$T^{\rm s}_{\mu\nu}(\eta,\bx)$. A flavour of their analysis can be
gained from considering a universe consisting of nothing but
cold dark matter and strings.  Then the density perturbation
obeys the equation
$$
\ddot \de + {\dot a \over a} \dot \de - {3\over 2}\({\dot a \over a}\)^2 \de
= 4\pi G (T^{\rm s}_{00}+T^{\rm s}_{ii}),
\label{\eCDMdRho}
$$
in the synchronous gauge.  This is a coordinate choice, which is
defined by setting the time-time and the time-space components of
the metric perturbation to zero [\ct{Wei72},\ct{KolTur90}].

Equation (\rf{\eCDMdRho}) is an inhomogeneous second order
ordinary differential equation, subject to boundary conditions
on $\de$ and $\dot\de$ at some initial time $\et_i$.
It is thus solvable by a Green's function method.  One finds
that the solution splits into two parts: $\de^I(\et)$, which depends
on the boundary conditions, and $\de^S(\et)$, which depends only
on $T^{\rm s}_+ = T^{\rm s}_{00}+T^{\rm s}_{ii}$.
Veeraraghavan and Stebbins termed these the `initial compensation' and the
`subsequent compensation'.  The combined energy-momentum of
the matter and the strings cannot be affected
on scales greater than the horizon: the large-scale ($k\et \ll 1$) density
perturbations in the strings are {\it compensated\/} by perturbations
in the matter.
On these large scales the compensation obeys a conservation law
[\ct{VeeSte90}]
$$
8\pi G(a^2\rho\,\de + T^{\rm s}_{00}) + 2\({\dot a \over a}\)\dot\de = 0.
\eqqno
$$
The initial compensation apparently
contributes a white noise term proportional to
$k^0$ to the power spectrum, since energy density has been taken
from the matter in order to create the string network.
This is partly cancelled by the subsequent compensation, because
the strings preferentially seed overdensities where the underdensities
are.  This cancellation fails when the strings move away from
their original locations.  Since the strings move relativistically,
inhomogeneities are generated on a scale
$k \sim \eta^{-1}$.  The net effect is to produce density
perturbations whose behaviour at late times
($\eta \gg k^{-1}$) is
$$
P(\bk) \sim  k.
\eqqno
$$
This is the Harrison--Zel'dovich form of the power spectrum,
which is scale invariant: that is, the r.m.s.\ mass fluctuation
in spheres the size of the horizon is constant [\ct{KolTur90}].
The power spectrum
of galaxy number density fluctuations is consistent with an
approximately scale-invariant spectrum,
so this is
a prerequisite for any successful theory of structure formation. The
scale-invariance of the spectrum is in fact a direct consequence
of the scaling behaviour of the network.  If the total length of string
per horizon volume is proportional to the horizon length, then
the string density goes as $\mu t^{-2}$, which, as we saw in section 5.3, is
a constant fraction of the total density,  $1/(6\pi G t^2)$
in the matter era.  The amplitude of the density perturbations is
proportional to the amount of string, and so $\de \sim G\mu$.

The full cold dark matter
(CDM) power spectrum has been
studied by Albrecht and Stebbins [\ct{AlbSte92a}], using
the AT simulations, who also incorporated theoretical
uncertainties about network evolution by modelling strings
with different amounts of
small-scale structure.  The resulting power spectra for one of
their models are
displayed in figure 6.3, along with the same
authors' calculation  of the
string-seeded hot dark matter (HDM) spectrum [\ct{AlbSte92b}].
For comparison,
the standard inflationary spectra are also displayed
[\ct{LytLid93}].  All are conventionally normalized so that
$\si_8$,
the variance of the mass fluctuations in spheres of
$8h^{-1}$ Mpc, is 1. We shall say more about this below.

Albrecht and Stebbins also looked for
specifically `stringy', planar or linear, features.  They
concluded that they
would not show up clearly in CDM; they would be swamped by the
general background of perturbations generated by strings
on smaller scales.  The situation is very different, however,
in the case of an HDM Universe
[\ct{AlbSte92b}].  There they found that stringy features
would indeed stand out rather clearly, essentially due to
the `free streaming' of the neutrinos on scales less than
$20h^{-1}$ Mpc.

The parameter $\si_8$ is a common measure of
the density fluc\-tu\-ations on galactic scales.
This is the
scale on which the density perturbations are just now going
non-linear; that is to say, the observed value $\sg$ based
on measurements of the galaxy correlation function is
unity.  However, this is not a direct measurement of $\si_8$
{\it per se}.  It seems likely that galaxy formation will
occur preferentially near the highest peaks of the density
distribution, in which case the galaxy distribution will be
more correlated that the mass distribution.  In other
words, the observed fluctuation in the galaxy distribution
is expected to be larger than that in the (mainly dark
matter) mass distribution.  It is usual to define a bias
parameter $b$ by
 $$
\sg = b\si_8.
\eqqno
 $$
Using the COBE normalization to calculate $\si_8$ then
gives an estimate of the required bias, namely
 $$
b = \si_8^{-1}.
\eqqno
 $$

The value of the dimensionless parameter
$G\mu$ required to yield the right magnitude of the
fluctuations depends on the assumptions made about
string evolution.  But assuming reasonable values for
the parameters, Albrecht and Stebbins found in the cold
dark matter case,
 $$
\mu_6 \approx 1.8 \;(h=1), \qquad
\mu_6 \approx 2.8\; (h=0.5),
\eqqno
$$
where $\mu_6 = 10^6 G\mu$, in
reasonable agreement with the COBE normalizations quoted above
(\rf{\eBSBMuCOBE},\rf{\eCFGTMuCOBE}).
(These figures come from one of three
possible scenarios analysed by Albrecht and Stebbins.)  In
the case of hot dark matter, the corresponding values were
 $$
\mu_6 \approx 2.0 \;(h=1), \qquad
\mu_6 \approx 4.0\; (h=0.5).
\eqqno
$$
Thus at $h=1$ the bias parameter at $8h^{-1}$ Mpc is approximately
1 for both CDM and HDM, but at $h=0.5$ CDM requires a bias in
the range 1--2, while HDM needs 2--4.
However, these figures should be treated with caution, for
there is still
considerable uncertainty about the correct basis of the
calculations of the power spectrum.

\subsection{Gravitational waves}

As we have seen, the primary mechanism for energy loss
from loops is gravitational radiation.  Long strings
also radiate significantly.  Although gravitational
waves cannot as yet be detected directly, this radiation
may have measurable indirect effects.  Consideration of
the known limits on the intensity of gravitational
radiation in the Universe has yielded an important bound
on the string parameter $G\mu$.

There are in fact two distinct ways of bounding $G\mu$ by
considering the emitted gravitational radiation --- from
Big Bang nucleosynthesis and from studies of the
millisecond pulsar.

The primordial abundance of helium in the Universe,
observed to be around 23\%, is a key datum in the Big Bang
nucleosynthesis scenario.  The amount of helium produced
depends on a balance between the rates of the various
processes involved and the rate of the Hubble expansion,
which in turn depends on the energy density.  If for
example there were one extra species of neutrino, that
would increase the rate of expansion and so increase the
helium abundance, to an unacceptably high value --- this
is the cosmological evidence for the limit on the number
of neutrino species [\ct{Oli+90}].

Energy in the form of gravitational waves would have
exactly the same effect.  The validity of the
nucleosynthesis scenario thus places a limit on the total
energy density of gravitational radiation present in the
Universe at the time of nucleosythesis.  This may be
expressed as a fraction of the critical density
[\ct{CalAll92}],
 $$
\Om\rms{gr,nuc} < 0.05.
\eqqno
$$
More recent estimates
[\ct{OliSte94},\ct{KerKra94}] give
 $$
\Om\rms{gr,nuc} < 0.016\ {\rm or}\ 0.007,
\label{\eNucBd}
$$
respectively.

If the gravitational radiation is generated by cosmic
strings, this translates into a limit on $G\mu$.  Bennett
and Bouchet [\ct{BenBou91}] obtained the
limit $ G\mu < 6\times 10^{-6}, $
or a weaker bound if the QCD transition is
inhomogeneous.  Caldwell and Allen [\ct{CalAll92}] have
done a careful numerical study and shown that the limit
may be reduced if the typical loops size parameter is
large (in our notation, if $\ka \gg 1$).  For $\ka>10$, say,
the bound could be significant.
On the other hand, the recent conclusions about the
expected scaling configuration [\ct{AusCopKib93}], in
particular the fact that the parameters $\ga$ and $\gb$
are likely to be smaller in the final scaling regime than
the simulations would suggest, imply some weakening of
the constraint.

It is easy to find an approximate expression for the
gravitational wave energy density.   Let us consider first
the contribution of loops. We saw in section 4.2 that each
individual loop contributes the same power, $\Ga G\mu^2$,
throughout its lifetime.  Thus what we have to calculate
is the total number of loops present at any given time.

During the radiation-dominated era, energy in
the form of gravitational radiation scales in exactly the
same way as the total density (except for a small
numerical factor due to the change in the total number
of massless particle states).  Thus the total
gravitational-wave energy gradually builds up throughout
this era, giving a logarithmic time-dependence.  The power
radiated depends on the total number density of loops,
given by (\rf{\eLooNum}).  The total energy emitted into
graviational radiation per unit volume between times $t_1$
and $t_1+\d t_1$ is then $\Ga G\mu^2 n\rms{loops}(t_1)\d
t_1$.  The gravitational radiation emitted at time $t_1$
is then redshifted by the expansion of the Universe, so
that its density at a later time $t$  is reduced by the
factor $(t_1/t)^2$.  Integrating over $t_1$ yields
 $$
\Om\rms{gr,nuc} \approx {64\pi\over3}G\mu\nu A(\ka)
\(g\rms{*,nuc}\over g\rms{*,form}\)^{1/3}
\ln{a\rms{nuc}\over a\rms{form}},
\eqqno
$$
where
 $$
A(\ka) = {2\over3} {\ka^{3/2}-1\over \ka-1};
\eqqno
$$
$a\rms{nuc}$ and $a\rms{form}$ are the
values of the scale factor at nucleosynthesis and at the
epoch of formation of cosmic strings, respectively; while
$g\rms{*,nuc}$, $g\rms{*,form}$ are the corresponding
values of the effective number of spin states of
relativistic particles.   Since
$\ln(a\rms{nuc}/a\rms{form})\approx 40$,  and
$g\rms{*,nuc}/g\rms{*,form}\approx 0.1$, this yields the
bound
 $$
G\mu\nu A(\ka) < 4\times10^{-5},
\eqqno
$$
or, using the bounds (\rf{\eNucBd}),
 $$
G\mu\nu A(\ka) < 1.4\times10^{-5}\ {\rm or}\ 6\times10^{-6},
\eqqno
$$
respectively.  In general, even the latter bound is not
very restrictive, though if $\nu$ and $\ka$ are both fairly
large, say $\nu\ap2$ and $\ka\ap10$, it becomes significant.
It may become much tighter in the near future as further
results on nucleosynthesis become available.

In principle, we should also include the
gravitational waves emitted by the long strings.
However, in the radiation era, the loops emit
proportionately much more gravitational radiation, except
perhaps when $\ka$ is small, in which case the bound is in
any case very weak.  At the very most, this could not
change the bound by more than a factor of two.

The second bound imposed by the consideration of
gravitational waves
[\ct{HogRee84},\ct{BenBou90}, \ct{BenBou91},\ct{CalAll92}]
derives from the astonishing precision of
timing measurements of the milli\-second pulsars, believed
to be extremely rapidly spinning neutron stars.
The first of these pulsars has now been
followed for over 10 years [\ct{Sti+90}].
The rate of emission of pulses
is extremely regular, though gradually slowing.  Now, any
gravitational waves propagating in the space between us
and the pulsar would cause fluctuations in the timing of
the observed pulses, called timing residuals.

The experiment is complicated by the fact that there are
other sources of fluctuations, such as intrinsic noise in
the pulsar, interstellar propagation effects, and even
instabilities in the clock used to time the pulsar.  However,
the observed residuals do place an important upper bound
on the
total amount of gravitational radiation that can be
present in the Universe today.

Unlike the previous case, this is not a limit on the
total gravitational-wave energy density, but rather a
limit on the waves within a particular wavelength
interval near $\la\approx T$ where $T$ is the period of
observation.  Let the energy density in gravitational
waves with wavelength between $\la$ and $\la+\d\la$ be
$\rh\rms{gr}(\la) \d\la/\la$.  We also write, as usual,
$\Om\rms{gr}(\la) = \rh\rms{gr}(\la)/\rh\rms{c}$.
Gravitational waves of cosmological origin are expected to
have a scale invariant spectrum, that is constant $\Om\rms{gr}$.
In this case the limit is
[\ct{Sti+90}]
 $$
\Om\rms{gr}(7.1\ {\rm yrs}) < 4\times 10^{-7} h^{-2}.
\label{\eMSPBdOne}
$$
More recent unpublished data quoted in reference [\ct{CalAll92}] give
 $$
\Om\rms{gr}(8.2\ {\rm yrs}) < 1\times 10^{-7} h^{-2}.
\label{\eMSPBdTwo}
$$

With reasonable assumptions about the spectrum, this yields
a much lower bound on the total energy density in gravitational
radiation than the one at the time of nucleosynthesis.  The
reason is that once we enter the era of matter
domination, the energy density of gravitational radiation,
like that of electromagnetic radiation, rapidly becomes a
small fraction of the total.  Since the epoch, $t\rms{eq}$,
of equal radiation and matter densities, the
contribution to $\Om\rms{gr}$ of radiation generated
during the radiation-dominated era has been decreasing
like $a^{-1}$, giving an overall reduction by a factor
of $7.5\times 10^{-5} h^{-2}$.

To calculate the expected power in the relevant
wavelength interval, one needs to know how loops
radiate.  A small loop of length $L$ oscillates
quasi-periodically, with period $L/2$.  It therefore
radiates gravitational waves at the fundamental frequency
$2/L$ and at all the harmonics $2n/L$.  In the early
studies it was assumed that essentially all the power is
radiated at the fundamental.  But studies by Vachaspati
and Vilenkin [\ct{VacVil85},\ct{AllShe92}] showed that in
fact a lot of power goes into high harmonics; the power
$P_n$ dies off with $n$ like $n^{-q}$ where $q\ap 4/3$.
This implies that the dominant contribution to the $\Og(T)$
comes from radiation emitted in the fairly recent past
and, for $q<2$, at rather high values of $n$.

{}From (\rf{\eMSPBdOne}), Bennett and Bouchet
[\ct{BenBou91}], assuming that all the power was radiated
in the lowest mode, derived the bound $ G\mu < 4\times10^{-5}$.
Caldwell and Allen [\ct{CalAll92}] found that allowing
for the dependence of $P_n$ on $n$
has a large effect.  For $h=1$, they derived a much tighter
bound $G\mu < 2\times 10^{-6}$, or a lower figure
for large values of $\ka$.
Using the more recent value (\rf{\eMSPBdTwo}), they
obtained an even lower bound,
 $$
G\mu < 7\times 10^{-7}.\qquad (h=1)
\eqqno
$$
However, for $h=0.5$ their limits are less stringent by
a factor of 10.

As in the case of the previous bound, however, the recent
reassessment of the scaling parameters [\ct{AusCopKib93}]
will significantly weaken the bound.
For $q=4/3$, the revised limit is
 $$
\mu_6 \lap 0.2 \(0.1\over \nu\)^{3/2} \(\Ga\over100\)^{1/2}
\(10\over \ka\)\(\ka-1\over\ka\)^{1/2} {1\over h^{7/2}}.
\eqqno
$$
If $h\approx 0.5$, this is quite easy to satisfy, but for
$h\approx 1$ it could become a significant bound.

All these limits are derived assuming a scaling solution
back to the earliest time that gravitational waves with
period 1--10 yr today could have been emitted.  This would
have been between redshifts of approximately $10^{4-5} h^{-2}$.
In models where strings form at the end of inflation,
such as those discussed in section 5.2, it is
conceivable that the strings re-enter the
horizon at around that time, or even afterwards, and still
have a small enough correlation length to seed the galaxy-scale perturbations
[\ct{Yok89}].  This means that the amount of gravitational radiation at these
periods is reduced, and so the pulsar bound is relaxed or
even eliminated.

\subsection{Cosmic rays}

Another of the outstanding mysteries of astrophysics is
the origin of the most energetic cosmic rays, with
energies in excess of $10^{20}$ eV.  The known mechanisms
for accelerating particles, for example by shock-wave
acceleration in active galactic nuclei, do not seem
capable of generating such large energies [\ct{QueNai92}].

Some years ago it was suggested [\ct{Bha89}] that
`evaporation' of particles from cusps on cosmic strings
might provide the explanation.  As mentioned in
section 3.2, cusps occur where $\bXp$ vanishes
instantaneously and, correspondingly, $\bXd^2 = 1$; the
string momentarily reaches the speed of light.  Near a
cusp, there are two overlapping, oppositely oriented
segments.  It seems likely that the interactions of the
underlying fields would convert some of the energy of the
string into quanta of these gauge or scalar fields.  Such
particles are of course unstable and would eventually
decay into quarks and leptons.  The quarks in turn
undergo the usual QCD fragmentation process, ending up
as jets of hadrons, mostly pions with a few nucleons.
Since the decaying particles have masses in excess of
$10^{24}$ eV, and are moving with relativistic speeds, some
of the decay products may easily have energies well in
excess of $10^{20}$ eV.

The processes involved are extremely complex and it is
difficult to calculate the expected high-energy
particle flux with any precision, but estimates
[\ct{Bha89},\ct{BhaRan90}] suggest that the
mechanism could at least be a signficant contributor to
the cosmic ray spectrum above $10^{19}$ eV, though this has
recently been questioned [\ct{GilKib94}].  (It has also been
suggested [\ct{BabPacSpe87}] that superconducting strings could
be responsible for the puzzling gamma-ray bursters.)

The key
question is: what is the fraction $f\rms{cr}$ of the
initial energy of a cosmic string loop that is converted
in this way into high-energy particles?  Comparison with
the observed limits yields an upper bound,
 $$
G\mu\nu f\rms{cr} \lap 1.7\times 10^{-9}.
\eqqno
$$
If cosmic strings are the explanation, then this should be
an equality.  Recalling that for GUT strings, $G\mu\sim
10^{-6}$ and $\nu\sim 0.1$, this means $f\rms{cr}\ap
10^{-2}$.

This seems a rather large fraction.  Firstly, it obviously
requires that the actual energy emitted in the form of energetic
particles by each cusp should scale in the same way as the total
length of string.  For a string of fixed width, this is not what
one would expect [\ct{GilKib94}].  It also requires cusp
evaporation to be a repeated event.
Ignoring energy loss mechanisms, the
motion of a small loop is periodic and any cusp will
recur once every period.  It is not clear, however, what
effect the energy loss by evaporation will have on the
cusp.  If it causes it to disappear permanently, so that
the loop evaporation process occurs only once or a few
times, then it is hardly conceivable that $f\rms{cr}$
could be as large as $10^{-2}$.  On the other hand, it is
possible that the effect is only to shift the position of
the cusp, in which case each oscillation produces a new
cusp; the cusp gradually eats into the string.  Even so,
a fraction $10^{-2}$ is hard to attain.

\section{Conclusions}

In this final section, we aim to assess the current state
of the theory of cosmic strings.

Cosmic strings come in many different guises.  As
we discussed in section 2, they are a feature of
many unified theories of fundamental forces.  They may
appear at a thermal phase transition with critical temperature $\Tc$
anywhere between the grand unfication scale, a few orders of
magnitude below the Planck mass, and the electroweak
transition at a few hundred GeV.  They may even appear in the
late stages of an inflationary epoch. The tension $T$ and the
energy per unit length $\mu$ are both of order $\Tc^2$, and
in simple cases $T=\mu$.  Thus the possible values of $\mu$
range over nearly thirty orders of magnitude.  As yet,
there is no firm indication that cosmic strings of any type
do actually exist, but equally they are far from being
ruled out and in some contexts they provide a very natural
explanation for the observations.

\subsection{GUT-scale strings}

The cosmic strings which have attracted the most attention
are those produced at a grand-unification transition, for
which the dimensionless parameter $G\mu$ is of order
$10^{-6}$ (i.e., $\mu_6\sim1$).  Such strings naturally
generate density perturbations and fluctuations in the
microwave background temperature of roughly the observed
order of magnitude.  As we emphasized in section 1, this is
one of the great attractions of the cosmic-string scenario.
In the most popular alternative theory, based on inflation,
the density perturbations are generated by quantum
fluctuations during the inflationary era.  In that case, to
achieve the right order of magnitude requires fine tuning of
a coupling constant.  The simplest inflationary models
with a single scalar inflaton field require that the
self-coupling $\la$ of the inflaton satisfy typically
$\la\le 10^{-10}$ [\ct{Sta82}--\ct{GutPi82}].
It is possible to avoid this requirement in extended
theories involving two scalar fields
[\ct{LaSte89}--\ct{AdaFre91}],
but even then there is a fine-tuning requirement for the
coupling-constant ratios [\ct{AdaFreGut91}].

The cosmic-string scenario is thus a very attractive one.
The question is: does it stand up to more detailed
scrutiny?  Can it provide a quantitative as well as a
qualitative understanding of structure formation in the
Universe?  How do its predictions compare with those of
alternative models, particularly the inflationary model?

In their simplest forms, both the inflationary and
cosmic-string models have essentially a single adjustable
parameter, the inflaton coupling $\la$ or $G\mu$,
respectively.  One important test is whether it is possible
to fit both the COBE results [\ct{Smo+92},\ct{Wri+92}] on the
fluctuations in the microwave background and the density
fluctuations on galactic scales as determined by
observations of the galaxy distribution.

Both models predict essentially the Harrison--Zel'dovich
spectrum of pertur\-bations: the power spectrum $P(k)$ of
fluctuations behaves as $P(k)\pt k$ for small $k$.  This is
a consequence of the scale-invariance of the physics
responsible for generation of the perturbations.  In the
inflationary model, the magnitude $\de\rh/\rh$ of the
density perturbations on each scale is the same at the
epoch when the appropriate wavelength is equal to the
horizon distance.  Similarly, in the cosmic string
scenario, the magnitude of the perturbations generated on
each scale, once it has come inside the horizon, is the
same.  Once inside, the behaviour of peturbations depends on
how the Universe is expanding.  Before $t\rms{eq}$, the time of
equal densities of matter and radiation, their amplitude
remains constant.  After $t\rms{eq}$, the perturbations grow in
proportion to the scale factor.  This introduces a
 characteristic scale in the problem, namely the
comoving size of the universe at the time $t\rms{eq}$.
This corresponds to a
wavelength $\la\sim 40 h^{-1}$ Mpc.  On scales less than this,
where the wavelength was equal to the horizon distance
at some time before matter-radiation equality, the perturbations cannot
grow.  Thus for larger values of $k$,
the power spectrum $P(k)$ starts to bend, eventually reaching
$P(k)\pt k^{-3}$.  This happens in both the inflationary
and cosmic-string models, but for cosmic strings the
characteristic scale is related to the scale size of the
string network at $t\rms{eq}$, rather than the horizon
distance, and is thus somewhat smaller.  The curvature of
$P(k)$ does not then become significant until we reach
lower scales.  For
a given value of $\si_8$, the perturbation on very large
scales is a little smaller, while there is quite a lot
more power on small scales, especially with HDM.

It is becoming common to normalize the perturbation spectrum not
by $\si_8$, but by fitting the COBE r.m.s.~at angular scales of
$10^\circ$.
In principle this fixes the inflaton coupling or, in the cosmic
string case, the magnitude of $G\mu$.  The calculations
still have rather large errors, but the range of values
predicted is generally from 1 to $3\times 10^{-6}$.  For
example, [\ct{Per93}] gives a value
 $$
G\mu = (1.7 \pm 0.7)\times 10^{-6},
\eqqno
 $$
while [\ct{BenSteBou92}] give
 $$
G\mu = (1.5 \pm 0.5)\times 10^{-6};
\eqqno
 $$
 The most recent and thorough calculations, albeit based on
a Minkowksi space string simulation
[\ct{Cou+94}], give
$$G\mu = (2.0\pm0.5)\times 10^{-6}.
\eqqno
$$
On the large scales probed by COBE, the main contribution
to the microwave background perturbations, both from strings
and inflation, comes from the Sachs-Wolfe effect.  In this case,
there are simple relations between the temperature fluctuation
$\de T/T$ and the perturbation in the Newtonian potential $\Phi$,
or equivalently the density perturbation, on the same scale.
In the inflationary scenario, the density perturbations are
generated at very early times and are
of `adiabatic' type (so called because there is no perturbation
in the entropy per baryon).  This means that there is a fixed
relation between the density perturbations in radiation and matter,
namely $(\de\rho/\rho)\rms{rad} = \frac{4}{3}(\de\rho/\rho)\rms{mat}$.
Thus, once the matter component comes to dominate the energy
density, we have
$$
{\de T\over T} = {1\over 3}{\de \rho \over \rho}.
\eqqno
$$
For strings, the perturbations are generated while the CMB
radiation is travelling through them, and the relation is
[\ct{PenSpeTur94}]
$$
{\de T\over T} = {1\over\ga}{\de \rho \over \rho},
\eqqno
$$
where, we recall, $\ga$ is the ratio between the Hubble radius
and the scale size of the string
network.  Thus, if the scale size were the horizon distance,
strings (and other defects)
could give as much as 6 times the temperature fluctuation
for a given amplitude of $\de\rho/\rho$.  However, for strings
$\ga$ seems to be between 1.5 and 2.5, so the relation
between the temperature and density fluctuations is much the
same for strings and for inflation.

This rough equivalence shows up in the calculations of the bias
for the two scenarios.  We recall that the bias $b$ was defined
as the ratio between the r.m.s.\ fluctuations in the number of
galaxies in $8h^{-1}$ Mpc spheres, or $\sg$, and the mass
fluctuations $\si_8$ in the same volume.
The COBE normalized inflationary cold dark matter (CDM)
theory predicts
 $$
b = (0.5\pm0.1)h^{-1}.
\eqqno
 $$
For a Hubble constant of 50 km s$^{-1}$ Mpc$^{-1}$
($h=0.5$), this is approximately 1.
The values predicted by the cosmic string scenario are
slightly larger, but also subject to relatively large
errors because of the shakiness of the calculations.
For the case $h=0.5$, [\ct{Per93}] quotes,
 $$
\eqalign{
b &= 1.1\;{\rm to}\;2.9 \qquad {\rm (CDM)},\cr
b &= 1.4\;{\rm to}\;4 \qquad\;\;\; {\rm (HDM)}.
}
\eqqno
 $$

More stringent tests are set by studying CMB perturbations on
a range of angular scales. Although inflation and cosmic strings
both predict an
approximately scale-invariant spectrum of density perturbations,
their CMB spectra are different.  For example, the standard
inflationary CDM model predicts a flat spectrum of CMB perturbations
on scales above about $2^\circ$, whereas the string spectrum is
expected to fall slightly, corresponding to a tilted
model with power spectrum $P(k) \sim k^{1.4}$
[\ct{BenSteBou92},\ct{Per93}].  The quadrupole term is expected to be
very much lower than the r.m.s.\ prediction from inflation, as the
density fluctuation responsible for it has not yet been fully
generated.  The COBE data is unfortunately not good enough to
distinguish between $n=1$ and $n=1.4$.  The measured quadrupole
[\ct{Ben+94}], at $6\pm 3 \;\mu K$
is however rather lower than can be comfortably accounted for
by cosmic variance around the prediction from an inflationary
Harrison--Zel'dovich
spectrum normalized at $10^\circ$, $Q\rms{rms-PS} = 19.9\pm1.6
\;\mu K$ [\ct{Gor+94}].  However, the measurement of the quadrupole
is subject to severe contamination by galactic emission and so this
discrepancy may not be significant.

Measurements have been reported of $\de T/T$ on the
1$^\circ$ angular scale.  The most accurate of these is
the MAX experiment, who report [\ct{MAX93}]
 $$
{\de T\ov T}\Big|_{1^\circ} \simeq (3.6\pm0.2)\times 10^{-5}.
\eqqno
 $$
This is in good accord with COBE normalized inflationary CDM,
with the standard HZ scale-invariant power spectrum.  The
only calculation on these scales for the string scenario
[\ct{Cou+94}] assumed that the
universe was reionized some time after $t\rms{d}$, for
example by an early generation of very massive stars.
This has the effect of smoothing the perturbations
on 1$^\circ$ scales, so direct comparison between strings and
inflation over this result is not yet possible.

In terms of distinguishing between inflation and cosmic
strings, it is currently more useful to look at the spectrum of
density perturbations.  If both models are normalized to
the same value of $\si_8$, the cosmic-string model will
predict substantially more power on small scales and less
on large scales.  In the case of a CDM universe, the power
on large scales predicted by inflation is already somewhat
too low, particularly for matching the observations of
large-scale structure [\ct{DavEfsFreWhi92}].  Clearly the
cosmic-string model is worse.  Similarly, because of the
excess power on small scales, it would predict random
small-scale velocities larger than those observed.  It
seems clear that the combination of cosmic strings and CDM
is not a viable model.  The standard inflationary CDM model
is itself in some difficulty, which has led to various
proposed alternatives, such as mixed hot and cold dark matter
[\ct{vDaSch92}--\ct{TayRow92}], or more
complex inflationary models with tilted spectra
[\ct{CenGneKofOst92}]; it is in any case certain that cosmic strings
and CDM do not offer an improvement here.

The situation is very different however in the case of hot
dark matter (HDM), where the addition of cosmic strings
definitely does improve the agreement with observation
[\ct{AlbSte92b}].  Cosmic strings and HDM seem to provide a
promising alternative to inflation and CDM, although there
are potential problems with the high bias required and
a deficit of very small scale power. Unfortunately the
uncertainties involved in the
calculations are still too large to allow definitive
predictions, primarily because of the difficulty of tying
down the parameters of an evolving string network.

\subsection{Lighter strings}

Once we get below the grand unification scale, it is really
misleading to talk of {\it the\/} cosmic string scenario:
there are many different possible scenarios depending on
the type of string --- global or local, superconducting or
not, Abelian or non-Abelian --- and on the
symmetry-breaking scale.

One of the most attractive ideas, in the sense that it does
least violence to the Standard Model, is the axion string
scenario.  This is rather tightly constrained, in that
the relevant symmetry, the Peccei-Quinn global U(1) symmetry,
has to be broken at a scale in the range $10^{10}$--$10^{11}$
GeV.  The strings would have all annihilated by today, having
been joined by domain walls at the QCD phase transition,
but they would have left an important relic in the form of a
population of non-relativistic axions, possibly with critical
density.  Thus the axion strings could have generated the dark
matter.  Experiments to detect dark matter axions are in progress
[\ct{Sik91}].  However, the difficulty of detecting such weakly
interacting particles suggestions that confirmation or otherwise
will be some time coming.

Another set of models are those involving
relatively light, super\-con\-duc\-ting strings, formed
at a transition not far above the electroweak scale.  Peter
[\ct{Pet94}] has emphasized that superconductivity is a
rather generic feature of cosmic string models.  Such
strings are probably not relevant to large-scale structure
formation, but may have significant astrophysical effects
[\ct{Chu+86},\ct{ChuVil88}].  Massive cosmic strings (superconducting
or not) may have important roles to play in the
generation of large-scale magnetic fields
[\ct{Vac91},\ct{BraDavMatTro92}] and perhaps in creating the
net baryon number in the universe, i.e., the
matter-antimatter asymmetry [\ct{BraDavHin91},\ct{DavEar93}].

There are many exciting possibilities in this area, but as
yet it is difficult to point to definitive tests of the
models.  Much more theoretical work is needed, in
particular on the evolution of a network of superconducting
or global strings, where the dynamics is influenced by the
long-range forces between strings.

\subsection{Summary}

Cosmic strings are predicted by many unified theories of
fundamental forces.  If they exist, they provide one of the
few direct links between the observational features of our
present universe and the dramatic events of the immediate
aftermath of the Big Bang.

They can be formed at a phase transition anywhere between
the GUT and electroweak scales.  Light cosmic strings would
only be observable if superconducting, since their gravitational
interactions are so weak.  Axion strings, formed at an intermediate
scale, have the exciting possibility of generating a substantial
density of cold dark matter in the form of axions.

The best developed of the many cosmic-string scenarios
concerns GUT-scale strings, which generate both density
perturbations and perturbations in the microwave background
of roughly the observed order of magnitude.  In a CDM
universe, they do not seem very attractive, but the
combination of strings and HDM is a very promising
alternative to the popular inflationary models. The idea of
combining both hot and cold dark matter has been
put forward to improve inflationary models with a
Harrison--Zel'dovich spectrum, but this type of mixed dark
matter (MDM) has yet to be tried with strings.
If the
predictions can be tied down more accurately by improved
theoretical understanding of the problem of network
evolution, then definitive tests of this model should be
possible soon.

\ack

We are grateful to Mark Bowick for sending us a copy of Figure 1.1, to Michael
Goodband for providing Figures 2.1 and to Paul Shellard for Figures 5.2.  We
thank Albert Stebbins for helpful comments, and Catherine Wolfe for reading
the manuscript.  We also wish to acknowledge the hospitality of the Isaac
Newton
Institute where this review was completed.

\beginnumreferences

\citeitem{QFT}\refbk{Itzykson C and Zuber J-B 1980}{Quantum Field
Theory}{(New York: McGraw-Hill)};\ \refbk{Ramond P 1981}{Quantum Field Theory:
a
Modern Primer}{(Reading MA: Benjamin-Cummings)};\  \refbk{Cheng T-P and Li L-F
1984}{Gauge Theories and Elementary  Particle Physics}{(Oxford: Oxford
Univ.~Press)}
\citeitem{SUSY}\refbk{Bailin D and Love A 1994}{Supersymmetric gauge theory and
string theory}{(Bristol: Institute of Physics)}
\citeitem{NLC}\refbk{de Gennes P 1974}{The Physics of
Liquid Crystals}{(Oxford: Clarendon)};\
\ref{Kl\'eman M 1989}{\RPP}{52}{555}
\citeitem{Kap89}\refbk{Kapusta J 1989}{Finite Temperature Field
Theory}{(Cambridge: Cambridge Univ.~Press)}
\citeitem{Mer79}\ref{Mermin N D 1979}{\RMP}{51}{591}
\citeitem{Kib76}\ref{Kibble T W B 1976}{\JPA}{9}{1387}
\citeitem{Bow+94}\ref{Bowick M J, Chandar L, Schiff E A and Srivastava A M
1994}{Science}{264}{943}
\citeitem{KolTur90}\refbk{Kolb E W and Turner M S
1990}{The Early  Universe}{(Redwood City: Addison-Wesley)}
\citeitem{PenWil65}\ref{Penzias A A and Wilson R W 1965}{\APJ}
{142}{419}
\citeitem{CobeT}\ref{Mather J P \etal 1994}{\APJ}{420}{439}
\citeitem{EfsSutMad90}\ref{Efstathiou G, Sutherland W J and Maddox S J
1990}{Nature}{348}{705}
\citeitem{BahPirWei87}\refbk{Bahcall J, Piran T and Weinberg S 1987}
{Dark Matter in the Universe}{(Singapore: World Scientific)}
\citeitem{Oli+90}\ref{Olive K, Schramm D, Steigman G and Walker T 1990}
{\PL}{236B}{454}
\citeitem{VilEve82}\ref{Vilenkin A and Everett A E
1982}{\PRL}{48}{1867}
\citeitem{ShaVil84}\ref{Shafi Q and Vilenkin A 1984}{\PR}{D29}{1870}
\citeitem{Hel}\refbk{Donnelly R J 1991}{Quantized Vortices in Helium
II}{(Cambridge: Cambridge University Press)}
\citeitem{DavShe89a}\ref{Davis R L and Shellard E P S 1989}{\PRL}{63}{2021}
\citeitem{LazSha82}\ref{Lazarides G and Shafi Q
1982}{\PL}{115B}{21}
\citeitem{Hig66}\ref{Higgs P W 1966}{\PR}{145}{1156}
\citeitem{NieOle73}\ref{Nielsen H B and Olesen P 1973}{\NP}{B61}{45}
\citeitem{JacReb79}\ref{Jacobs L and Rebbi C 1979}{\PR}{B19}{4486}
\citeitem{BogVai76}\ref{Bogomol'nyi E B and Vainshtein A I 1976} {Sov. J.
\NP}{23}{588}  \orig{\YF}{23}{1111}{1976}
\citeitem{Abr57}\ref{Abrikosov A A 1957}{\JETP}{5}{1174}
\orig{\ZETF}{32}{1442}{1957}
\citeitem{Dav90}\ref{Davis R L 1990}{\MPL}{A5}{853}
\citeitem{Dav91}\ref{Davis R L 1991}{\MPL}{A6}{73}
\citeitem{Bog76}\ref{Bogomol'nyi E B 1976}{Sov. J. \NP}{24}{449}
\orig{\YF}{24}{861}{1976}
\citeitem{HilHodTur88}\ref{Hill C T, Hodges H M and Turner M S
1988}{\PR}{D37}{263}
\citeitem{TyuFatSch75}\ref{Tyupkin Y S, Fateev V A and Schwarz A S 1975}
{\JETPL}{21}{42}\orig{\ZETFPR}{21}{91}{1975}
\citeitem{AraFreGoe75}\ref{Arafune J, Freund P G O and Goebel C J 1975}
{\JMP}{16}{433}
\citeitem{Col88}\refbk{Coleman S 1988}{Aspects of Symmetry}{(Cambridge:
Cambridge University Press)}
\citeitem{Pre87}\refbk{Preskill J 1987}{Architecture of Fundamental
Interactions at Short Distances}{eds. Ramond P and Stora R (Amsterdam:
North-Holland) pp 235--338}
\citeitem{GodMan86}\ref{Goddard P and Mansfield P 1986}{\RPP}{49}{725}
\citeitem{NasSen83}\refbk{Nash C and Sen S 1983}{Topology and Geometry for
Physicists}{(New York: Academic Press)}
\citeitem{Ste57}\refbk{Steenrod N 1957}{Topology of Fibre
Bundles}{(Princeton:  Princeton University Press)}
\citeitem{HinKib85}\ref{Hindmarsh M and Kibble T W B 1985}{\PRL}{55}{2398}
\citeitem{HinThe86}\refbk{Hindmarsh M 1986}{Cosmic Strings and
Beads}{(PhD thesis, Univ. of London)}
\citeitem{AryEve87}\ref{Aryal M and Everett A E 1987}{\PR}{D35}{3105}
\citeitem{VacAch91}\ref{Vachaspati T and Ach\'ucarro A 1991}{\PR}{D44}{3067}
\citeitem{Vac92}\ref{Vachaspati T 1992}{\PRL}{68}{1977}
\citeitem{Pre92}\ref{Preskill J 1992}{\PR}{D46}{4218}
\citeitem{Hin92}\ref{Hindmarsh M 1992}{\PRL}{68}{1263}
\citeitem{Hin93a}\ref{Hindmarsh M 1993}{\NP}{B392}{461}
\citeitem{Ach+92}\ref{Ach\'ucarro A, Kujken K, Perivolaropoulos L
and Vachaspati T 1992}{\NP}{B388}{435}
\citeitem{Gib+92}\ref{Gibbons G, Ortiz M E, Ruiz Ruiz F and Samols T M 1992}
{\NP}{B385}{127}
\citeitem{JamPerVac93a}\ref{James M, Perivolaropoulos L and Vachaspati T
1993}{\PR}{D46}{R5232}
\citeitem{JamPerVac93b}\ref{James M, Perivolaropoulos L and Vachaspati T
1994}{\NP}{B395}{534}
\citeitem{HinJam94}\ref{Hindmarsh M and James M 1994}{\PR}{D49}{6109}
\citeitem{BarVacBuc94}\ref{Barriola M, Vachaspati T and Bucher M
1994}{\PR}{D50}{2819}
\citeitem{HilKagWid88}\ref{Hill C T, Kagan A L and Widrow L M
1988}{\PR}{D38}{1100}
\citeitem{AlfWil89}\ref{M G Alford and F Wilczek 1989}{\PRL}{62}{1071}
\citeitem{Hin89}\ref{Hindmarsh M 1989}{\PL}{225B}{127}
\citeitem{ZelKobOku75}\ref{Zel'dovich Ya B, Kobzarev I Yu and Okun' L B 1975}
{\JETP}{40}{1} \orig{\ZETF}{67}{3}{1974}
\citeitem{KibLazSha82}\ref{Kibble T W B, Lazarides G and Shafi G
1982}{\PR}{D26}{435}
\citeitem{Bai81}\ref{Bais F A 1981}{\PL}{98B}{437}
\citeitem{Vil82}\ref{Vilenkin A 1982}{\NP}{B196}{240}
\citeitem{GodOli78}\ref{Goddard P and Olive D I 1978}{\RPP}{41}{1357}
\citeitem{BarVil89}\ref{Barriola M and Vilenkin A 1989}{\PRL}{63}{341}
\citeitem{LazShaWal82}\ref{Lazarides G, Shafi Q and Walsh T F 1982}
{\NP}{B195}{157}
\citeitem{PreVil93}\ref{Preskill J and Vilenkin A 1993}{\PR}{D47}{2324}
\citeitem{Raj82}\refbk{Rajaraman R 1982}{Solitons and Instantons}
{(Amsterdam: North Holland)}
\citeitem{Ary+86}\ref{Aryal M, Everett A E, Vilenkin A and Vachaspati T 1986}
{\PR}{D34}{434}
\citeitem{VacVil87}\ref{Vachaspati T and Vilenkin A 1987}{\PR}{D35}{1131}
\citeitem{Wit84}\ref{Witten E 1984}{\NP}{B249}{557}
\citeitem{HawHinTur88}\ref{Haws D, Hindmarsh M and Turok N
1988}{\PL}{209B}{255}
\citeitem{DavShe88a}\ref{Davis R L and Shellard E P S 1988}{\PL}{207B}{404; }
\ref{\dash\ 1988}{\PL}{209B}{485}
\citeitem{SupCon}\refbk{Tinkham M 1975}{Introduction to
Superconductivity}{(New York: McGraw-Hill)}
\citeitem{BabPirSpe88}\ref{Babul A, Piran T and Spergel D N 1988}
{\NP}{B202}{307}
\citeitem{AmsLag88}\ref{Amsterdamski P and Laguna-Castillo P 1988}
{\PR}{D37}{877}
\citeitem{Pet92}\ref{Peter P 1992}{\PR}{D45}{1091}
\citeitem{Pet93}\ref{Peter P 1993}{\PR}{D47}{3169}
\citeitem{Zha87}\ref{Zhang S 1987}{\PRL}{59}{2111}
\citeitem{Eve88}\ref{Everett A E 1988}{\PRL}{61}{1807}
\citeitem{Sch82}\ref{Schwarz A S 1982}{\NP}{B208}{141}
\citeitem{SchTyu82}\ref{Schwarz A S and Tyupkin Y S 1982}{\NP}{B209}{427}
\citeitem{Alf+91}\ref{Alford M, Benson K, Coleman S and March Russell J 1991}
{\NP}{B349}{439}
\citeitem{Eve93}\ref{Everett A E 1993}{\PR}{D47}{R1277}
\citeitem{BucLoPre92}\ref{Bucher M, Lo Hoi-Kwong and Preskill J 1992}
{\NP}{B386}{3}
\citeitem{Chu+91}\ref{Chuang I, D\"urrer R, Turok N and
 Yurke B 1991}{Science}{251}{1336}
\citeitem{CarDeGMat63}\ref{Caroli C, de Gennes P and Matricon J 1963}
{\PL}{9}{307}
\citeitem{JacRos81}\ref{Jackiw R and Rossi P 1981}{\NP}{B190}{681}
\citeitem{Wei81}\ref{Weinberg E 1981}{\PR}{D24}{2669}
\citeitem{Anom}\ref{Bell J S and Jackiw R 1969}{\NC}{60A}{47};\ \ref{Adler S
1969}{\PR}{177}{2426};\ \ref{Bardeen W A 1969}{\PR}{184}{1848}
\citeitem{Wid88}\ref{Widrow L M 1988}{\PR}{D38}{1684}
\citeitem{HilLee88}\ref{Hill C T and Lee K 1988}{\NP}{B297}{765}
\citeitem{CalHar85}\ref{Callan C G and Harvey J 1985}{\NP}{B250}{427}
\citeitem{LazSha85}\ref{Lazarides G and Shafi Q 1985}{\PL}{157}{123}
\citeitem{BarMat87}\ref{Barr S M and Matheson A 1987}{\PL}{198B}{146};\
\ref{\dash\ 1987}{\PR}{D36}{2905}
\citeitem{HilWid87}\ref{Hill C T and Widrow L M 1987}{\PL}{189B}{17}
\citeitem{Dav87b}\ref{Davis R L}{\PR}{D36}{2267}
\citeitem{Hin88}\ref{Hindmarsh M 1988}{\PL}{200B}{429}
\citeitem{GUT}\ref{Georgi H and Glashow S L 1974}{\PRL}{32}{438};\ \ref{Pati
J and Salam A 1974}{\PR}{D10}{275};\ \ref{Ross G G 1981}{\RPP}{44}{655}
\citeitem{GeoGla74}\ref{Georgi H and Glashow S 1974}{\PRL}{32}{438}
\citeitem{SO10GUT}\refbk{Georgi H 1974}{Particles and Fields 1974}{ed.\
Carlson C (New York: American Institute of Physics) pp 575--582;\ }
\ref{Fritzsch H and Minkowski P 1975}{\APNY}{93}{193}
\citeitem{Seesaw}\refbk{Gell-Mann M, Ramond P and Slansky R
1979}{Supergravity}{eds.\ van Nieuwen\-huisen P and Freedman D
(Amsterdam: North-Holland)}
\citeitem{AmaDeBFur91}\ref{Amaldi U, de Boer W and Furstenau
1991}{\PL}{260}{447}
\citeitem{Chu+86}\ref{Chudnovsky E M, Field G, Spergel D N and Vilenkin A
1986}{\PR}{D34}{944}
\citeitem{OstThoWit86}\ref{Ostriker J, Thompson C and Witten E
1986}{\PL}{180B}{231}
\citeitem{HilSchWal87}\ref{Hill C T, Schramm D N and Walker T
1987}{\PR}{D36}{1007}
\citeitem{FieVil87}\ref{Field G and Vilenkin A 1987}{Nature}{326}{772}
\citeitem{MalBut89}\ref{Malaney R A and Butler M N 1989}{\PRL}{62}{117}
\citeitem{Yaj87}\ref{Yajnik U 1987}{\PL}{184B}{229}
\citeitem{PecQui77}\ref{Peccei R D and Quinn H 1977}{\PRL}{38}{1440; }
\ref{\dash 1977}{\PR}{D16}{1791}
\citeitem{Wei78}\ref{Weinberg S 1978}{\PRL}{40}{223}
\citeitem{Wil78}\ref{Wilczek F 1978}{\PRL}{40}{279}
\citeitem{GroPisYaf81}\ref{Gross D J, Pisarski R D and Yaffe L G
1981}{\RMP}{53}{43}
\citeitem{Sik91}\refbk{Sikivie P 1991}{TASI 91 (Proceedings of the
Theoretical Advanced Study Institute, Boulder, CO)}{pp 399--420}{}
\citeitem{Dic+78}\ref{Dicus D A, Kolb E W, Teplitz V L and
Wagoner R V 1978}{\PR}{D18}{1829}
\citeitem{EliOli87}\ref{Ellis J and Olive K 1987}{\PL}{193B}{525}
\citeitem{Dav86}\ref{Davis R L 1986}{\PL}{180B}{225}
\citeitem{HarSik87}\ref{Harari D and Sikivie P 1987}{\PL}{195B}{361}
\citeitem{DavShe89b}\ref{Davis R L and Shellard E P S 1989}{\NP}{B234}{167}
\citeitem{Wil82}\ref{Wilczek F 1982}{\PRL}{49}{1549}
\citeitem{JoyTur94}\ref{Joyce M and Turok N 1994}{\NP}{B416}{389}
\citeitem{BibDva90}\ref{Bibilashvili T M and Dvali G R 1990}{\PL}{248B}{259}
\citeitem{DvaSen94}\ref{Dvali G R and Senjanovi\'c G 1994}{\PRL}{72}{9}
\citeitem{For74}\ref{F\"orster D 1974}{\NP}{B81}{84}
\citeitem{GodGolRebTho73}\ref{Goddard P, Goldstone J, Rebbi C and Thorne
C 1973}{\NP}{B56}{109}
\citeitem{NamGot70}\refbk{Nambu Y 1970}{Symmetries and Quark
Models}{ed. Chand R (New York: Gordon and Breach)}
\citeitem{GreSchWit87}\refbk{Green M B, Schwartz J and Witten E
1987}{Superstring Theory}{(Cambridge: Cambridge University Press)}
\citeitem{DifGeo}\refbk{Eisenhart
L P 1925}{Riemannian Geometry}{(Princeton: Princeton University Press)};\
\refbk{Spivak M 1979}{A Comprehensive Introduction to Differential
Geo\-metry}{(Berkeley: Publish or Perish)}
\citeitem{MaeTur88}\ref{Maeda K and Turok N 1988}{\PL}{202B}{376}
\citeitem{Gre88a}\ref{Gregory R 1988}{\PL}{206B}{199}
\citeitem{BarHoc89}\ref{Barr S M and Hochberg D 1989}{\PR}{D39}{2308}
\citeitem{GreHawGar90}\ref{Gregory R, Haws D and Garfinkle D 1990}{\PR}{D42}
{343}
\citeitem{Mor89}\refbk{Morikawa M 1989}{Effective Action and Radiative
Reaction of Cosmic Strings}{(Vancouver: UBC preprint)}
\citeitem{Gre91}\ref{Gregory R 1991}{\PR}{D43}{520}
\citeitem{SilMai93}\ref{Silveira V and Maia M D 1993}{\PL}{174A}{280}
\citeitem{Lar93}\ref{Larsen A L 1993}{\PL}{181A}{369}
\citeitem{KibTur82}\ref{Kibble T W B and Turok N 1982}{\PL}{116B}{141}
\citeitem{Kib85}\ref{Kibble T W B 1985}{\NP}{B252}{227 [E {\bf B261} 461]}
\citeitem{GarVac87}\ref{Garfinkle D and Vachaspati T 1987}{\PR}{D36}{2229}
\citeitem{Bur85}\ref{Burden C 1985}{\PL}{164B}{277}
\citeitem{Tur83b}\ref{Turok N 1983}{\PL}{123B}{387}
\citeitem{Vil90}\ref{Vilenkin A 1990}{\PR}{D41}{3038}
\citeitem{Car90}\ref{Carter B 1990} {\PR}{D41}{3869}
\citeitem{TurBha84}\ref{Turok N and Bhattacharjee P 1984}{\PR}{D29}{1557}
\citeitem{Tho88}\ref{Thompson C 1988}{\PR}{D37}{283}
\citeitem{Vil81b}\ref{Vilenkin A 1981}{\PR}{D24}{2082}
\citeitem{Eve81}\ref{Everett A E 1981}{\PR}{D24}{858}
\citeitem{Vil91}\ref{Vilenkin A 1991}{\PR}{D43}{1060}
\citeitem{VilVac87a}\ref{Vilenkin A and Vachaspati T 1987}{\PRL}{58}{1041}
\citeitem{CopHinTur87}\ref{Copeland E, Hindmarsh M and Turok N
1987}{\PRL}{58}{1910}
\citeitem{SpePirGoo87}\ref{Spergel D N, Piran T and
Goodman J 1987}{\NP}{B291}{647}
\citeitem{Car89}\ref{Carter B 1989}{\PL}{224B}{61; }\ref{\dash\ 1989}
{\PL}{228B}{466}
\citeitem{DavShe89c}\ref{Davis R L and Shellard E P S 1989}{\NP}{B323}{209}
\citeitem{Nie80}\ref{Nielsen N K 1980}{\NP}{B167}{249}
\citeitem{NieOle87}\ref{Nielsen N K and Olesen P 1987}{\NP}{B291}{829}
\citeitem{KalKle}\refbk{Duff M J 1987}{Architecture of Fundamental
Interactions at Short Distances}{eds. Ramond P and Stora R (Amsterdam:
North-Holland) pp 819--903}
\citeitem{CopHawHinTur88}\ref{Copeland E, Haws D,
Hindmarsh M and Turok N  1988}{\NP}{B304}{908}
\citeitem{KalRam74}\ref{Kalb M and Ramond P 1974}{\PR}{D9}{2273}
\citeitem{LunReg76}\ref{Lund F and Regge T 1976}{\PR}{D14}{1524}
\citeitem{VilVac87b}\ref{Vilenkin A and Vachaspati T 1987}{\PR}{D35}{1138}
\citeitem{DavShe88b}\ref{Davis R L and Shellard E P S 1988}{\PL}{214B}{219}
\citeitem{She87}\ref{Shellard E P S 1988}{\NP}{B283}{264}
\citeitem{MorMyeReb88a}\ref{Moriarty K, Myers E and Rebbi C 1988}{\PL}
{207B}{411}
\citeitem{MorMyeReb88b}\refbk{Moriarty K, Myers E and Rebbi C 1988}
{Cosmic Strings: the Current Status}{eds. Accetta F S and Krauss
L M (Singapore: World Scientific) pp 11--24}
\citeitem{MatMcC88}\refbk{Matzner R and McCracken J 1988}
{Cosmic Strings: the Current Status}{eds. Accetta F S and Krauss
L M (Singapore: World Scientific) pp 32--41}
\citeitem{LagMat90}\ref{Laguna P and Matzner R 1990}{\PR}{D41}{1751}
\citeitem{She88}\refbk{Shellard E P S 1988}
{ Cosmic Strings: the Current Status}{eds. Accetta F S and Krauss
L M (Singapore: World Scientific) pp 25--31}
\citeitem{Vil81a}\ref{Vilenkin A 1981}{\PR}{D23}{852}
\citeitem{Got85}\ref{Gott J R 1985}{\APJ}{288}{422}
\citeitem{His85}\ref{Hiscock W A 1985}{\PR}{D31}{3288}
\citeitem{Lin85}\ref{Linet B 1985}{\GRG}{17}{1109}
\citeitem{FutGar88}\ref{Futamase T and Garfinkle D 1988}{\PR}{D37}{2086}
\citeitem{HinWra90}\ref{Hindmarsh M and Wray A 1990}{\PL}{251B}{498}
\citeitem{Vic87}\ref{Vickers J A G 1987}{\CQG}{4}{1}
\citeitem{Gre89}\ref{Gregory R 1989}{\PR}{D39}{2108}
\citeitem{MosPol87}\ref{Moss I G and Poletti S 1987}{\PL}{199B}{34}
\citeitem{HarSik88}\ref{Harari D and Sikivie P 1988}{\PR}{D37}{3438}
\citeitem{CohKap88}\ref{Cohen A G and Kaplan D B 1988}{\PL}{215B}{67}
\citeitem{Gre88b}\ref{Gregory R 1988}{\PL}{215B}{663}
\citeitem{Wei72}\refbk{Weinberg S 1972}{Gravitation and Cosmology}
{(New York: Wiley)}
\citeitem{HogRee84}\ref{Hogan C and Rees M 1984}{Nature}{311}{109}
\citeitem{VacVil85}\ref{Vachaspati T and Vilenkin A 1985}{\PR}{D31}{3052}
\citeitem{Dur89}\ref{D\"urrer R 1989}{\NP}{B328}{238}
\citeitem{Vac87}\ref{Vachaspati T 1987}{\PR}{D35}{1767}
\citeitem{CopHawHin90}\ref{Copeland E, Haws D and Hindmarsh M 1990}{\PR}
{D42}{726}
\citeitem{QuaSpe90}\ref{Quashnock J and Spergel D N 1990}{\PR}{D42}{2505}
\citeitem{Mit+89}\ref{Mitchell D, Turok N, Wilkinson R B and Jetzer P 1989}
{\NP}{B315}{1; }\ref{Wilkinson R B, Mitchell D and Turok N 1990}
{\NP}{B332}{131}
\citeitem{Sak90}\ref{Sakellariadou M 1990}{\PR}{D42}{354}
\citeitem{Hin90a}\ref{Hindmarsh M 1990}{\PL}{251B}{28}
\citeitem{BatShe94}\ref{Battye R A and Shellard E P S 1994}{\NP}{B423}{260}
\citeitem{Vac86}\ref{Vachaspati T 1986}{\NP}{B277}{593}
\citeitem{FroGar90}\ref{Frolov V P and Garfinkle D 1990}{\PR}{D42}{3980}
\citeitem{Dav85}\ref{Davis R L 1985}{\PR}{D32}{3172}
\citeitem{Sak91}\ref{Sakellariadou M 1991}{\PR}{D44}{3767}
\citeitem{HagSik91}\ref{Hagmann C and Sikivie P 1991}{\NP}{B363}{247}
\citeitem{SreThe87}\ref{Srednicki M and Theisen S 1987}{\PL}{189B}{397}
\citeitem{Bra87}\ref{Brandenberger R H 1987}{\NP}{B293}{812}
\citeitem{BraDavMat88}\ref{Brandenberger R, Davis A-C and Matheson A
1988}{\NP}{B307}{909}
\citeitem{Per+91}\ref{Perkins W, Perivolaropoulos L, Davis A-C,
Brandenberger R and Matheson A 1991}{\NP}{B353}{237}
\citeitem{FewKay93}\ref{Fewster C and Kay B 1993}{\NP}{B399}{89}
\citeitem{Lin86}\ref{Linet B 1986}{\PR}{D33}{1833}
\citeitem{ConSca}\ref{'t Hooft G 1988}{\CMP}{118}{685;\ }
\ref{Deser S and Jackiw R 1988}{\CMP}{118}{495;\ }
\ref{Gibbons G, Ruiz Ruiz F, and Vachaspati T 1990}{\CMP}{127}{295}
\citeitem{RubCal}\ref{Rubakov V
1981}{\JETPL}{33}{644}\orig{\ZETFPR}{33}{658}{1981};\ \ref{\dash\ 1982}
{\NP}{B203}{311};\ \ref{Callan C G 1982}{\PR}{D25}{2141};\
\ref{\dash\ 1982}{\PR}{D26}{2158}
\citeitem{AlfRusWil89}\ref{Alford M, March Russell J and Wilczek F 1989}
{\NP}{B328}{140}
\citeitem{Per+90}\ref{Perivolaropoulos L, Matheson A, Davis A-C and
Brandenberger R 1990}{\PL}{245B}{556}
\citeitem{Ma93}\ref{Ma C P 1993}{\PR}{D48}{530}
\citeitem{BucGol94}\ref{Bucher M and Goldhaber A 1993}{\PR}{D49}{4167}
\citeitem{BraDavMat89}\ref{Brandenberger R, Davis A-C and Matheson A
1989}{\PL}{218B}{304}
\citeitem{Sak67}\ref{Sakharov A 1967}{\JETPL}{5}{24}\orig{\ZETFPR}
{5}{32}{1967}
\citeitem{BhaKibTur82}\ref{Bhattarcharjee P, Turok N and Kibble T W B
1982}{\PL}{119B}{95}
\citeitem{KawMae88}\ref{Kawasaki M and Maeda K 1988}{\PL}{208B}{84}
\citeitem{BraDavHin91}\ref{Brandenberger R, Davis A-C and
Hindmarsh M 1991}{\PL}{263B}{239}
\citeitem{DavEar93}\ref{Davis A-C and Earnshaw M 1993}{\NP}{B394}{21}
\citeitem{DolJac74}\ref{Dolan L and Jackiw R 1974}{\PR}{D9}{3320}
\citeitem{Wei74}\ref{Weinberg S 1974}{\PR}{D9}{3357}
\citeitem{ArnYaf94}\ref{Arnold P and Yaffe L G 1994}{PR}
{D49}{3003}
\citeitem{Sha93}\ref{Shaposhnikov M 1993}{\PL}{316B}{112}
\citeitem{RudSri93}\ref{Rudaz S and Srivastava A M 1993}{\MPL}{A8}{1443}
\citeitem{Col77}\ref{Coleman S 1977}{\PR}{D15}{2927}
\citeitem{CalCol77}\ref{Callan C and Coleman S 1977}{\PR}{D16}{1762}
\citeitem{Lin83}\ref{Linde A D 1983}{\NP}{B216}{421}
\citeitem{Lan92}\refbk{Langer 1992}{Solids Far from
Equilibrium}{ed. Godr\`eche C (Cambridge: Cam\-bridge University Press) pp
297--363}
\citeitem{HinBraDav94}\ref{Hindmarsh M, Davis A-C and
Brandenberger R  1994}{\PR}{D49}{1944}
\citeitem{VacVil84}\ref{Vachaspati T and Vilenkin A 1984}{\PR}{D30}{2036}
\citeitem{LeePro91}\ref{Leese R and Prokopec T 1991}{\PL}{260B}{27}
\citeitem{Gin60}\ref{Ginzburg V L 1960}{Sov.\ Phys.\ Sol.\
State}{2}{1824}\orig{Fiz.\ Tverd.\ Tela}{2}{2031}{1960}
\citeitem{HinStr94}\refbk{Hindmarsh M and Strobl K 1994}{Statistical
Properties of Strings}{Cambridge/ Sussex preprint\ }{DAMTP--94--56,
SUSX--TP--94/72, hep--th/9410094}
\citeitem{Kib86a}\ref{Kibble T W B 1986}{\PL}{166B}{311}
\citeitem{LazSha84}\ref{Lazarides G and Shafi Q 1984}{\PL}{148B}{35}
\citeitem{VisOliSec87}\ref{Vishniac E T, Olive K A and Seckel D 1987}
{\NP}{B289}{717}
\citeitem{Yok88}\ref{Yokoyama J 1988}{\PL}{212B}{273}
\citeitem{Kib87}\refbk{Kibble T W B 1987}{Cosmology and Particle Physics,
Lectures at GIFT XVIIth International Seminar on Theoretical Physics}
{eds.\ Alvarez E, Dominguez Tenreiro R,
Ib\'a\~nez Cabanell J M and Quir\'os M (Singapore: World Scientific), pp
171--208}
\citeitem{Vil84a}\ref{Vilenkin A 1984}{\PRL}{53}{1016}
\citeitem{Kib86b}\ref{Kibble T W B 1986}{\PR}{D33}{328}
\citeitem{VacEveVil84}\ref{Vachaspati T, Everett A and Vilenkin A 1984}
{\PR}{D30}{2046}
\citeitem{Vil81}\ref{Vilenkin A 1981}{\PRL}{46}{1169}\ [E\ {\bf 46} 1496]
\citeitem{CopKibAus92}\ref{Copeland E, Kibble T W B and Austin D 1992}
{\PR}{D45}{R1000}
\citeitem{AusCopKib93}\ref{Austin D, Copeland E and Kibble T W B 1993}
{\PR}{D48}{5594}
\citeitem{AlbTur85}\ref{Albrecht A and Turok N 1985}{\PRL}{54}{1868}
\citeitem{AlbTur89}\ref{Albrecht A and Turok N 1989}{\PR}{D40}{973}
\citeitem{BenBou88}\ref{Bennett D P and Bouchet F R 1988}{\PRL}{60}{257}
\citeitem{BenBou89}\ref{Bennett D P and Bouchet F R 1989}{\PRL}{63}{2776}
\citeitem{BenBou90}\ref{Bennett D P and Bouchet F R 1990}{\PR}{D41}{2408}
\citeitem{AllShe90}\ref{Allen B and Shellard E P S 1990}{\PRL}{64}{119}
\citeitem{SheAll90}\refbk{Shellard E P S and Allen B 1990}{The Formation
and Evolution of Cosmic Strings}{eds.\ Gibbons G W, Hawking S W and Vachaspati
T  (Cambridge: Cambridge University Press) pp 421--448}
\citeitem{SmiVil87}\ref{Smith A G and Vilenkin A 1987}{\PR}{D36}{990}
\citeitem{SakVil90}\ref{Sakellariadou M and Vilenkin A 1990}
{\PR}{D42}{349}
\citeitem{Ben86a}\ref{Bennett D P 1986}{\PR}{D33}{872}
\citeitem{Ben86b}\ref{Bennett D P 1986}{\PR}{D34}{3592}
\citeitem{KibCop91}\ref{Kibble T W B and Copeland E 1991}{Physica
Scripta}{T36}{153}
\citeitem{Emb}\ref{Embacher F 1992}{\NP}{B387}{129; 163;\ }
\ref{\dash\ 1994}{\PR}{D49}{5030}
\citeitem{AllCal91}\ref{Allen B and Caldwell R 1991}{\PR}{D43}{2457}
\citeitem{QuaPir91}\ref{Quashnock J and Piran T 1991}{\PR}{D43}{2497}
\citeitem{Aus93}\refbk{Austin D 1993}{Small Scale Structure on
Cosmic Strings}{University of Sussex preprint SUSX--TP--93/3--2,
hep-th/9305028}
\citeitem{AusCopKib94}\refbk{Austin D, Copeland E and Kibble T W B
1994}{Characteristics of Cosmic String Configurations}{University of
Sussex/Imperial College preprint SUSX--TH--93/3/6, IMPERIAL/TP/93--94/45,
hep-ph/9406379}
\citeitem{Car90b}\refbk{Carter B 1990}{The Early Universe and  Cosmic
Structure}{Proc. Xth Moriond Astrophysics Meeting, eds.~Blanchard  A and Tran
Thanh Van J (Gif sur Yvette: Editions Fronti\`eres)}
\citeitem{ChuVil88}\ref{Chudnovsky E and Vilenkin A 1988}{\PRL}{61}{1043}
\citeitem{Explo}\ref{Ostriker J A and Cowie L L 1981}{\APJ}{243}{L127};\
\ref{Carr B J and Ikeuchi S}{\MNRAS}{213}{497};\  \ref{Vishniac E T, Ostriker
J A and Bertschinger E 1985}{\APJ}{291}{399} \citeitem{LevFreSpe92}\ref{Levin
J J, Freese K and Spergel D N 1992}{\APJ}{389}{464}
\citeitem{Kim87}\ref{Kim J E 1987}{Phys. Rep.}{150}{1}
\citeitem{Sik82}\ref{Sikivie P 1982}{\PRL}{48}{1156}
\citeitem{She90}\refbk{Shellard P 1990}{The Formation and Evolution  of
Cosmic Strings\ }{eds.\ Gibbons G W, Hawking S W and Vachaspati T
(Cambridge: Cambridge University Press) pp 107--115}
\citeitem{AbbSik83}\ref{Abbott L F and Sikivie P 1983}{\PL}{120B}{133}
\citeitem{DinFis83}\ref{Dine M and Fischler W 1983}{\PL}{120B}{137}
\citeitem{PreWisWil83}\ref{Preskill J, Wise M B and Wilczek F
1983}{\PL}{120B}{127}
\citeitem{Vil84b}\ref{Vilenkin A 1984}{\APJ}{282}{L51}
\citeitem{Gar85}\ref{Garfinkle D 1985}{\PR}{D32}{1323}
\citeitem{Vil86}\ref{Vilenkin A 1986}{Nature}{322}{613}
\citeitem{Hin90b}\refbk{Hindmarsh M B 1990}{The Formation and Evolution of
Cosmic Strings}{eds.\ Gibbons G W, Hawking S W and Vachaspati T (Cambridge:
Cambridge University Press) pp 527--542}
\citeitem{GLens}\refbk{Schneider P, Ehlers J and Falco E E 1992}
{Gravitational Lensing}{(Berlin: Springer Verlag)}
\citeitem{CowHu87}\ref{Cowie L L and Hu E M 1987}{\APJ}{318}{L33}
\citeitem{Hu90}\ref{Hu E M 1990}{\APJ}{360}{L7}
\citeitem{HogNar84}\ref{Hogan C and Narayan R 1984}{\MNRAS}{211}{575}
\citeitem{Pac86}\ref{Paczy\'nski B 1986}{Nature}{319}{567}
\citeitem{KaiSte84}\ref{Kaiser N and Stebbins A 1984}{Nature}{310}{391}
\citeitem{Ste88}\ref{Stebbins A 1988}{\APJ}{327}{584}
\citeitem{Hin94}\ref{Hindmarsh M B 1994}{\APJ}{431}{534}
\citeitem{BouBenSte88}\ref{Bouchet F R, Bennett D P  and Stebbins A 1988}
{Nature}{335}{410}
\citeitem{Rea+88}\ref{Readhead A C S \etal 1988}{\APJ}{346}{566}
\citeitem{SteVee94}\refbk{Stebbins A and Veeraraghavan S 1994}
{Beyond the Small-Angle Approximation for MBR Perturbations from
Seeds}{preprint FERMILAB--PUB--94/047--A, astro--ph 9406067 (to appear
in \PR\ D)}
\citeitem{BenSteBou92}\ref{Bennett D P, Stebbins A and
Bouchet F 1992} {\APJ}{399}{L5}
\citeitem{Smo+92}\ref{Smoot G \etal 1992}{\APJ}{396}{L1}
\citeitem{Ben+92}\ref{Bennett C L \etal 1992}{\APJ}{396}{L7}
\citeitem{Cou+94}\ref{Coulson D, Ferreira P, Graham P and Turok N 1994}
{Nature}{368}{27}
\citeitem{MoePerBra94}\ref{Moessner R, Perivolaropoulos L and Brandenberger
R 1994}{\APJ}{425}{365}
\citeitem{Tur85}\ref{Turok N 1985}{\PRL}{55}{1801}
\citeitem{SilVil84}\ref{Silk J and Vilenkin A 1984}{\PRL}{53}{1700}
\citeitem{VacVil91}\ref{Vachaspati T and Vilenkin A 1990}{\PRL}{67}{1057}
\citeitem{VeeSte90}\ref{Veeraraghavan S and Stebbins A 1990}{\APJ}{365}
{37}
\citeitem{PenSpeTur94}\ref{Pen U L, Spergel D N and Turok N 1994}
{\PR}{D49}{692}
\citeitem{AlbSte92a}\ref{Albrecht A and Stebbins A 1992}{\PRL}{68}{2121}
\citeitem{AlbSte92b}\ref{Albrecht A and Stebbins A 1992}{\PRL}{69}{2615}
\citeitem{LytLid93}\ref{Lyth D H and Liddle A R 1993}{Phys.~Rep.}{231}{1}
\citeitem{CalAll92}\ref{Caldwell R R and Allen B 1992}{\PR}{D45}{3447}
\citeitem{OliSte94}\refbk{Olive K and Steigman G 1994}
{On the Abundance of Primordial Helium}
{Ohio State Univ.\ preprint OSU--TA--6/94, astro--ph 9406067}
\citeitem{KerKra94}\ref{Kernan PJ and Krauss LM 1994}{\PRL}{72}{3309}
\citeitem{BenBou91}\ref{Bennett D P and Bouchet F R 1991}{\PR}{D43}{2733}
\citeitem{Sti+90}\ref{Stinebring D R, Ryba M F, Taylor J
H and Romani R W  1990}{\PRL}{65}{285}
\citeitem{AllShe92}\ref{Allen B and Shellard E P S 1992}{\PR}{D45}{1898}
\citeitem{Yok89}\ref{Yokoyama J 1989}{\PRL}{63}{712}
\citeitem{QueNai92}\ref{Quenby J J and Naidu K 1992}{\NP\ {\rm(}Proc.
Supp.{\rm)}}{B1}{1}
\citeitem{Bha89}\ref{Bhattacharjee P 1989}{\PR}{D40}{3968}
\citeitem{BhaRan90}\ref{Bhattacharjee P and Rana N C 1990}{\PL}{246B}{365}
\citeitem{GilKib94}\ref{Gill A J and Kibble T W B 1994}{\PR}{D50}{3660}
\citeitem{BabPacSpe87}\ref{Babul A, Paczy\'nski B and Spergel D 1987}{\APJ}
{316}{L49};\ \ref{Paczy\'nski B 1988}{\APJ}{335}{L49}
\citeitem{Sta82}\ref{Starobinsky A A 1982}{\PL}{117B}{175}
\citeitem{Haw82}\ref{Hawking S W 1982}{\PL}{115B}{295}
\citeitem{GutPi82}\ref{Guth A H and Pi S-Y 1982}{\PRL}{49}{1110}
\citeitem{LaSte89}\ref{La D and Steinhardt P 1989}{\PRL}{62}{3760}
\citeitem{LucMat85}\ref{Lucchin F and Matarrese S 1985}{\PR}{D32}{1316}
\citeitem{SteAcc90}\ref{Steinhardt P and Accetta F S 1990}{\PRL}{64}{2740}
\citeitem{AdaFre91}\ref{Adams F C and Freese K 1991}{\PRL}{64}{2740}
\citeitem{AdaFreGut91}\ref{Adams F C, Freese K and Guth A H
1991}{\PR}{D43}{965}
\citeitem{Wri+92}\ref{Wright E L \etal 1992}{\APJ}{396}{L13}
\citeitem{Per93}\ref{Perivolaropoulos L 1993}{\PL}{298B}{305}
\citeitem{Ben+94}\refbk{Bennett C \etal 1994}{Cosmic Temperature Fluctuations
from Two Years of COBE DMR Observations}{COBE preprint, astro-ph/9401012}
\citeitem{Gor+94}\ref{G\'orski K M \etal 1994}{\APJ}{430}{L89}
\citeitem{MAX93}\ref{Meinhold P \etal 1993}{\APJ}{409}{L1;\ }\ref{Gunderson J O
\etal 1993}{\APJ}{413}{L1}
\citeitem{DavEfsFreWhi92}\ref{Davis M, Efstathiou G, Frenk C S and
White S D M 1992}{Nature}{356}{489}
\citeitem{vDaSch92}\ref{van Dalen A and Schaefer R K 1992}{\APJ}{398}{33}
\citeitem{DavSumSch92}\ref{Davis M, Summers F J and Schlegel D 1992}
{Nature}{359}{393}
\citeitem{TayRow92}\ref{Taylor A N and Rowan-Robinson M 1992}{Nature}{359}
{396}
\citeitem{CenGneKofOst92}\ref{Cen R, Gnedin N Y, Kofman L A and Ostriker
J P 1992}{\APJ}{399}{L11}
\citeitem{Pet94}\ref{Peter P 1994}{\PR}{D49}{5052}
\citeitem{Vac91}\ref{Vachaspati T 1991}{\PL}{265B}{258}
\citeitem{BraDavMatTro92}\ref{Brandenberger R H, Davis A-C, Matheson
A M and Trodden M 1992}{\PL}{293B}{287}

\figures

\figcap{1.1}{Defect formation at the isotropic-nematic phase transition
in a liquid crystal.  As the sample cools, bubbles
of the ordered, nematic, phase nucleate ($a$), grow ($b$), and merge
($c$), resulting in the formation of a network of line disclinations
at the bubble boundaries ($d$).  The scale of the network then increases
as strings straighten and loops collapse ($e$).  (Reproduced by permission of
the authors and publisher from Bowick M J, Chandar L, Schiff E A and Srivastava
A M 1994 {\it Science} {\bf 264} 943.)}

\figcap{2.1}{Field ($a$) and energy density ($b$) profiles of the global
string.  The shading in ($a$) shows the phase of the complex scalar
field changing from 0 (lightest) to $2\pi$ (darkest) around the string.
The sharp point in the energy density is an artefact of the
discretization.}

\figcap{2.2}{A multiply-connected vacuum manifold $\M$.  Two paths with
base point $\phi_0$ are shown.  The dashed line represents the
identity class, for it is contractible to the base point, while the solid
line is not.}

\figcap{2.3}{A disconnected isotropy group, consisting of two parts
$H_0$ and $H_1$ contained in $G$.  The path shown begins and ends
at the same point in $G/H$, and thus forms a non-contractible loop in
$\M$.}

\figcap{2.4}{The field configurations of the two SO(3) $\to\{{\bf 1}\}$
string solutions. The arrows indicate the magnitudes and directions
in internal space of each field, relative to the axes shown.  Type
(ii) has lower energy.}

\figcap{2.5}{Monopoles joined by strings.  Monopoles exist if $G$
is multiply connected: the diagram shows the gauge transformations
on a large sphere around the monopole. If $H$ is disconnected, as here,
the monopoles may get joined by more than one string.}

\figcap{3.1}{A simple loop trajectory over one period of
oscillation.  Note the cusps at $t=0$ and $t=8$.}

\figcap{3.2}{The generic behaviour of the effective mass per unit length
$\mu$ and the tension $T$ of a superconducting string with current $I$.
At the critical current $I_{\rm c}$, $\d T/\d \mu$ becomes
positive and the current-carrying condensate becomes unstable.}

\figcap{3.3}{Intercommuting strings.  Two segments travelling in opposite
directions have just crossed.  The total winding number in planes $A$
and $B$ is 2 and 0 respectively, allowing the strings to reconnect the
other way.}

\figcap{3.4}{The formation of kinks after intercommuting.}

\figcap{5.1}{The formation of global strings.  Bubbles of the low
temperature phase nucleate ($a$), collide ($b$) and merge ($c$).
As this happens, the field is pulled out of the central minimum,
in a direction determined by the initial values of the field inside the
bubbles.  If the initial phases are widely distributed around the
vacuum manifold, as here, a string forms.}

\figcap{5.2}{Two boxes of string taken from the radiation ($a$) and the
matter ($b$) era simulations of Allen and Shellard.  Both are
horizon-sized cubes taken from near the end of the simulation.}

\figcap{6.1}{Gravitational lensing by cosmic string.  The space-time is
conical: a wedge $AO_1O_2A'O'_1O'_2$ is cut out and the sides identified.
The observer at $O_1O_2$ can see two images of a quasar beyond the
string, separated by an angle $\psi$, given in equation (\rf{\eLenAng}).}

\figcap{6.2}{The formation of a wake by a string.  In the string rest frame,
matter (assumed collisionless) approaches with velocity $-\bi{v}$.  In
the conical coordinate system, the particles move in straight lines.
When projected onto a plane, the particle trajectories are
bent behind the string into a wake with opening angle $8\pi G\mu$.}

\figcap{6.3}{The power spectra of density perturbations from strings
(dashed) compared to inflation (solid), for both cold and hot dark
matter, normalized to unity at 8 Mpc ($h=1$).  The lines
 with the most power on small scales are for CDM.  (Reproduced by permission of
the authors and publisher from Albrecht A and Stebbins A 1992 {\it\PRL} {\bf
69}
2615.)}

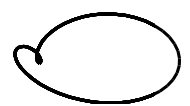
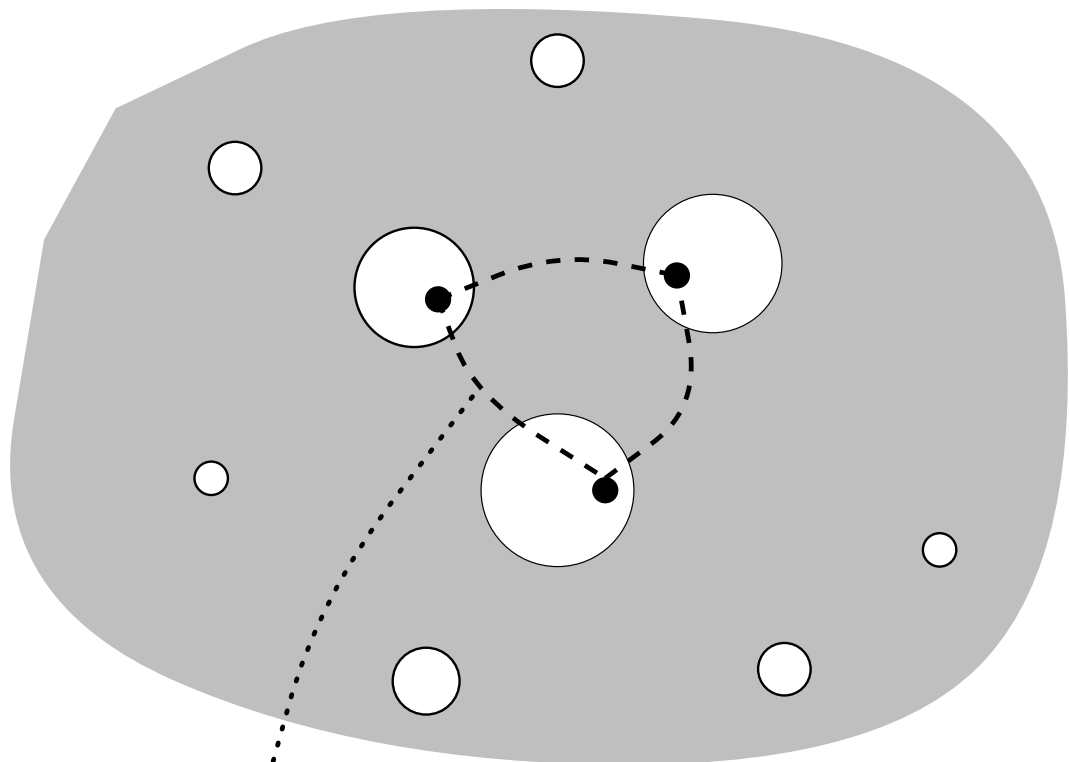
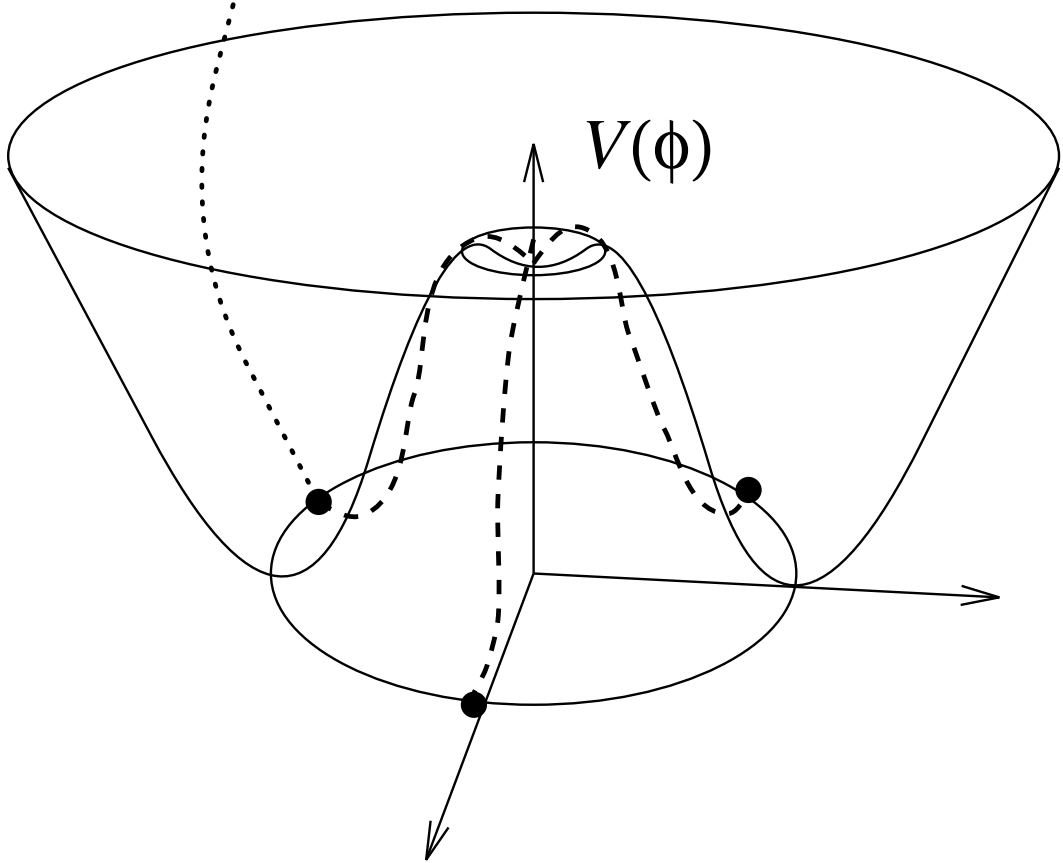
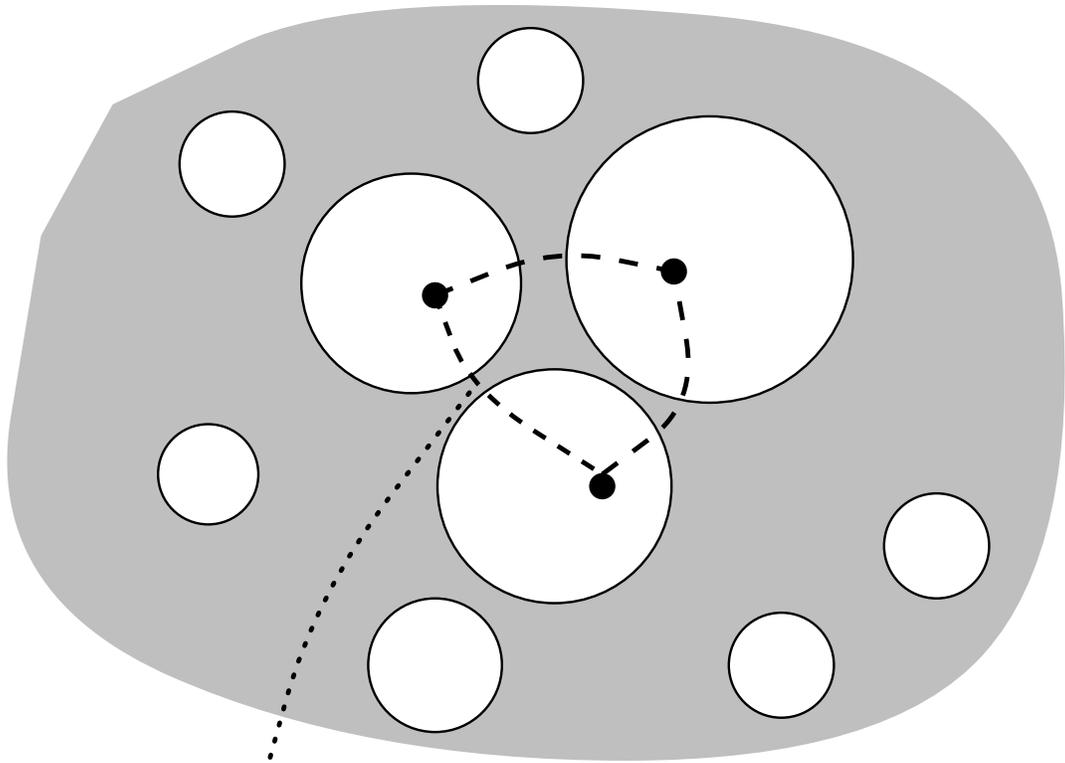
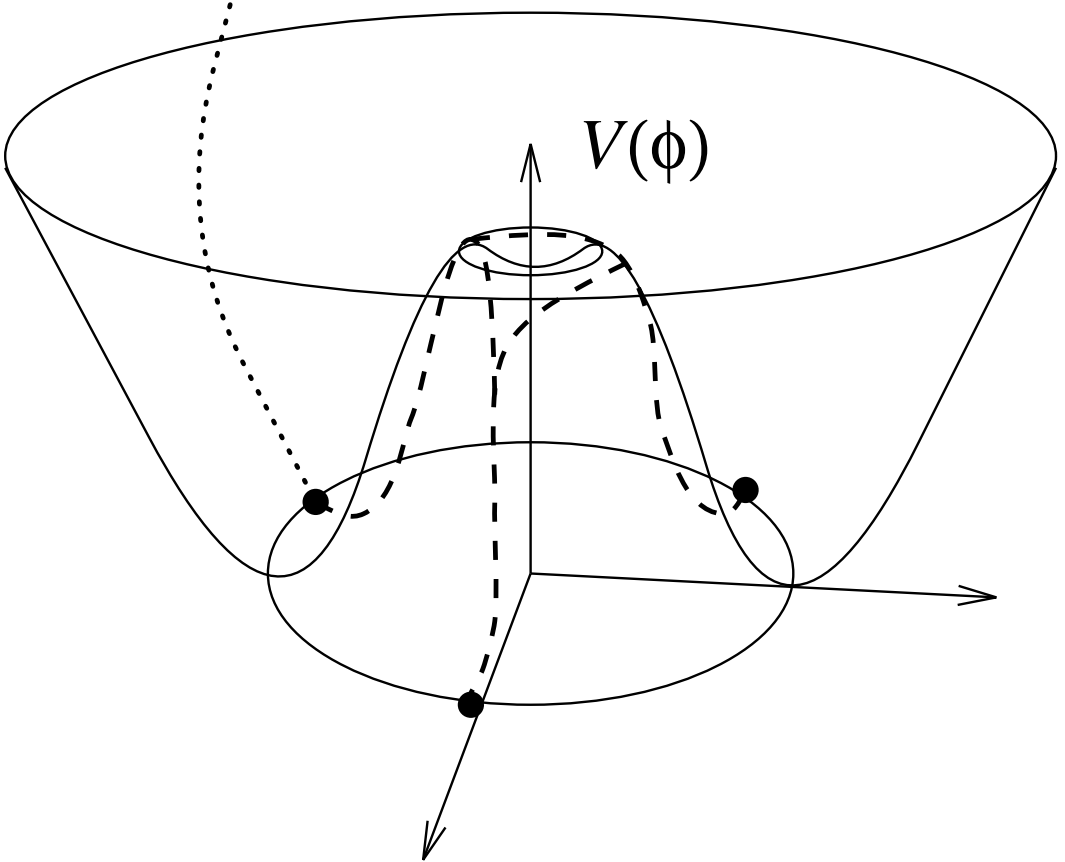
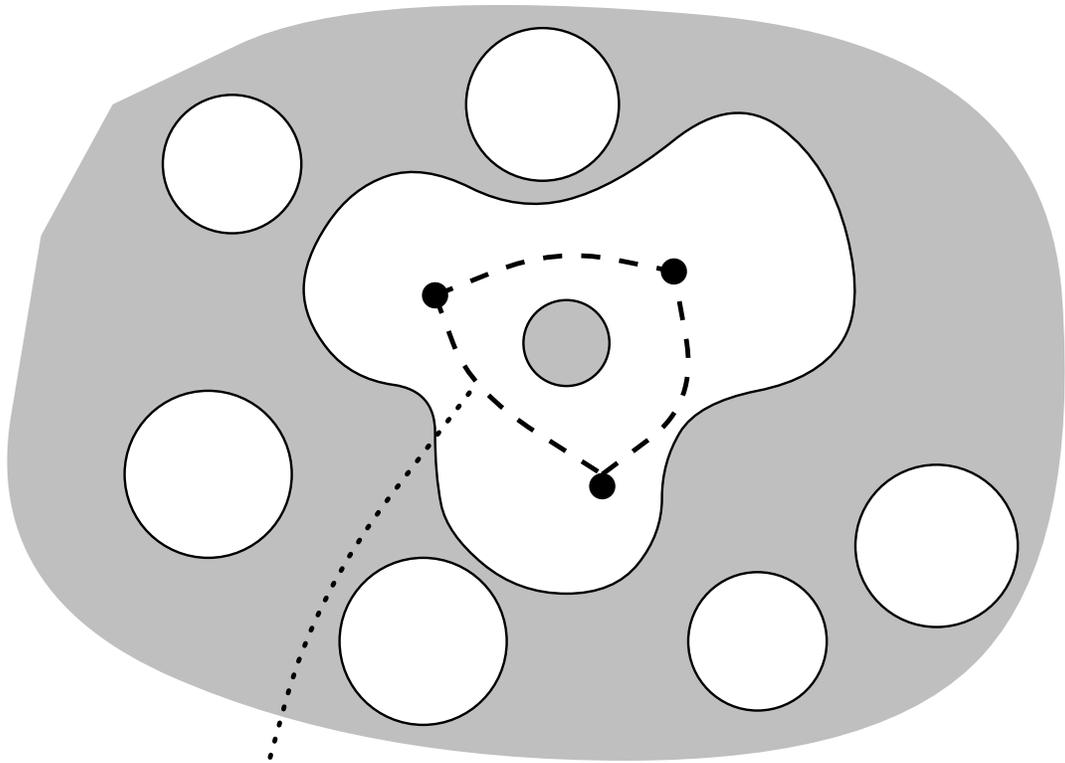
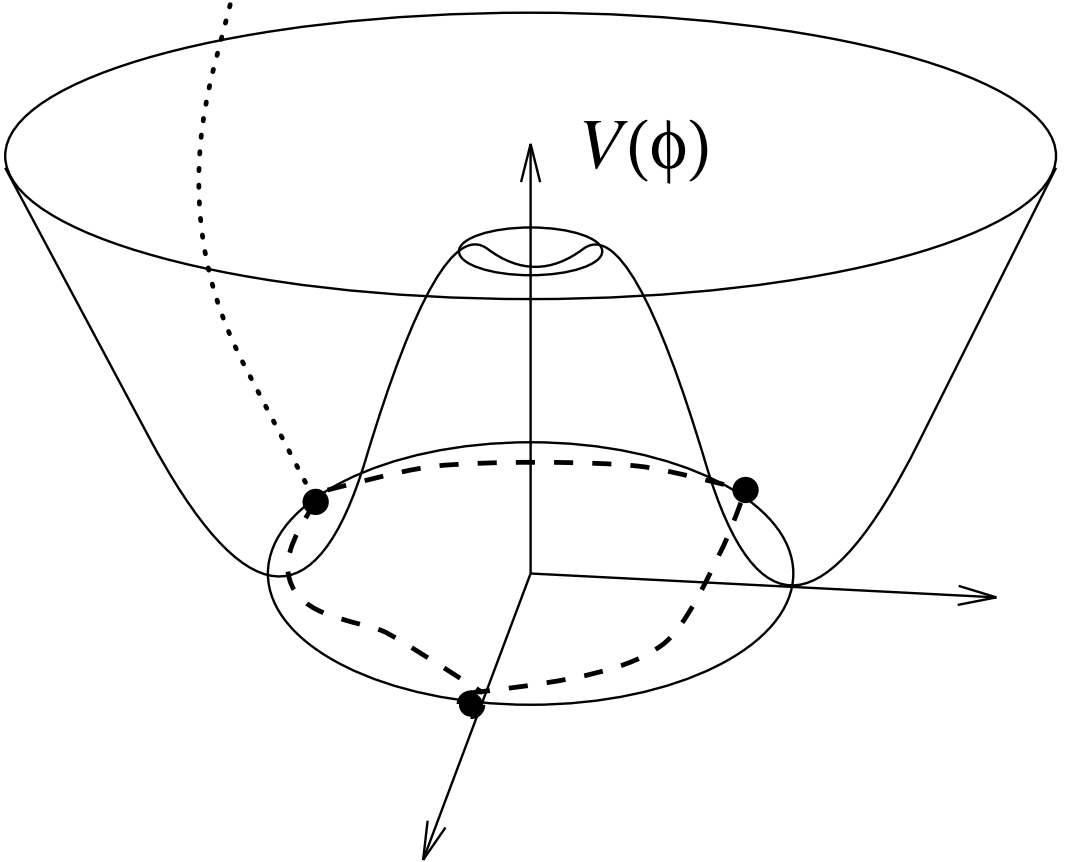
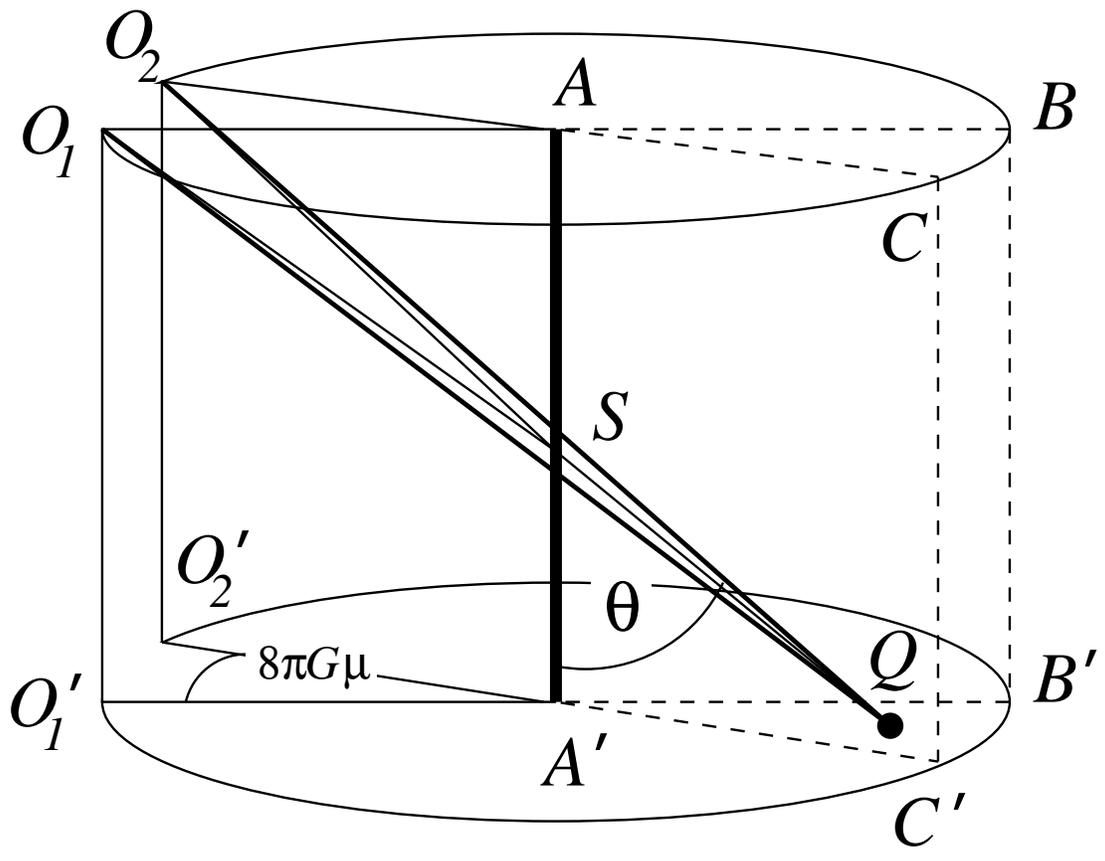
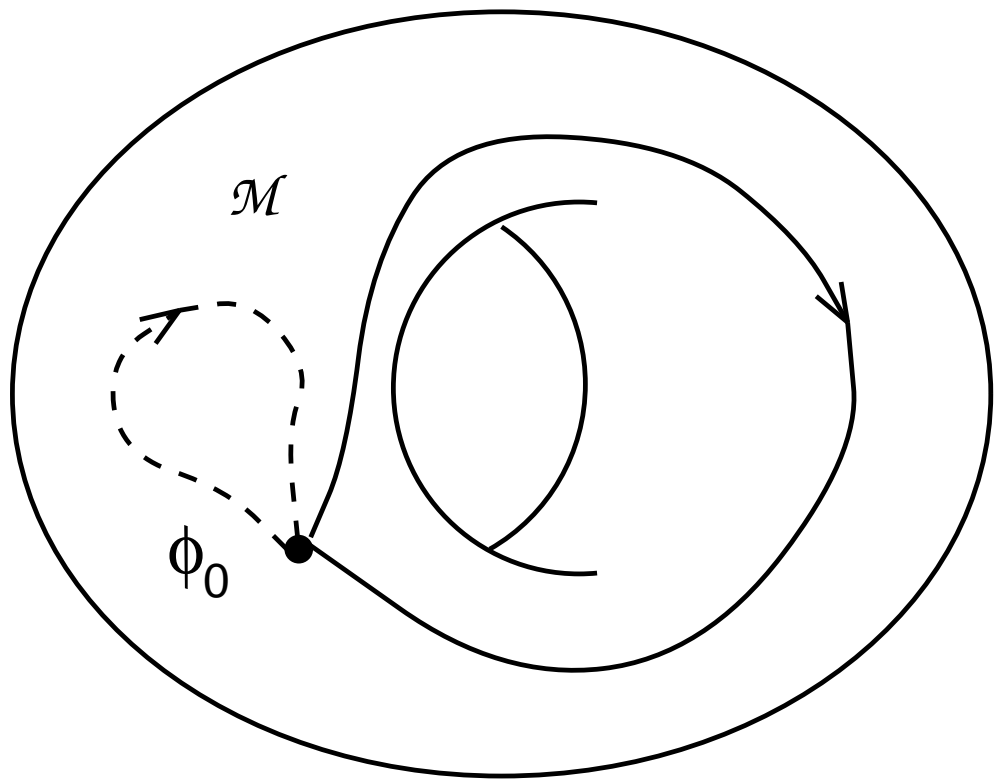
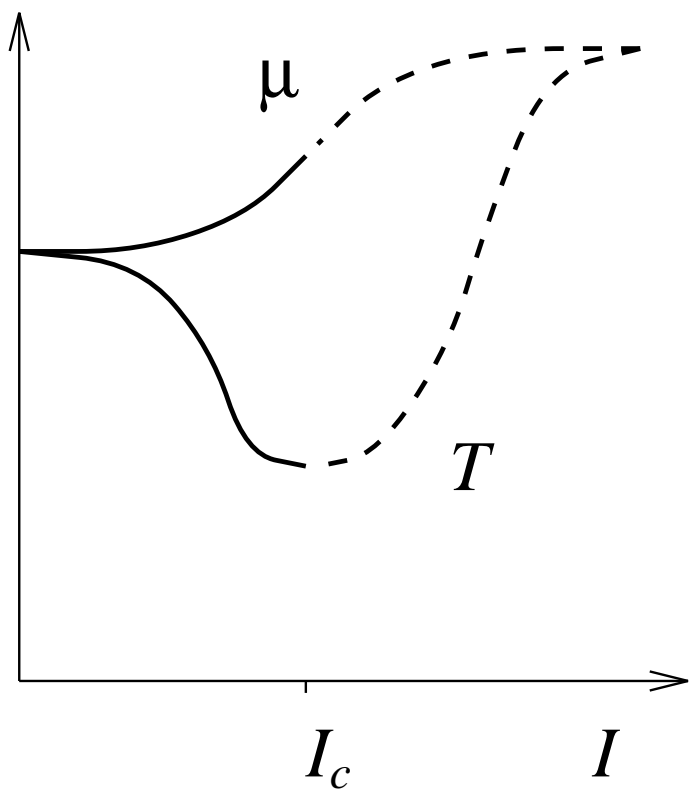
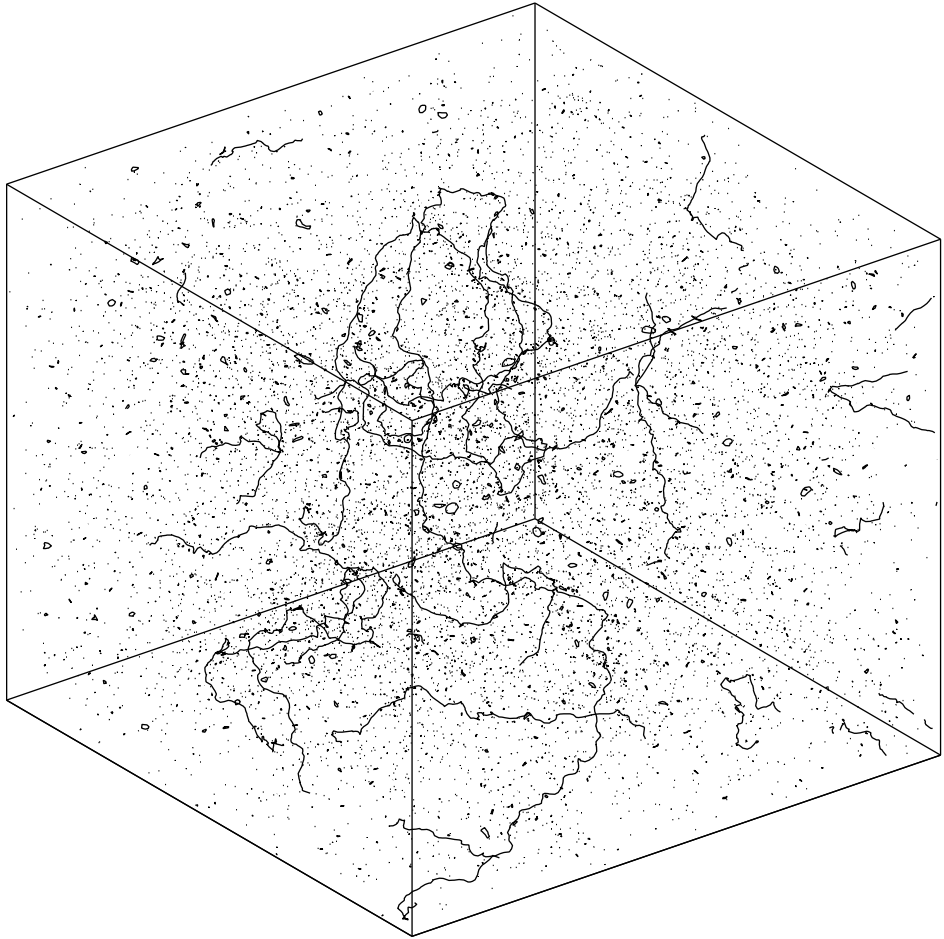
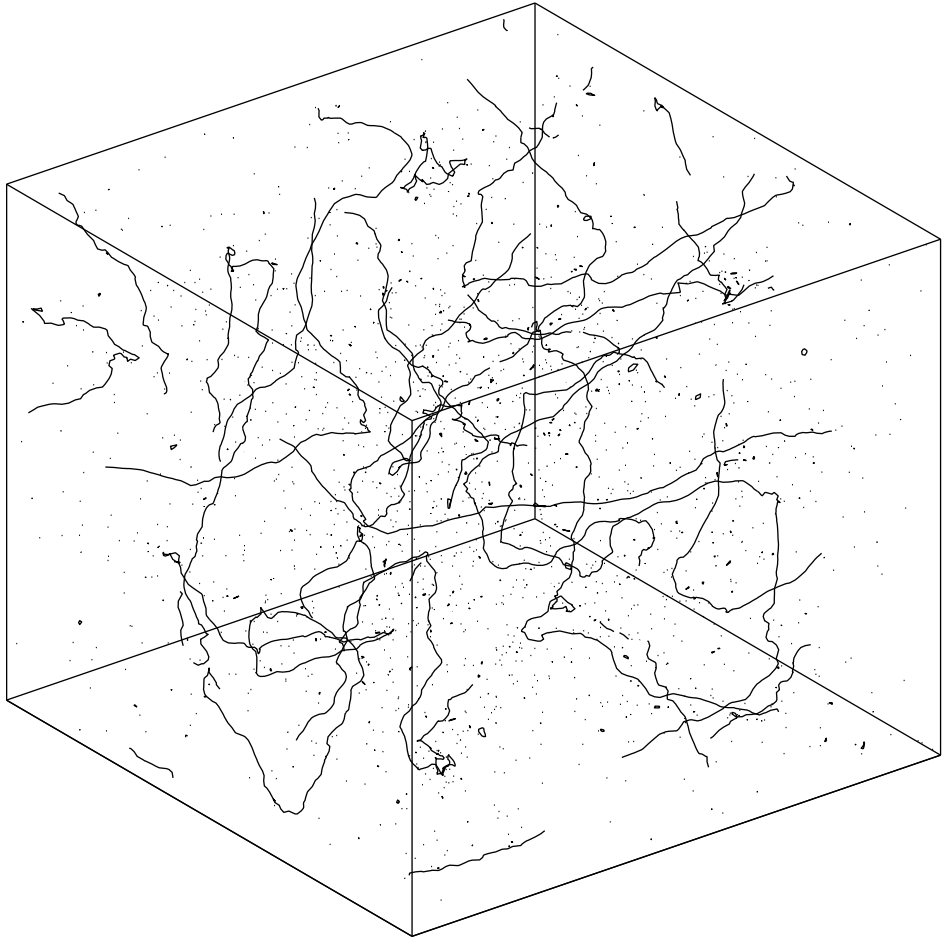
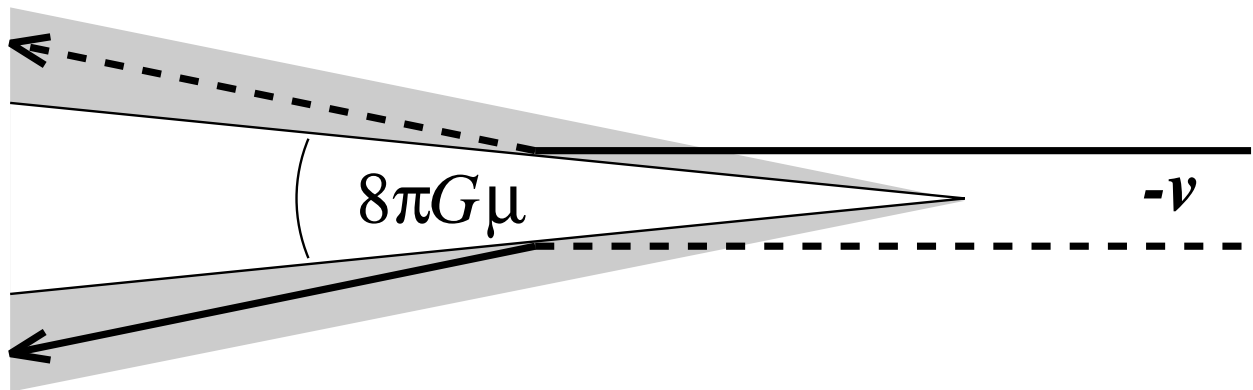
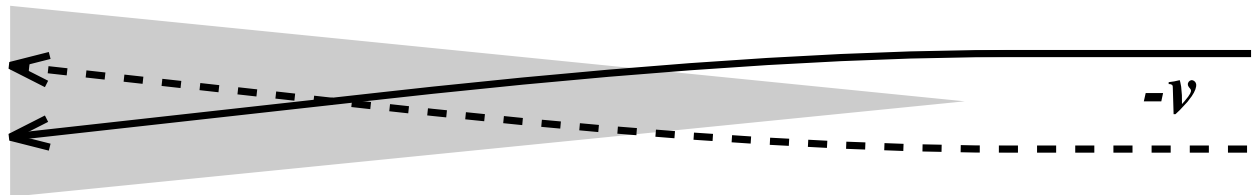
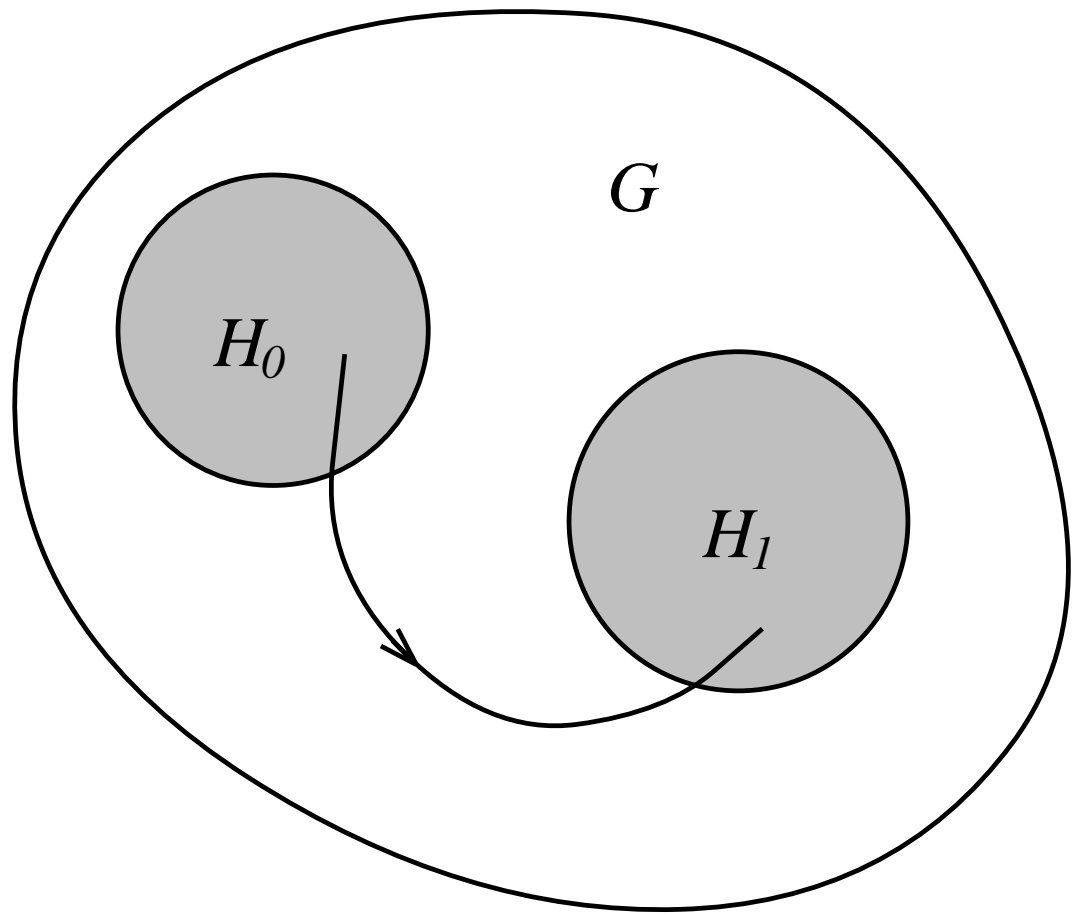
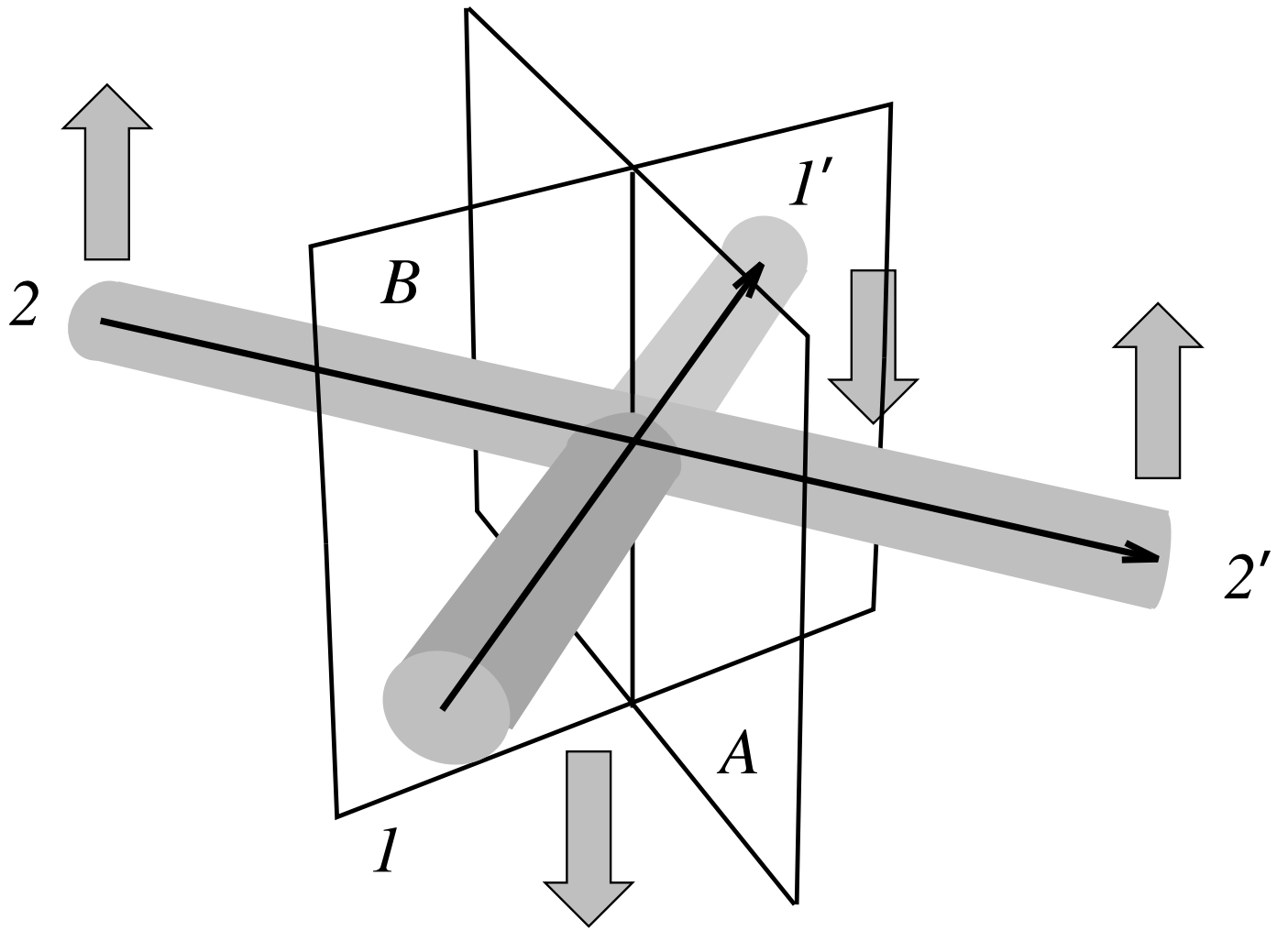
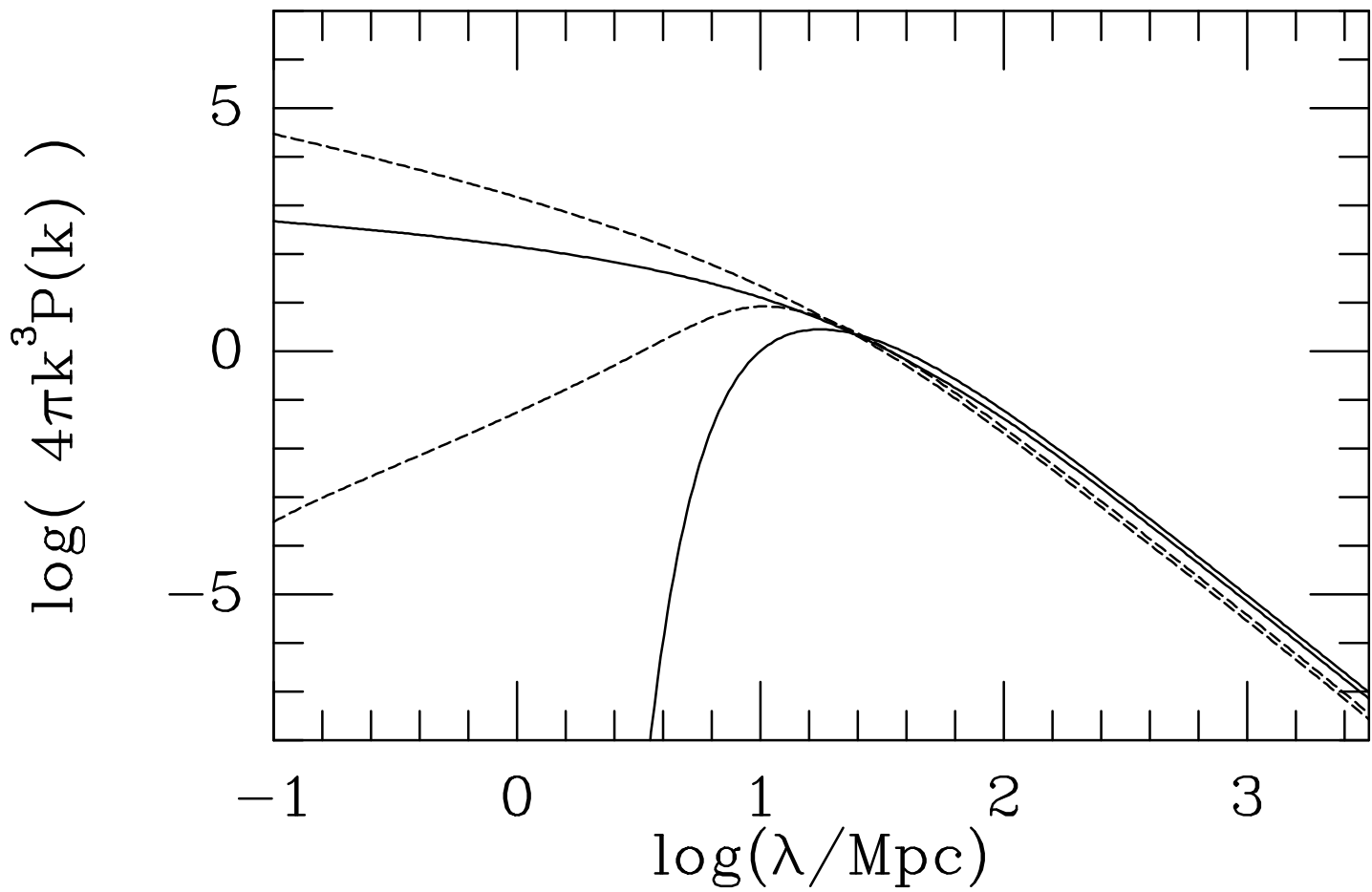
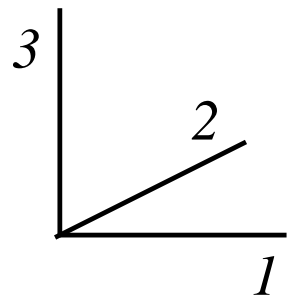
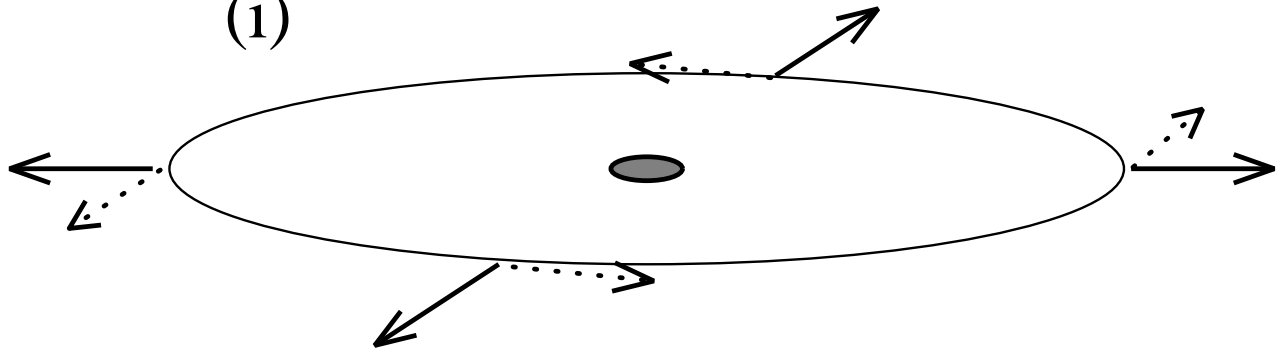
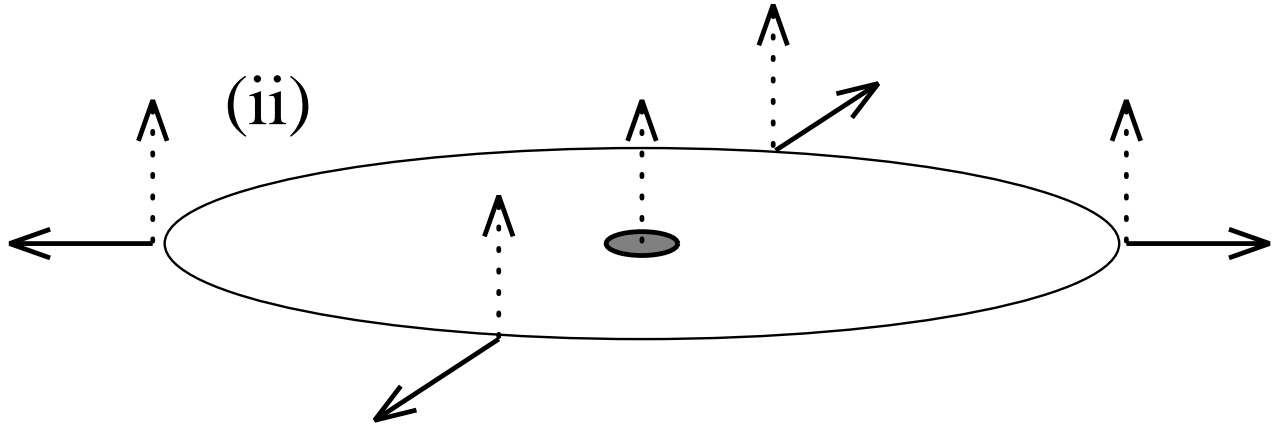
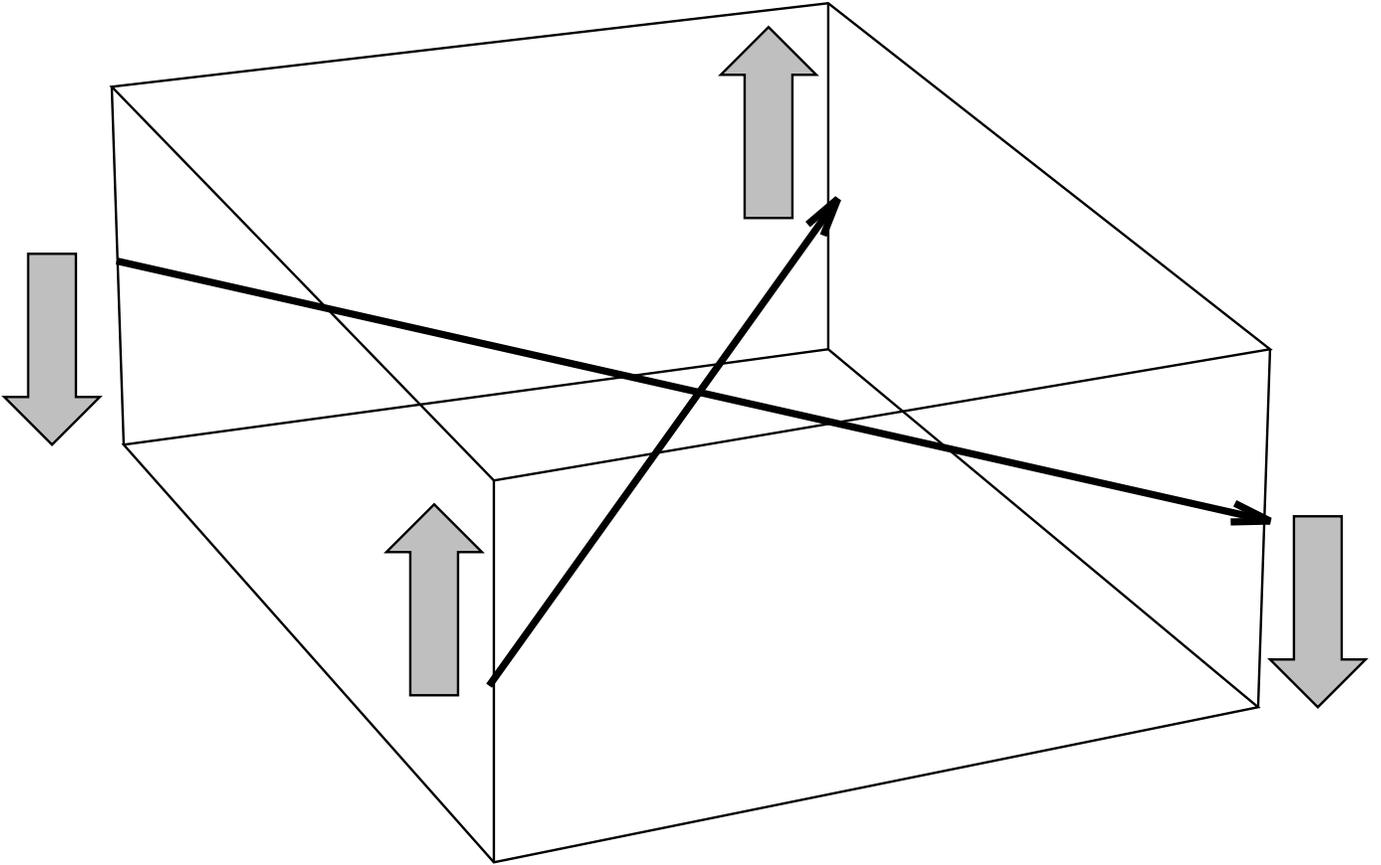
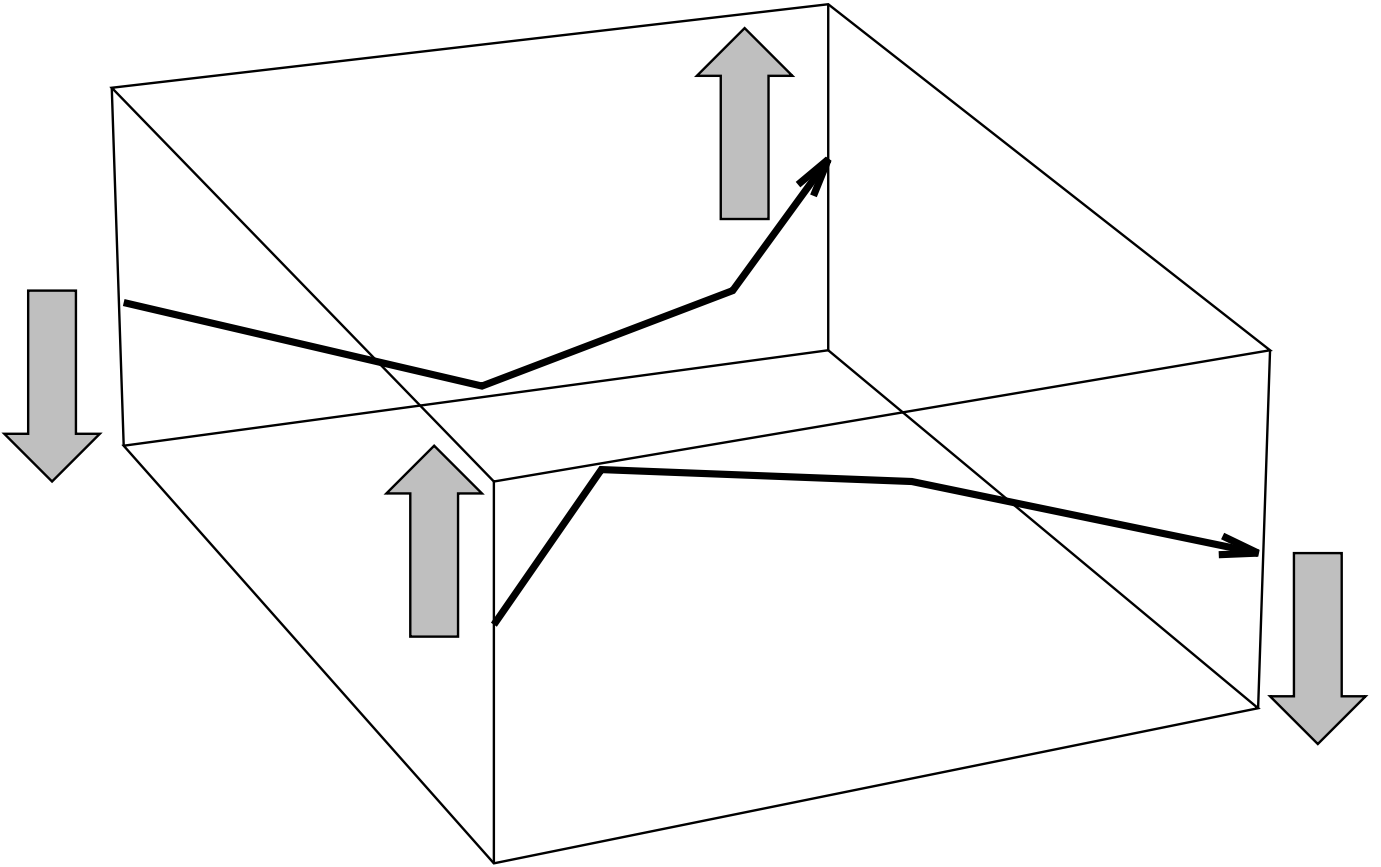
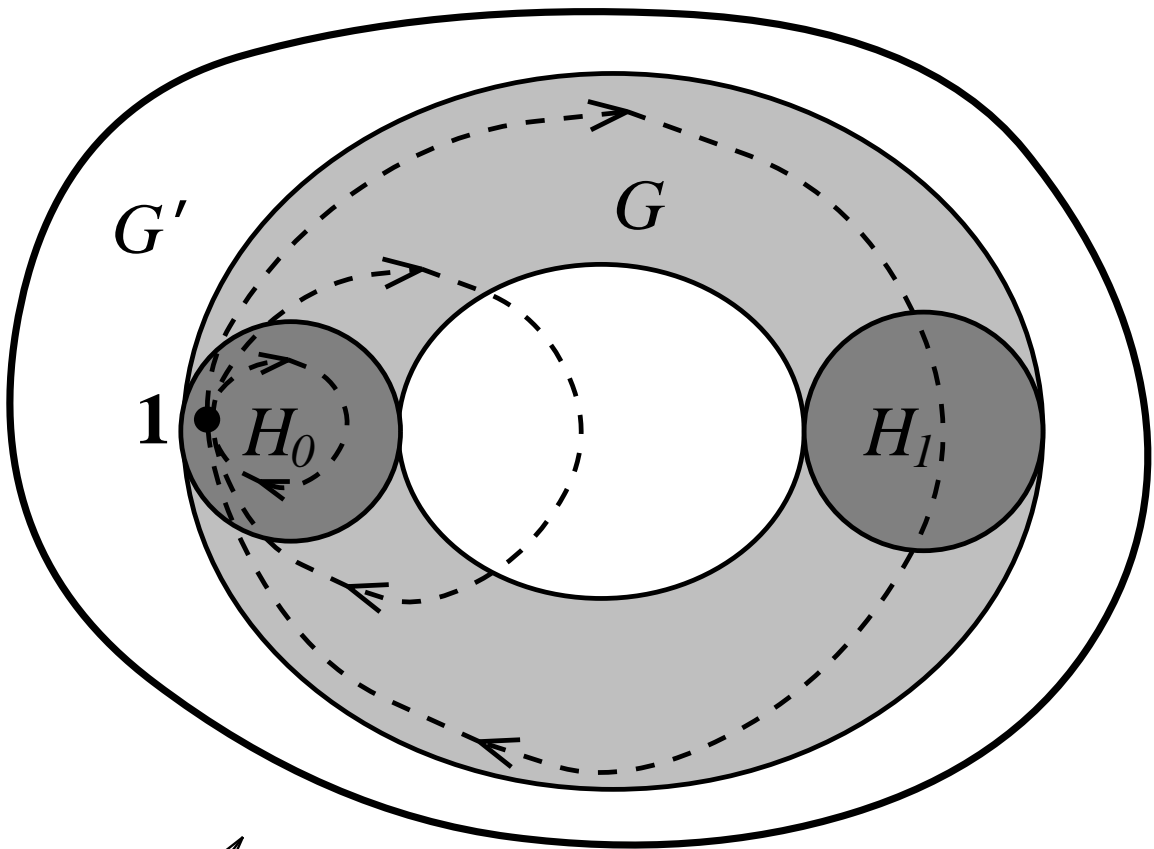
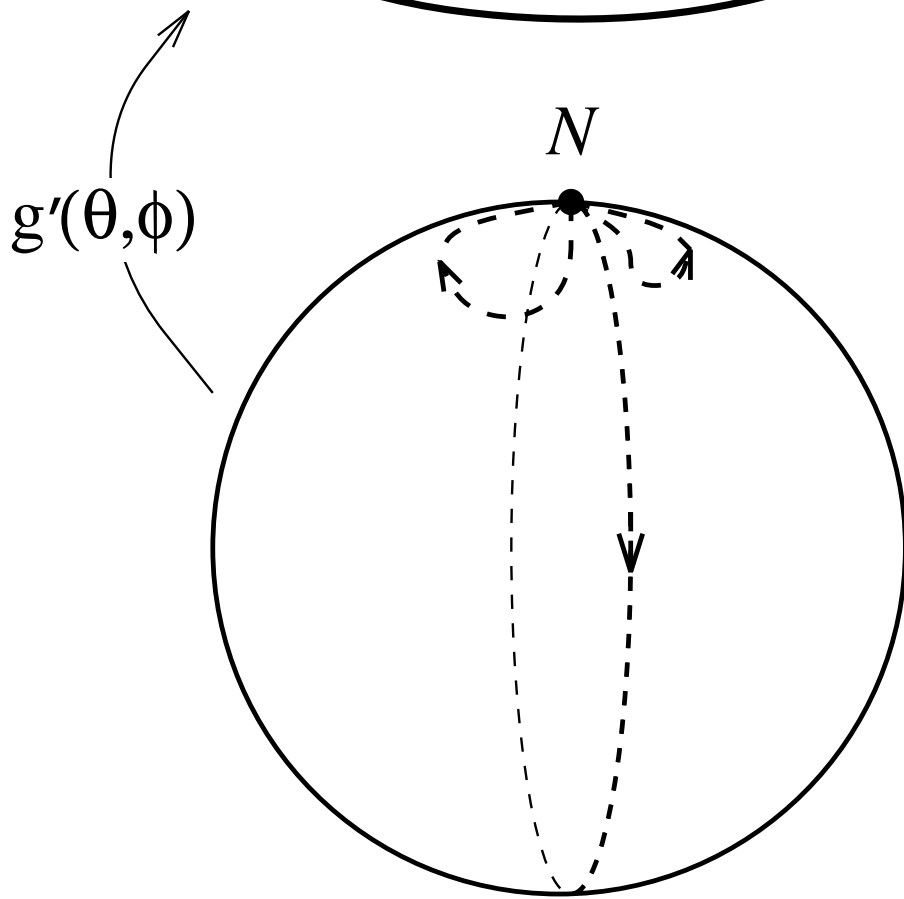
\bye